\documentclass[orivec,runningheads]{llncs}
\usepackage[T1]{fontenc}
\usepackage{graphicx}
\usepackage[ruled,linesnumbered]{algorithm2e}
\usepackage{cite}

\usepackage{amsthm}

\usepackage{enumitem}
\setlist[itemize]{itemsep=0.25em}

\usepackage{mathtools}
\usepackage{amssymb}
\usepackage{mathbbol}
\usepackage{caption}

\newcommand{\GAME}{\mathsf{G}}
\providecommand{\gamedes}[1]{\noindent\underline{\textbf{Game $\GAME_{#1}$:}}}

\DeclareSymbolFontAlphabet{\AMSmathbb}{AMSb}
\DeclareSymbolFontAlphabet{\BBOLmathbb}{bbold}

\ExplSyntaxOn
\bool_new:N \l__mathbb_first_bool

\cs_new_protected:Nn \__mathbb_char:n
{
  \bool_if:NF \l__mathbb_first_bool { , }
  \bool_set_false:N \l__mathbb_first_bool
  \int_compare:nTF { \char_value_uccode:n {`#1} = `#1 }
    { \AMSmathbb{#1} }
    { \BBOLmathbb{#1} }
}

\RenewDocumentCommand{\mathbb}{m}
{
  \bool_set_true:N \l__mathbb_first_bool
  \clist_map_function:nN {#1} \__mathbb_char:n
}
\ExplSyntaxOff

\usepackage{adjustbox}
\usepackage[dvipsnames, table]{xcolor}
\usepackage[svgnames]{xcolor}
\usepackage[operators,sets,asymptotics, probability, advantage, lambda]{cryptocode} % for protocols
\createprocedureblock{procb}{center,boxed}{}{}{}

\usepackage{stmaryrd} %llbracket
\usepackage{multirow} 
\usepackage{makecell}

\usepackage[hidelinks]{hyperref}
\usepackage[capitalise]{cleveref}

\makeatletter
\let\SC@oldappendix\appendix
\renewcommand\appendix{%
  \SC@oldappendix
  \renewcommand*{\theHsection}{appendix.\Alph{section}}%
  \renewcommand*{\theHsubsection}{\theHsection.\arabic{subsection}}%
  \renewcommand*{\theHsubsubsection}{\theHsubsection.\arabic{subsubsection}}%
}
\makeatother
\crefname{theorem}{Theorem}{Theorems}
\Crefname{theorem}{Theorem}{Theorems}
\crefname{lemma}{Lemma}{Lemmas}
\Crefname{lemma}{Lemma}{Lemmas}
\crefname{corollary}{Corollary}{Corollaries}
\Crefname{corollary}{Corollary}{Corollaries}
\crefname{assumption}{Assumption}{Assumptions}
\Crefname{assumption}{Assumption}{Assumptions}
\crefname{claim}{Claim}{Claims}
\Crefname{claim}{Claim}{Claims}

\newcommand{\prover}{\mathcal{P}}
\newcommand{\verifier}{\mathcal{V}}
\newcommand{\simulator}{\mathcal{S}}
\newcommand{\adversary}{\mathcal{A}}
\newcommand{\state}{\mathsf{st}}

\newcommand{\todo}[1]{\textcolor{blue}{\textbf{TODO: #1}}}
\newcommand{\yeoh}[1]{\textcolor{blue}{\textbf{yeoh: #1}}}
\newcommand{\shalti}[1]{\textcolor{pink}{\textbf{shalti: #1}}}

\renewcommand{\todo}[1]{{\errmessage{Error: \noexpand is disabled}}}
\renewcommand{\yeoh}[1]{{\errmessage{Error: \noexpand is disabled}}}
\renewcommand{\shalti}[1]{{\errmessage{Error: \noexpand is disabled}}}

\newcommand{\inlinecomment}[1]{\textcolor{blue}{//#1}}

\newcommand{\vk}{\mathsf{vk}}
\newcommand{\aux}{\mathsf{aux}}
\newcommand{\keygen}{\mathsf{KeyGen}}
\newcommand{\sign}{\mathsf{Sign}}
\newcommand{\verify}{\mathsf{Ver}}
\newcommand{\deriveaddress}{\mathsf{DeriveAdd}}
\renewcommand{\advantage}{\mathsf{Adv}}
\renewcommand{\adversary}{\mathcal{A}} %This cannot be changed.
\newcommand{\eoa}{\mathsf{EOA}}
\newcommand{\hash}{\mathsf{H}}

\newcommand{\indicator}[1]{\mathbf{1}\{{#1}\}}

\newcommand{\mlwe}{\mathsf{MLWE}}
\newcommand{\msis}{\mathsf{MSIS}}
\newcommand{\sis}{\mathsf{MSIS}}
\newcommand{\lwe}{\mathsf{MLWE}}

\newcommand{\name}{\text{Northstar}}
\newcommand{\batchgeneral}{\mathsf{GB}}
\newcommand{\ntt}{\mathsf{NTT}}
\renewcommand{\negl}{\mathsf{negl}(\lambda)}
\newcommand{\NTT}{\mathsf{NTT}}
\newcommand{\INTT}{\mathsf{INTT}}
\newcommand{\coeff}{\mathsf{Coeff}}

\newcommand{\protocol}{\prod}
\newcommand{\polyoracle}[1]{\llbracket{#1}\rrbracket}
\newcommand{\nttcoeffrelation}{\mathsf{NTT}\leftrightarrow\mathsf{Coeff}}
\newcommand{\sk}{\mathsf{sk}}
\newcommand{\pk}{\mathsf{pk}}
\newcommand{\DS}{\mathsf{DS}}
\newcommand{\extract}{\mathsf{Ext}}
\newcommand{\extractorkey}{\mathsf{ek}}

\newcommand{\zksimulator}{\mathsf{SIM}}
\newcommand{\ppgen}{\mathsf{PPGen}}
\newcommand{\nizk}{\mathsf{NIZK}}
\newcommand{\prove}{\mathsf{Prove}}
\newcommand{\publicparam}{\mathsf{pp}}
\newcommand{\provingkey}{\mathsf{nizkpk}}
\newcommand{\verificationkey}{\mathsf{nizkvk}}

\newcommand{\rlwe}{\mathbb{R}\mathbb{L}\mathbb{W}\mathbb{E}}
\newcommand{\rgsw}{\mathbb{R}\mathbb{G}\mathbb{S}\mathbb{W}}
\newcommand{\bsk}{\mathsf{bsk}}
\newcommand{\ksk}{\mathsf{ksk}}
\newcommand{\lut}{\mathsf{lut}}

\newcommand{\commitgen}{\mathsf{CKeyGen}}
\newcommand{\commitkey}{\mathsf{ctk}}
\newcommand{\commit}{\mathsf{Com}}
\newcommand{\open}{\mathsf{Open}}
\newcommand{\commitment}{\mathsf{cm}}
\newcommand{\commitscheme}{\mathsf{COM}}

\newcommand{\numasset}{n_{\mathsf{asset}}}
\newcommand{\token}{\mathsf{Tk}}

\newcommand{\tokenname}{\mathsf{NorthstarToken}}

\usepackage[framemethod=TikZ]{mdframed}
\usepackage{xcolor}
\usepackage{environ}
\usepackage{varwidth}
\definecolor{titleBg}{RGB}{200, 220, 255}
\definecolor{frameColor}{RGB}{0, 0, 0}
\newlength{\contentwidth}
\NewEnviron{titledbox}[1]{%
    \setbox0=\hbox{%
        \begin{varwidth}{50cm}%
            \BODY
        \end{varwidth}%
    }%
    \setlength{\contentwidth}{\dimexpr\wd0+20pt\relax}%
    \begin{mdframed}[
        userdefinedwidth=\contentwidth,
        align=center,
        linewidth=1pt,
        linecolor=frameColor,
        backgroundcolor=white,
        roundcorner=0pt,
        innertopmargin=15pt,
        innerbottommargin=10pt,
        innerleftmargin=10pt,
        innerrightmargin=0pt,
        singleextra={%
            \node[
                fill=titleBg,
                draw=none,
                inner sep=5pt,
                rounded corners=2pt,
                anchor=west,
                xshift=10pt
            ] at (P-|O) {\bfseries #1};%
        },
        firstextra={%
            \node[
                fill=titleBg,
                draw=none,
                inner sep=5pt,
                rounded corners=2pt,
                anchor=west,
                xshift=10pt
            ] at (P-|O) {\bfseries #1};%
        },
    ]
        \BODY
    \end{mdframed}%
}

\usepackage{esvect}
\renewcommand{\vec}[1]{\vv{#1}}

\begin{document}
\title{$\name$: Efficient Verifiable Negacyclic Polynomial Arithmetic with Applications to Encrypted Lattice-based Multi-Asset ERC-20 and Verifiable FHE}
\titlerunning{$\name$}

\author{
Yeoh Wei Zhu\inst{1}\thanks{Corresponding author.} \and
Shaltiel Eloul\inst{1}\thanks{Corresponding author.} \and
Yash Satsangi\inst{1} \and
Rob Otter\inst{1}
}

\authorrunning{Zhu et al.}

\institute{Global Technology Applied Research, JPMorganChase \\ London, UK}

\maketitle              % typeset the header of the 

\begin{abstract}
Lattice-based cryptography can be used to construct a variety of post-quantum cryptographic schemes, ranging from standard cryptographic primitives such as digital signatures and encryption, to fully homomorphic encryption (FHE). However, the verification of correct lattice computations remains relatively expensive, particularly for computationally intensive FHE operations and for resource-constrained environments such as on-chain verification via smart contracts. We propose $\name$, an efficient technique for proving negacyclic arithmetic natively.
When integrated with existing STARK proving frameworks, $\name$ achieves the best overall performance across prover time, verifier time, and proof size compared to prior state-of-the-art approaches for verifiable FHE constructions. Furthermore, $\name$ is used to construct the first verifiable encrypted multi-asset ($\sim 1000$ assets) variant of ERC-20 that
is efficiently verifiable by a smart contract, and is practically deployable in the current blockchain ecosystem.

% \keywords{Blockchain  \and Tokenization \and Multi-asset \and Post Quantum}
\end{abstract}

\section{Introduction}
Classical public-key cryptography based its security on the presumed computational intractability of problems such as integer factorization \cite{rsa1978} and the discrete logarithm \cite{diffie_hellman}. However, Shor's algorithm \cite{Shor1995-mn} renders both problems solvable in polynomial time on a sufficiently powerful quantum computer, thereby undermining the security foundations of widely deployed cryptographic schemes. This vulnerability has motivated extensive research into post-quantum cryptographic constructions. Among the most prominent candidates are those based on lattice problems such as Learning with Errors (LWE) and the Short Integer Solution (SIS) problem \cite{regev_lattice}, which are conjectured to remain hard even against quantum adversaries.
For applications requiring functionalities beyond basic encryption or signature, it becomes important to provide efficient verifiability schemes performed over these cryptographic primitives. Some of the standard approaches to achieving verifiability include zero-knowledge proofs (ZKP) based on sigma protocols \cite{hazay2010sigma}, general-purpose succinct proof systems (zkSNARKs \cite{snark_bitansky2012extractable,groth2016size}, zkSTARKs \cite{ben2018scalable}), and techniques derived from secure multi-party computation (MPC) \cite{ishai2007zero}. Among these, general-purpose succinct proof systems have gained considerable traction, enabling a range of applications including zero-knowledge virtual machines (zkVMs)\cite{Laub2026risc0,cairo}, zkRollups \cite{chaliasos2025towards}, and verifiable fully homomorphic encryption (FHE) \cite{zama_TremblayThibault2025}.

In lattice-based cryptographic schemes, it is required that certain ring element has small norm when represented in the coefficient domain of the underlying polynomial ring, as this norm bound constitutes the core hardness assumption for both MLWE and MSIS. On the other hand, efficient polynomial ring multiplication is computed by first transforming operands into the Number Theoretic Transform (NTT) domain and then performing component-wise multiplication. Consequently, the proof system must encode polynomials in both the NTT and coefficient representations and establish their consistency. We denote this consistency relation as $\nttcoeffrelation$ throughout the paper.
The straightforward approach to verify the correctness of the $\nttcoeffrelation$ is to perform the NTT computation in-circuit via butterfly \cite{cooleytukey} operations. However, such in-circuit non-native arithmetic incurs substantial overhead as evidenced by the STARK
-based approach \cite{zama_TremblayThibault2025}. Recently, \cite{hasteboot} identified the inherent inefficiency of encoding butterfly operations within the univariate or multivariate polynomial setting of STARK-based ZkVM systems, and proposed instead to verify the $\nttcoeffrelation$ relation via a multilinear sumcheck protocol, by expressing the NTT as a multilinear encoding of a matrix–vector product.
In this work, we demonstrate how to achieve comparable (or better) performance improvements within the STARK-based univariate proof framework. Importantly, our technique can be applied as a modular extension to STARK enabling efficient verifiable negacyclic polynomial multiplication, analogous to how the LogUp argument \cite{habock2022multivariate} is used to extend STARK with efficient lookup table functionality.

\subsubsection{Contribution.}
We propose $\name$, a general technique for efficiently establishing the equivalence between two trace polynomials that encode the NTT representation and the coefficient representation of the same underlying quotient ring polynomial. The technique integrates seamlessly with existing univariate STARK-based proving systems without requiring significant modifications to their core architecture. The encoding admits an unstructured layout: the representations may be placed at arbitrary positions within the trace row, provided that (i) the placement is known to the verifier and (ii) the trace polynomial contains the complete representation. We summarize the contributions as follows:

\begin{enumerate}
    \item \textbf{$\name$: A Novel PIOP for $\mathcal{R}_{\nttcoeffrelation}$ using univariate sumcheck.} We propose a scalable and efficient PIOP for the relation $\mathcal{R}_{\nttcoeffrelation}$ that enables efficient arithmetic operations over quotient rings $R_q = \mathbb{Z}_q[X]/(X^N+1)$. Our approach avoids in-circuit butterfly operations and directly equates two trace polynomials, the encoding the coefficient representation and the NTT representation of the same underlying quotient ring. 
    \item \textbf{Best Overall Verifiable TFHE.} Using the proposed PIOP, we construct an efficient verifiable TFHE programmable bootstrapping operation (PBS). When compared to the previous state-of-the-art approach, our approach manages to be efficient in prover time, verifier time, and communication simultaneously while the previous approach is inefficient in at least one metric across different PCS configurations.
    \item \textbf{Novel Encrypted Multi-Asset ERC-20 / ERC-1155.} To further showcase the practicality of $\name$, we construct
    the first encrypted multi-asset ERC-20 that can simultaneously encode  $\sim 1000$ assets while remaining both efficiently verifiable on-chain and efficiently provable locally.
\end{enumerate}

\subsection{Related Works.}

\textbf{Generic Arithmetic Circuit.} 
The correct computation of FHE can be generically proven by encoding the computation steps as an arithmetic circuit representations such as QAP/R1CS \cite{Gennaro2013,BenSasson2013}.  
The work in \cite{zama_TremblayThibault2025} follows the generic approach and constructs a verifiable TFHE scheme with bootstrapping, but the overhead of in-circuit NTT results in a substantial prover running time of nearly half an hour for a verifiable PBS operation, making it impractical.

\textbf{Specialized Protocols.}
Recent work by \cite{Cascudo2025} uses lookup-based techniques and multilinear sumcheck to tackle the problem of verifiable FHE, but it falls shorts of supporting bootstrapping and only supports the leveled CKKS FHE scheme \cite{ckks_Cheon2017}.
For the TFHE scheme \cite{Chillotti2019}, the work of \cite{hasteboot} proposes a verifiable FHE with bootstrapping support that achieves efficiency in only some partial metrics depending on the polynomial commitment scheme (PCS) configuration. 
However, these techniques work over multilinear sumcheck protocol and are non-native to the univariate/multivariate-based proving systems \cite{ben2018scalable,deepfri,circlestark2024}. 
While it is possible to relate the committed univariate polynomial and the  multilinear polynomial through a scheme like \cite{dualpcs2024}, this incurs additional overhead which is undesirable and the provided construction works over KZG commitment \cite{kzg10}. 
% \yeoh{give a more convincing reason, since matrix vector multiplication should be ok over univariate sumcheck as well.}

\textbf{Other Approaches.} 
Instead of proving the computation using a proving system, there exists another line of work \cite{Gama2025,Garg2024,fhesnark} that homomorphically computes the proof through HE. However, this approach leaks at least one bit of plaintext information, as discussed in \cite{Aranha2024} which violates the standard IND-CPA security of existing FHE schemes.
Zero-Knowledge Proof (ZKP) based on sigma protocols \cite{hazay2010sigma} has been used to prove various relations of the quotient ring polynomial for lattice-based cryptographic constructions \cite{LNPS21,matrict_2_2022,qpadl}. However, sigma protocols are generally not succinct and non-native arithmetic relations cannot be efficiently verified using sigma protocols. It is also possible to achieve similar private-computation objectives efficiently using trusted execution environment (TEE) \cite{li2024sok}, though TEE has many practical design issues that compromise its integrity and privacy guarantees \cite{tee_fail,rauscher2026telescope,de2025badram}.

\section{Technical Overview}
\subsection{NTT and Coefficients Relations: Previous Approaches}
We require both the NTT domain representation for linear-time multiplication and the coefficient-domain representation to check for a short norm in a typical verifiable lattice-based cryptographic construction. Let $\vec f_{\ntt}$ and $\vec f_{\coeff}$ be the NTT-domain vector and coefficient-domain vector respectively.

\textbf{Emulating the Butterflies.} One straightforward way to check that the claimed NTT representation and coefficient representation have the correct relation is to simply check that $\vec f_{\NTT}= \ntt(\vec f_{\coeff})$ in an arithmetization circuit. However, it is generally regarded as a concretely expensive operation to compute in an arithmetization circuit format as noted in \cite{hasteboot}. It can be seen that for the application of verifiable FHE, computing using in-circuit butterfly operations lead to a substantial performance penalty as demonstrated in \cite{zama_TremblayThibault2025}. 

\textbf{Sumchecking Matrix Multiplication.} The authors of \cite{hasteboot} propose using a multilinear sumchecking technique to perform the relation check by expressing the NTT transform as a matrix-vector multiplication $$\tilde f_{\NTT}(Y) = \sum_{x \in \{0,1\}^{\log N}}\tilde f_{\coeff}(x) \cdot \tilde E(Y,x)$$ where $\tilde f$ represents the multilinear extension \cite{thaler2022proofs} of $f$, evaluation point matrix $E(Y,X)=\omega^{(2Y+1) \cdot X}$ and its multilinear extension $\tilde E$,
% \shalti{$\tilde E$?}
and primitive $2N$-th root of unity $\omega$. The matrix multiplication is a straightforward evaluation at the twisted points (see \cref{sec:negacyclic_background}). However, the matrix construction of $E$ and the matrix evaluation to obtain the $\NTT$ form can be improved.

\subsection{$\name$: Proving $\nttcoeffrelation$ with Univariate Sumcheck}
\textbf{Initial Solution using Random Evaluation and Sumcheck. } To provide support for the verifiable $\nttcoeffrelation$ relation in the univariate setting, we start by looking at the univariate sumcheck over a multiplicative subgroup from Aurora \cite{BenSasson2019}, which has the following form:
$$f(X)= \frac{\beta}{|H|} + R(X)X + Q(X)Z_H(X)
$$
where $\deg(f) = d \geq |H|$, $\deg(R) \leq |H|-2$ and $\deg(Q) \leq d - |H|$. Using \cref{lemma:uni_sumcheck}, the above shows that the sum over the multiplicative subgroup H, $\sum_{h_i \in H} f(h_i)=\beta$ is found at the constant term scaled by a factor $\frac{1}{|H|}$ where $Z_H(X)$ is a divisor , $Q(X)$ is a quotient and $R(X)$ is a remainder polynomial.

 Given the quotient ring polynomial $f_{\mathcal{R}_q}(X) = c_0+c_1X^1+\cdots+c_{N-1}X^{N-1}$, the coefficient $\vec f_{\coeff}$ and NTT $\vec f_{\ntt}$ representations are encoded as the trace polynomial $f_{\coeff}(X)$ and $f_{\ntt}(X)$ respectively where $j \in \{\coeff,\ntt\}, f_{j}(h_i)= \vec f_{j}[i]$. Let $f_{\coeff}(X)$ and $f_{\ntt}(X)$ be committed under any PCS. We now check equality by sampling a random challenge point $\alpha$ and then check the following:
$$
f_{\coeff}(X) \cdot \sigma_\alpha(X) = \frac{f_{\NTT}(\alpha)}{|H|} + R(X)X + Q(X)Z_H(X)
$$
where $\sigma_{\alpha}(X)$ is the trace polynomial whose $i$-th trace row encodes $\alpha^{i-1}$. It can be seen that if we use the modified trace domain $H'=\{\psi\phi^{i-1}\}_{i \in [|H|]}$, then by using \cref{lemma:uni_sumcheck} and
% $\sum_{h_i \in H} f_{\coeff}(h_i) \cdot \sigma_\alpha(h_i) = $
$\sum_{h_i \in H} f_{\coeff}(h_i)\cdot\sigma_\alpha(h_i) = \sum_{j=0}^{N-1} c_j\cdot\alpha^{j} = f_{\mathcal{R}_q}(\alpha)= f_{\ntt}(\alpha)$, the above equality is correct. The soundness follows from the Schwartz-Zippel lemma. However, the above approach requires the use of non-standard twisted trace domain $H'$ and the approach does not work when we transition to zero-knowledge setting \cite{zkstark_habock2024note} in which the trace polynomial is typically masked as $\tilde f_{\ntt}(X)= f_{\ntt}(X) + m(X)Z_H(X)$ and we are left with an extra term $m(\alpha)Z_H(\alpha)$. 

\textbf{Equality through polynomial evaluation and Lagrange evaluation on NTT representations.}
We now work with the standard trace domain $H$ and make use of interpolation to close the gap above.
Based on Lagrange interpolation detailed in \cref{sec:background}, we have the following identity
$$
f_{\mathcal{R}_q}(X) = \sum_{h_i \in H} {f}_{\NTT}(h_i)\cdot \ell_{\psi  \phi^{i-1}}(X) 
$$
where $\ell_{\psi \phi^{i-1}}(X)$ is the Lagrange interpolation basis.
Therefore, we can rewrite the sumcheck as follow:
\begin{align}
\label{eq:ntt_sumcheck}
    f_{\coeff}(X) \cdot \sigma_\alpha(X) - f_{\NTT}(X) \cdot \Lambda_{\alpha}(X) = 0 + R(X)X + Q(X)Z_H(X)
\end{align}
where $\Lambda_{\alpha}$ is the polynomial that interpolates over $H$ such that $\Lambda_{\alpha}(h_i) = \ell_{\psi h_i}(\alpha)$. The soundness follows from Schwartz-Zippel lemma where $f_{\coeff}$ and $f_{\NTT}$ are committed to before the challenge $\alpha$ is sampled. We now check that $R(X)$ is low-degree and test the equality \cref{eq:ntt_sumcheck} at a random point as described in the univariate sumcheck protocol from \cite{BenSasson2019}.

\textbf{Adding Zero-Knowledge and Batching.} 
The approach described above can be made zero knowledge with the standard algebraic masking technique. 
Given a committed masking polynomial $m(X)$ with a claimed sum $\sum_{h_i \in H} m(h_i)=\beta_{m}$, the verifier samples a masking randomization factor $\rho$, and then the prover and verifier perform a sumcheck protocol over the following:
{
\setlength{\abovedisplayskip}{2pt}
\setlength{\belowdisplayskip}{2pt}
\setlength{\abovedisplayshortskip}{0pt}
\setlength{\belowdisplayshortskip}{0pt}
\begin{align*}
\label{eq:ntt_sumcheck_zk}
    \rho (\tilde f_{\coeff}(X) \cdot \sigma_\alpha(X) - \tilde f_{\NTT}(X) \cdot \Lambda_{\alpha}(X)) + m(X) = \frac{\beta_m}{|H|} + R(X)X + Q(X)Z_H(X).
\end{align*}
}

\textbf{Generalized Placement and Batching.} We further show that batching multiple quotient ring polynomials into one trace polynomial and enforcing the $\nttcoeffrelation$ relation for multiple trace polynomials can be done efficiently using a random linear combination at \cref{sec:poly_stacking}. Our technique does not require the sequential ordering of the encoded polynomial coefficients and supports arbitrary, publicly known placement within the trace rows, enabling greater flexibility when designing the constraints.

\subsection{Extension for STARK framework and Applications to Verifiable FHE and Verifiable Encrypted ERC-20}
At the end of $\name$ protocol, the verifier needs to check the equality in \cref{eq:ntt_sumcheck} on an (out-of-domain) random point $z$ and verify that $R(X)$ is a low-degree polynomial. Since the polynomials are committed using a FRI-based univariate PCS in the univariate STARK framework, evaluations on a specified random point $z$ are given as claim and checked through the deep quotient mechanism \cite{deepfri} in the STARK framework which reduces the verification to low-degree testing. Meanwhile, the low degree property of $R(X)$ can be efficiently checked using the FRI low-degree testing in the STARK framework through random-linear-combination batching \cite{cryptoeprint:2022/1216} together with the existing low-degree polynomials. Therefore, $\name$ incurs minimal overhead in practice. The overview is given in \cref{fig:architecture_overview}.
To showcase the practicality of $\name$, we construct a new verifiable (T)FHE scheme and a new verifiable encrypted ERC-20 scheme. 

{
\setlength{\intextsep}{2pt}
\setlength{\textfloatsep}{2pt}
\setlength{\floatsep}{2pt}
\begin{figure}
    \centering
    \includegraphics[width=0.6\linewidth]{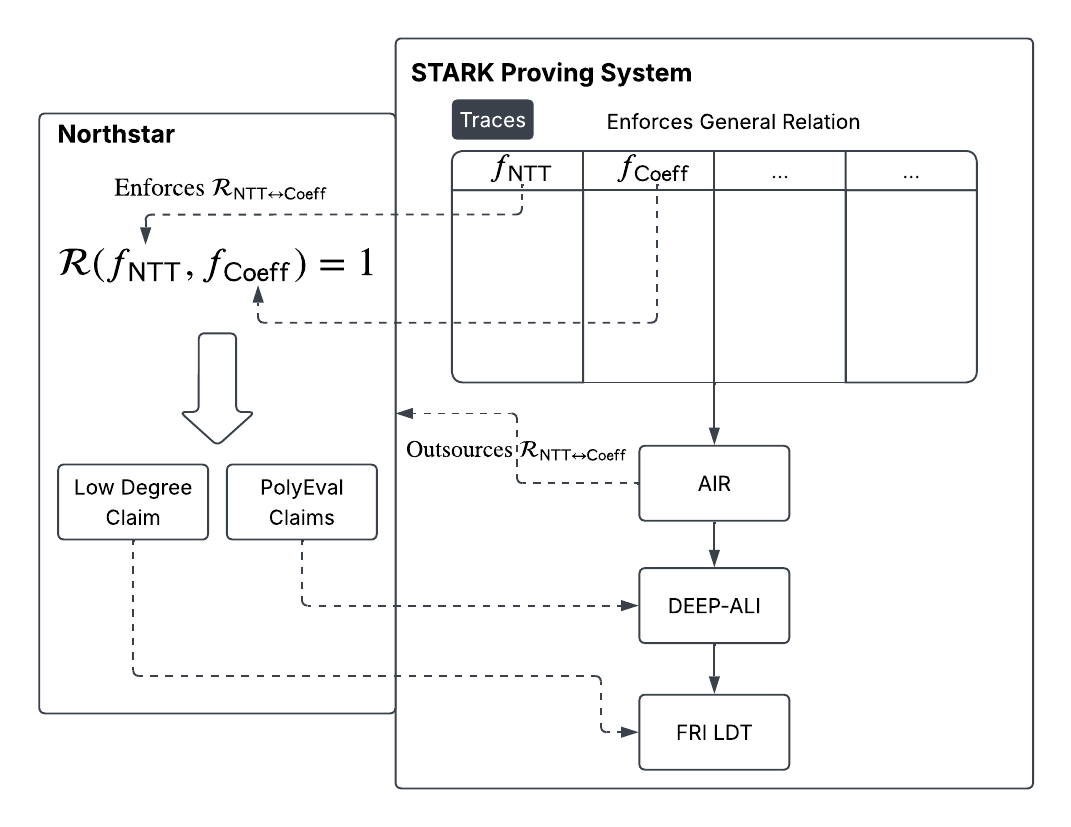}
    \caption{Architecture Overview for Verifiable Computing using $\name$.}
    \label{fig:architecture_overview}
\end{figure}
}

\textbf{Verifiable FHE.} We show that using $\name$ together with the Plonky3 STARK proving system yields overall improvements in verifiable FHE task. Concretely, we achieve for the first time a short prover time, verifier time and proof size simultaneously where the previous state-of-the-art (SoTA) \cite{hasteboot} exhibits inefficiency in at least one of these metrics under different PCS configurations.

\textbf{Encrypted Multi-Asset ERC-20.} To the best of our knowledge and for the first time, we show a construction of a confidential multi-asset token (encrypted multi-asset ERC-20 or encrypted ERC-1155) that supports transactions involving thousand(s) of assets at the same time while simultaneously possess performance profile similar to that of a single-asset transaction.

\section{Preliminaries}
\label{sec:background}
\textbf{Syntax.} We use $[n]$ to denote the set $\{1,2,\ldots,n\}$, the syntax for $A[i]$ to denote accessing the $i$-indexed list element of $A$, and we use $[a,b]$, $(a,b)$, $(a,b]$, $[a,b)$, to represent close, open, and half open intervals respectively. Let $\indicator{\cdot}$ denote the indicator function.
Let $\mathbb{F}$ be a field, and let $H = \langle \omega \rangle = \{\omega^i : i \in [0,n_H-1]\}= \{h_1,\ldots,h_{n_H}\}$ be a multiplicative subgroup of order $|H|$ where $\omega$ is the primitive $|H|$ root of unity and $n_H = |H|$. We consider $|H|=2^k$ for some positive integer $k$. 
Let $Z_H(X)=X^{n_H}-1$ be a vanishing polynomial on points specified in $\{h_i : h_i \in H\}$ such that $Z_H(h_i)=0$ and $Z_H(x) \neq 0$ where $x \notin H$.  

An \textit{indexed relation} $\mathcal{R}$ is a set of tuples $(\mathbb{i}, \mathbb{x}; \mathbb{w})$ where $\mathbb{i}$ is the public index fixed at setup time, $\mathbb{x}$ is the public statement, and $\mathbb{w}$ is the private witness. The language $\mathcal{L(\mathcal{R}})$ associated to a relation $\mathcal{R}$ is the set of pairs $(\mathbb{i}, \mathbb{x})$ such that $\exists \mathbb{w}, (\mathbb{i}, \mathbb{x}; \mathbb{w}) \in \mathcal{R}$. We use $f(X)$ to denote a polynomial and sometimes omit $(X)$ and use $f$ if the context is clear. An oracle for a polynomial $f$ is denoted by $\polyoracle{f}$.
We recall polynomial interactive oracle proofs (PIOPs) in the Appendix \ref{app:piop_background}.

\begin{lemma}[Schwartz-Zippel\cite{Schwartz1980,Zippel1979}] 
\label[lemma]{lemma:schwartz-zippel}
Let $g : \mathbb{F}^{m} \rightarrow\mathbb{F}$ be a non-zero $m$-variate polynomial of total degree at most $d$ over a field $\mathbb{F}$. Then for any finite set $S \subseteq\mathbb{F}$, $\Pr[g(r)=0 : r \sample S^m] \leq d/|S|$. 
\end{lemma}

In this paper, we consider the univariate case where $m=1$. The polynomial can be evaluated at a random point $r$ using Horner's method in $O(d)$ time.

\noindent\textbf{Lagrange Interpolation. } We recall the Lagrange interpolation using Lagrange basis polynomials for a polynomial of degree at most $n-1$. Let $((x_1,y_1),\ldots, (x_{n},y_{n})) \in (\mathbb{F}\times\mathbb{F})^{n}$ with distinct x-coordinates where $x_i \neq x_j$ for $i\neq j$. Then, there exists a unique polynomial $p(X) \in \mathbb{F}[X]$ of degree at most $n-1$ such that $\forall i\in [n], p(x_i)=y_i$. The polynomial can be interpolated using the formula $p(X)=\sum_{i=1}^{n} y_i \ell_i(X)$ where $\ell_i$
is the Lagrange basis polynomial $\ell_i(X)=\prod_{1\leq j \leq n, j \neq i} \frac{X-x_j}{x_i-x_j}$. The evaluation of the interpolating polynomial can be computed in $O(n)$ field operations 
using the method from \cite{Berrut2004} 
by expressing the interpolation as the barycentric second form $p(X)= (\sum_{i=1}^{n} \frac{w_i}{X-x_i}y_i)$/$(\sum_{i=1}^{n} \frac{w_i}{X-x_i})$ where $w_i=\frac{1}{\prod_{j\neq i} (x_i-x_j)}$. Note that the Lagrange basis polynomial can be rewritten as $\ell_i(X)=\ell(X)\cdot \frac{w_i}{X-x_i}$ where $\ell(X)=(X-x_1)(X-x_2)\ldots(X-x_n)$. For $\{x_1, \ldots,x_n\} = H$, $\ell(X)=X^{|H|}-1$.
The preprocessing of $w_i$ terms can be done in $O(n^2)$ time during setup given the terms $\{x_i\}_{i \in [n]}$ is fixed as is the case for our construction. The inversion can be computed faster using  batch inversion.
For notational convenience, we use another equivalent notation for the Lagrange basis polynomial $\ell_{x_i}(X) = \prod_{x_j\neq x_i} \frac{X-x_j}{x_i-x_j}$ when the interpolating set $\{(x_i,y_i)\}_{i\in [n]}$ is clear.

\subsection{Negacyclic Polynomial Multiplication and NTT}
\label{sec:negacyclic_background}

% Zero-indexing
Let $f(X)$ and $g(X)$ be polynomials in the quotient ring $\mathcal{R}_q=\mathbb{Z}_q[X]/(X^N+1)$ with degree $\leq N-1$. Then, the product $p(X)=f(X) \cdot g(X) = \sum_{k=0}^{N-1} p_k X^{k}$ is the negacyclic convolution defined as 
$p_k = \sum_{i=0}^{k} f_ig_{k-i}- \sum_{i=k+1}^{N-1} f_ig_{k+N-i} \mod q.$ 
For a fully splitting ring where $q \equiv 1 \mod 2N$, the negacyclic convolution can be computed using pointwise multiplication in the NTT domain. Let $\phi$ be an $N$-th primitive root of unity and $\psi$ be a $2N$-th primitive root of unity for $\mathbb{Z}_q$, and $\phi=\psi^2$. The NTT representation of a polynomial $f(X)$ can be defined as $\vec f_{\ntt}=(\widehat f_{1},\cdots,\widehat f_{N})$ where $\widehat f_i = f(\psi \phi^{i-1})= f(\psi^{2(i-1)+1})$. Then, the product can be computed as $f(X)\cdot g(X)= \INTT(\vec f_{\NTT} \circ  \vec g_{\NTT})$ where $\circ$ represents pointwise multiplication and $\INTT$ is the inverse NTT transform. The NTT transform can be computed in $O(N \log N)$ steps using Fast-Fourier transform style algorithms such as Cooley-Tukey \cite{cooleytukey} for NTT and the Gentleman-Sande \cite{gentlemansande} for INTT butterfly operations.

\subsection{Univariate Sumcheck for Multiplicative Cyclic Group}

It is known that $\sum_{h_i \in H} f(h_i)=|H|\cdot f(0)$ if $\deg(f(X))\leq |H|-1$ and $H$ is a multiplicative subgroup (Remarks 5.6 at \cite{BenSasson2019}). The aforementioned property is used in \cite{Lipmaa2022,Zhang2020} to construct zkSNARK proving systems. The property is captured by the following two lemmas. For completeness, we provide proofs for the lemmas below in the appendix since it is not found in \cite{BenSasson2019}.

\begin{lemma}[Sum and Scaled Constant Coefficient] 
\label[lemma]{lemma:sum_over_h_constant}
Let $f \in \mathbb{F}[X]$ be a degree $\leq |H|-1$ polynomial, then $\sum_{h_i \in H} f(h_i)=|H| \cdot f(0)$.
\end{lemma}

\begin{lemma}[Higher Degree Sum Relation] 
\label[lemma]{lemma:uni_sumcheck}
Let $f \in \mathbb{F}[X]$ be a polynomial with degree $d \geq |H|$ , $\sum_{h_i \in H} f(h_i)=\beta$ iff there exists $R(X) \in \mathbb{F}[X]$ and $Q(X) \in \mathbb{F}[X]$, such that $f(X)= \frac{\beta}{|H|} + R(X)X + Q(X)Z_H(X)$ where $\deg(R) \leq |H|-2$ and $\deg(Q) \leq d - |H|$.
\end{lemma}

\subsection{STARK Overview: AIR, DEEP-ALI and Low-Degree Testing}
\label{sec:background_stark}
We give a STARK overview based on \cite{cryptoeprint:2022/1216}, in particular we briefly discuss the DEEP algebraic linking (DEEP-ALI) \cite{deepfri} and how it can be used to prove the satisfiability of algebraic intermediate representations (AIR). An AIR is defined over the trace domain $H$ with generator $\omega$ and each $h_i=\omega^{i-1}$ in $H$ encodes a ``row'' of $W$ columns witnesses forming the trace polynomials $(p_1(X),\ldots,p_W(X))$ where $p_j(h_i)$ encodes the $j$-th column witnesses for the $i$-th row.
The polynomials are committed using hash-based Merkle-tree over the FRI evaluation domain $D$.
The constraint is defined to be $CP_1, \ldots, CP_{C}$, and the satisfiability of the AIR is defined by $\forall i \in [C], \forall x \in a_i \cdot H_i, CP_i(p_1(x),\ldots,p_W(x),p_1(\omega x), \allowbreak \ldots, p_W(\omega x))=0$ where $H_i$ is a subgroup of H, $a_i \cdot H_i$ is a specified coset, and only neighboring rows are considered.
In DEEP-ALI, the verifier samples a random point $z \sample \mathbb{F}\setminus(D\cup H)$ outside the FRI evaluation domain $D$ and trace domain $H$. 
Let the claimed evaluations of a polynomial be $(v_{i,1},\ldots v_{i,m})$ at points $(z_1,\ldots,z_m)$, and let $V_i(X)= \sum_{j \in [m]} v_{i,j} \cdot \ell_{z_j}(X)$ be the unique polynomial that interpolates the claim. 
The prover then sends the claimed evaluations $(z, v_{i,1}), (\omega \cdot z, v_{i,2})$ for $i \in [W]$ of the witness polynomials $p_i(X)$ together with claimed evaluations $v$ of $q_{\lambda}(X)$ at $z$. 
Then, the verifier checks using the claimed evaluations at point $z$ on the identity $q_{\lambda}(X)= \sum_{i\in[C]} \lambda^{i-1} \frac{CP_i(p_1(X),\ldots,p_{W}(X),p_1(\omega X), \ldots, p_{W}(\omega X))}{Z_{a_i \cdot H_i}(X)}$ where $\lambda$ is a challenge, $Z_{a_i\cdot H_i}$ is the vanishing polynomial over the coset $a_i\cdot H_i$ using the claimed evaluations. Finally, the validity of the claimed evaluations is checked by running batched FRI for the proximity of the evaluation quotients $\forall i \in [W],\frac{p_i(X)-V_i(X)}{(X-z)(X-\omega z)}$ and $\frac{q_{\lambda}(X)-v}{X-z}$ to low-degree polynomials.

\section{$\name$: Efficient PIOP for the relation $\mathcal{R}_{\nttcoeffrelation}$ }

A key challenge in proving verifiable lattice-based cryptographic constructions is that we need to have access to the coefficients for enforcing norm shortness and have access to NTT representation to perform polynomial multiplication efficiently. While we can verify the cryptographic computation using a generic proving framework like zkSTARK \cite{Lubarov2026Plonky3,ben2018scalable,ethstark}, relating the coefficient representation to the NTT representation is costly via in-circuit arithmetic. We introduce a new relation $\mathcal{R}_{\nttcoeffrelation}$ and propose a protocol $\protocol_{\nttcoeffrelation}$ that efficiently checks the relation. We construct and analyze protocols in the PIOP abstraction where the polynomials are provided as oracles and rely on the compilation described in \cref{sec:compilation} to replace polynomial oracles with a concrete PCS.
We start with the simple case where the quotient ring polynomial $f(X)$ have its coefficient and NTT representations encoded as $f_{\coeff}$ and $f_{\ntt}$ respectively using the AIR encoding from \cref{sec:background_stark}.
We define the relation $\mathcal{R}_{\nttcoeffrelation}$  formally below.

\begin{definition} [Relation $\mathcal{R}_{\nttcoeffrelation}$]
\label{def:ntt_relation}
For a trace domain $H$ encoded quotient ring polynomial $f(X) = \sum_{i=0}^{N-1} f_iX^{i} \in \mathbb{Z}_q[X]/(X^N+1)$ such that $\forall i \in [N], f_{\coeff}(h_i)= f_{i-1}$, and $h_i \in H$, the relation $\mathcal{R}_{\nttcoeffrelation}$ is defined as
\begin{align*}
\mathcal{R}_{\nttcoeffrelation} = \left\{ 
\begin{array}{c}
     \bigl( (N,H,\psi, \polyoracle{f_{\ntt}}, \polyoracle{f_{\coeff}}) ; (f_{\ntt}, f_{\coeff})\bigr): \\
    % f_{\coeff}(h_i) = \sum_{i=0}^{N-1} f_iX^i, f_{\ntt} = (\widehat f_1,\ldots,\widehat f_{N}), \\
    \forall i \in [N],  f_{\ntt}(h_i) = \sum_{j \in [N]} f_{\coeff}(h_j)\cdot(\psi^{2(i-1)+1})^{j-1}; \\
    |H| > 1; N=|H|
\end{array}
\right\}
\end{align*}
where $\psi$ is a $2N$-th primitive root of unity for $\mathbb{Z}_q$, $H$ is the trace domain set and $h_j \in H$. 
\end{definition}
The relation above checks the simple case where the encoding is ordered and exact, the generalization to arbitrary placement is given in \cref{sec:poly_stacking}. 
We construct a PIOP protocol $\protocol_{\nttcoeffrelation}$ for $\mathcal{R}_{\nttcoeffrelation}$ using similar framework as \cite{BenSasson2019}. The protocol is described in \cref{proto:proto_ntt} and we define challenge and interpolating polynomials $\sigma_\alpha(X)$ and $\Lambda_\alpha(X)$ below. 

\begin{definition}[Challenge Polynomial $\sigma_{\alpha}(X)$ and Lambda Lagrange Basis Polynomial $\Lambda_{\alpha}(X)$]
\label{def:challenge_lagrange_poly}
For a trace domain H, a challenge $\alpha \in \mathbb{F}$, a $2N$-th primitive root of unity $\psi$ and an $N$-th primitive root of unity $\phi$, we denote the challenge polynomial as $\sigma_{\alpha}(X)$ where $\forall i \in [|H|], \sigma_{\alpha}(h_i)= \alpha^{i-1}$ and $h_i \in H$. In addition, we denote the lambda Lagrange basis polynomial as $\Lambda_\alpha(X)$ where  $\forall i\in [N],\Lambda_{\alpha}(h_i) = \ell_{\psi \phi^{i-1}}(\alpha) = \ell_{x_{i}}(\alpha)$ and $\ell_{x_{i}}(X)= \prod_{1\leq j \leq |H|, j \neq i} \frac{X-x_j}{x_i-x_j}$ is the Lagrange basis for interpolation. 
\end{definition}

{
\setlength{\intextsep}{0pt}
\setlength{\textfloatsep}{0pt}
\setlength{\floatsep}{0pt}
\begin{figure}
    \begin{adjustbox}{minipage=1.176\linewidth,scale=1, center}
    \begin{titledbox}{Protocol $\protocol_{\nttcoeffrelation}$ for $\mathcal{R}_{\nttcoeffrelation}$}
    \begin{pcvstack}
    \begin{enumerate}
        \item $\prover$ sends $\polyoracle{f_{\ntt}}, \polyoracle{f_{\coeff}}$
        \item $\verifier$ sends challenge $\alpha \sample \mathbb{F}$
        \item $\prover$ sends $\polyoracle{R}, \polyoracle{Q}$
        \item $\verifier$ samples $z \sample \mathbb{F}$, queries $\polyoracle{f_{\ntt}}, \polyoracle{f_{\coeff}}, \polyoracle{Q}, \polyoracle{R}$ at point $z$, and computes $\sigma_\alpha(z)$, $\Lambda_{\alpha}(z), Z_H(z)$
    \end{enumerate}
    $\verifier$ accepts if $f_{\coeff}(z)\cdot \sigma_\alpha(z) - f_{\ntt}(z)\cdot \Lambda_{\alpha}(z)= R(z)z + Q(z)Z_H(z)$
    \\\phantom{accepts if} $ \wedge \deg(R) \leq |H|-2 \wedge \deg(Q)\leq (2N-2)-|H| \wedge N=|H| \wedge |H|>1$
    \end{pcvstack}
    \end{titledbox}
    \end{adjustbox}
    \caption{
    PIOP Protocol $\protocol_{\nttcoeffrelation}$ for $\mathcal{R}_{\nttcoeffrelation}$ 
    }
    \label{proto:proto_ntt}
\end{figure}
}

\begin{theorem}
\label{theo:main_nongeneral}
The protocol $\protocol_{\nttcoeffrelation}$ described in \cref{proto:proto_ntt} is complete, and knowledge-sound PIOP for the relation $\mathcal{R}_{\nttcoeffrelation}$ (\cref{def:ntt_relation}).
\end{theorem}

\begin{proof}
\textbf{Completeness.} We first show completeness. Let $\bigl((N,H,\psi, \polyoracle{f_{\ntt}}, \polyoracle{f_{\coeff}}) ; \allowbreak (f_{\ntt}, f_{\coeff})\bigr) \in \mathcal{R}$ and therefore $\forall i \in [N],  f_{\ntt}(h_i) = \sum_{j \in [N]} f_{\coeff}(h_j)\cdot(\psi^{2(i-1)+1})^{j-1}$.
Recall that for $f(X) = \sum_{i=0}^{N-1} f_iX^{i} \in \mathbb{Z}_q[X]/(X^N+1)$, we have the encoding $\forall i \in [N], f_{\coeff}(h_i)= f_{i-1}$.

By definition, NTT representation can be written as $\vec f_{\ntt}=(\widehat f_{1},\cdots,\widehat f_{N})=(f_{\ntt}(h_1),\ldots, f_{\ntt}(h_N))$ where $\widehat f_i = f(\psi \phi^{i-1})= f(\psi^{2(i-1)+1})= \sum_{j=0}^{N-1} f_j(\psi^{2(i-1)+1})^{j} = \sum_{j \in [N]} f_{\coeff}(h_j)\cdot(\psi^{2(i-1)+1})^{j-1}$. Now, we show that $\sum_{h_i \in H} f_{\coeff}(h_i)\cdot \sigma_\alpha(h_i) - f_{\ntt}(h_i)\cdot \Lambda_{\alpha}(h_i) =0$ below.

\begin{align*}
&\sum_{h_i \in H} [f_{\coeff}(h_i)\cdot \sigma_\alpha(h_i) - f_{\ntt}(h_i)\cdot \Lambda_{\alpha}(h_i)]  \\
&= \sum_{h_i \in H} f_{\coeff}(h_i)\cdot \sigma_\alpha(h_i) - \sum_{h_i \in H} f_{\ntt}(h_i)\cdot \Lambda_{\alpha}(h_i) \\
&= \sum_{i \in [N]} f_{i-1} \cdot \alpha^{i-1} - \sum_{i\in [N]}  [\sum_{j \in [N]} f_{\coeff}(h_j)\cdot(\psi^{2(i-1)+1})^{j-1}] \cdot \ell_{\psi \phi^{i-1}}(\alpha) \\
&= \sum_{i \in [N]} f_{i-1} \cdot \alpha^{i-1} - \sum_{i\in [N]}  f(\psi^{2(i-1)+1}) \cdot \ell_{\psi^{2(i-1)+1}}(\alpha) \\
&= f(\alpha) - f(\alpha) = 0.
\end{align*}

Applying \cref{lemma:uni_sumcheck} and $\sum_{h_i \in H} [f_{\coeff}(h_i)\cdot \sigma_\alpha(h_i) - f_{\ntt}(h_i)\cdot \Lambda_{\alpha}(h_i)] = \beta = 0 $, the polynomial $p(X) = f_{\coeff}(X)\cdot \sigma_\alpha(X) - f_{\ntt}(X)\cdot \Lambda_{\alpha}(X)$ can be re-written as $p(X) = R(X)X + Q(X)Z_H(X)$ since degree of $p(X) = 2|H|-2 \geq |H|$ when $|H| > 1$. Since the polynomials on both side are equal, the verifier will always accept at the end of execution when $|H|>1$ and $N=|H|$. 

{$\space$} 

\textbf{Knowledge-Soundness.} It is sufficient to show $\delta$-soundness since we can apply \cref{remark:knowledge_soundness}, and the protocol will be $\delta$-knowledge sound as required.

%  ; (f_{\ntt}, f_{\coeff})
If $\bigl( (N,H,\psi, \polyoracle{f_{\ntt}}, \polyoracle{f_{\coeff}})\bigr) \notin \mathcal{L}(\mathcal{R_{\nttcoeffrelation}})$, assume that $N=|H|$ and $|H|>1$ is verified during protocol or during setup, then the verifier should reject with $1-\delta$ probability when $\exists i\in [N],\widehat f_i \neq f(\psi^{2(i-1)+1})$ where $(\widehat f_{1},\cdots,\widehat f_{N})=(f_{\ntt}(h_1),\ldots, f_{\ntt}(h_N))$.

Note that the derivation from \textit{Completeness} proof is now $\sum_{h_i \in H} [f_{\coeff}(h_i)\cdot \sigma_\alpha(h_i) - f_{\ntt}(h_i)\cdot \Lambda_{\alpha}(h_i)] = f(\alpha)-f'(\alpha)$ where $f'(X)$ is the polynomial interpolated in the general case. Since there exists at least a point $X=\psi^{2(i-1)+1}$ for some $i \in [N]$ where $\widehat f_i =f'(\psi^{2(i-1)+1}) \neq f(\psi^{2(i-1)+1})$ 
, therefore $f(X) \neq f'(X)$. 
We now show that $f(\alpha)-f'(\alpha)$ is zero with negligible probability. Let $\bar f(X)=f(X)-f'(X)$ be a non-zero polynomial, we now apply Schwartz-Zippel (\cref{lemma:schwartz-zippel}) to obtain that $\Pr[\bar f(\alpha)=0] \leq \frac{N-1}{|\mathbb{F}|}$ where $\alpha \sample \mathbb{F}$.

From \cref{lemma:uni_sumcheck}, we can rewrite $f_{\coeff}(X)\cdot \sigma_\alpha(X) - f_{\ntt}(X)\cdot \Lambda_{\alpha}(X)= R(X)X + Q(X)Z_H(X)$ iff $\sum_{h_i \in H} [f_{\coeff}(h_i)\cdot \sigma_\alpha(h_i) - f_{\ntt}(h_i)\cdot \Lambda_{\alpha}(h_i)]=0$ where $\deg(R)\leq |H|-2$ and $\deg(Q) \leq (2N-2)-|H|$. Let $P_{LHS}(X)$ be the polynomial at the left hand side and $P_{RHS}(X)$ be the polynomial at the right hand side. We can use Schwartz-Zippel lemma to conclude that $\Pr[P_{LHS}(z) - P_{RHS}(z)=0: P_{LHS}(X) \neq P_{RHS}(X) \wedge z \sample \mathbb{F}] \leq \frac{ 2N-2}{|\mathbb{F}|}$.
Thus, we conclude that $\protocol_{\nttcoeffrelation}$ is knowledge sound for $\delta = \frac{3N-3}{|\mathbb{F}|}$ using union bound as required. 
\end{proof}

\textbf{Prover Running-Time. }
The prover has already preprocessed the  barycentric weights for the Lagrange interpolation which initially takes $O(N^2)$. The prover then takes as inputs, the coefficients and its NTT representations for quotient ring polynomial $f$, prepares the polynomial oracles $\polyoracle{f_{\ntt}}, \polyoracle{f_{\coeff}}$  in linear time $O(N)$ through interpolation technique described in \cref{sec:background}. Similarly, after the verifier returns the evaluation challenge $\alpha$, it can construct the polynomials $\sigma_{\alpha}(X), \Lambda_{\alpha}(X)$ in linear time and subsequently constructs the resulting sumchecked polynomial $f_{\coeff}(X)\cdot \sigma_\alpha(X) - f_{\ntt}(X)\cdot \Lambda_{\alpha}(X)$ in time $O(N^2)$ dominated by schoolbook-based polynomial multiplication or $O(N \log N)$ through NTT-based polynomial multiplication. Since $Z_H(X)$ is sparse with two non-zero terms, the long division algorithm takes linear time in the degree when preparing the oracle $\polyoracle{Q},\polyoracle{R}$ for $f_{\coeff}(X)\cdot \sigma_\alpha(X) - f_{\ntt}(X)\cdot \Lambda_{\alpha}(X)= Q(X)Z_H(X)+R(X)X$. Therefore, the prover time for the PIOP protocol is dominated by $O(N \log N)$ for computing the polynomial multiplication in the online phase.

\textbf{Verifier Running-Time.} 
The verifier needs to evaluate $Z_H(X)$ which is a closed-form evaluation of $X^N-1$ computable in $O(\log N)$ time using the square-and-multiply algorithm. The verifier is then required to compute $\sigma_\alpha(z)$ and $\Lambda_{\alpha}(z)$. The verifier computes the $N$ consecutive power terms $\alpha^{i-1}$ and evaluates $\sigma_\alpha(z)$ using the preprocessed Lagrange interpolation using the barycentric second form in $O(N)$ time. For the chosen interpolation parameter, we have $\ell(X)=X^N+1$ which can be computed using the same technique as before. The set of  $\{\ell_{\psi^{2(i-1)+1}}(\alpha)\}_{i\in [N]}$ is first computed in $O(N)$ time  using the formula $\ell_i(X)=\ell(X)\cdot \frac{w_i}{X-x_i}$ where $\ell(X)$ is as computed as before and $w_i$ is precomputed, then the evaluated Lagrange interpolation basis polynomial is then used to compute $\Lambda_{\alpha}(z)$ in $O(N)$ time. The equality check is then done with a fixed number of field operations. Therefore, the verifier runs in time $O(N)$ which can be made to run in time $O(\mathsf{polylog}(N))$ using the technique below.

\subsubsection{Delegating Evaluation of $\Lambda_{\alpha}$ and $\sigma_\alpha$}
The verifier in the protocol from \cref{proto:proto_ntt} runs in linear time because the verifier needs to prepare and evaluate the  $\Lambda_{\alpha}$ and $\sigma_\alpha$ on some random $\alpha$. To have the verifier run in sublinear time, the verifier can delegate the evaluation to the prover and simply verify the prover-provided polynomials are well-formed. To verify the well-formedness of $\sigma_\alpha(X)$, the verifier uses a similar technique from \cref{sec:background_stark} and verifies the following constraint
% CP_{\sigma,\alpha}(X) =
$$ \sigma_{\alpha}(X)\alpha-\sigma_\alpha(\omega X) = 0, \forall X\in H\setminus\{\omega^{n_H-1}\} \wedge \sigma_\alpha(\omega^0)=\alpha^0$$
The constraint checks that $\sigma_{\alpha}(X)$ starts with $\alpha^0$ and each next row entry is multiplied by $\alpha$. For the well-formness of $\Lambda_{\alpha}(X)$, the following constraint:
$$\Lambda_{\alpha}(X)  \cdot (\alpha-\widehat x(X))-(\alpha^N+1)\cdot \widehat w(X)=0, \forall X\in H$$
is checked where $\widehat w(X)$ is the preprocessed Lagrange weight and $\widehat x(X)$ is the preprocessed polynomial entry such that $\widehat x(h_i)=x_i$ and $\widehat w(h_i)=w_i$. 
The constraints checks $\forall h_i \in H, \Lambda_{\alpha}(h_i)= \ell_{\psi^{2(i-1)+1}}(\alpha)=\ell(\alpha) \cdot \frac{w_i}{\alpha - x_i}$ where $\ell(\alpha)=\alpha^N+1$. 
The constraints described can be efficiently enforced by STARK in $O(\mathsf{polylog}(N))$ verifier time.
The use cases in \cref{sec:application} do not make use of this optimization. 

\begin{remark}
The verifier in the PIOP protocol from \cref{proto:proto_ntt} can run in sublinear time by delegating the linear-time task using the above technique. The technique described above applies to the generalized case presented in the next section.
\end{remark}

\subsubsection{Adding Zero-Knowledge}

The $\name$ approach can be made zero-knowledge with the standard algebraic masking technique as used in \cite{BenSasson2019,Chiesa2020}. 
Given a committed masking polynomial $m(X)$ with the claimed sum $\sum_{h_i \in H} m(h_i)=\beta_{m}$, the verifier samples a masking randomization factor $\rho$, and the prover and verifier perform a univariate sumcheck protocol over the following:
\begin{align*}
    \rho (\tilde f_{\coeff}(X) \cdot \sigma_\alpha(X) - \tilde f_{\NTT}(X) \cdot \Lambda_{\alpha}(X)) + m(X) = \frac{\beta_m}{|H|} + R(X)X + Q(X)Z_H(X)
\end{align*}
where $\tilde f(X)$ is the masked polynomial $\tilde f(X) = f(X) + m'(X)Z_H(X)$ for a random masking polynomial $m'(X)$ and $R(X)$ and $Q(X)$ are new decomposed polynomials with respect to the changes made. 
We refer to \cite{BenSasson2019, Chiesa2020} for more details and sketch the transformation properties below.
The modified protocol preserves the knowledge soundness (with negligible loss) and completeness of the protocol. 
These polynomial encodings, masked as $\tilde f = f + m'(X)Z_H(X)$, are protected against up to $b$ queries in $\mathbb{F}\setminus H$ because their evaluations are $b$-wise independent (guaranteed by $\deg (m') \ge b$). Together with the statement mask $m(X)$, which randomizes the decomposed polynomials $R(X), Q(X)$, this yields perfect zero knowledge in the IOP model with a query bound $b$ and a query checker that rejects any query to point in $H$. We omit the ZK transformation for the generalized version since it proceeds similarly with the stated transformation.

\subsection{$\mathcal{R}^{\batchgeneral}_{\nttcoeffrelation}$: Quotient Ring Stacking and Generalized Placement}
\label{sec:poly_stacking}
We now extend the Northstar technique to work over general encoding of the quotient ring polynomials into a trace polynomial. Let the quotient ring polynomials of degree at most $N-1$ be $g_1(X),g_2(X),\ldots,g_M(X)$ for some positive integer $M$. Let the size of the trace domain be $n_H = |H| \geq M\cdot N $.
Let the trace polynomial $g_{\coeff}$ encodes a batched polynomial coefficients representations and trace polynomial $g_{\ntt}$ encodes batched NTT representations.

For each polynomial block $j \in [M]$, we define the generalized mapping as $\varphi_j: [N] \rightarrow [n_H]$, that maps $i \mapsto \varphi_j(i)$. To handle trace polynomial that can hold extra trace rows $n_H \geq M\cdot N$, we define $\varphi_{M+1} : [n_H - M \cdot N] \rightarrow [n_H]$. We denote the image of the mapping as $I_j := \{\varphi_j(i)\}_{i \in[N]}$ and for $I_{M+1} :=\{\varphi_{M+1}(i)\}_{i \in[n_H-M\cdot N]}$. The mapping set defined as $\{\varphi_{j}\}_{j\in[M+1]}$ is disjoint $I_{j} \cap I_{k} = \emptyset$ for $j\neq k$ and covering $\bigcup_{j \in [M+1]} I_j = [n_H]$. We denote $I^{\{\NTT,\coeff\}}$ for $\varphi^{\{\ntt,\coeff\}}$ for the NTT and coefficient representation mappings respectively.

\begin{definition} [Relation $\mathcal{R}^{\batchgeneral}_{\nttcoeffrelation}$]
\label{def:ntt_relation_general}
Let the mapping set $\{\varphi_{j}\}_{j\in [M+1]}$ be disjoint and covering as defined above. For a trace domain $H=\{h_1,\ldots, h_{n_H}\}$ encoded quotient ring polynomial $g_j(X) = \sum_{i=0}^{N-1} c_{j,i}X^{i} \in \mathbb{Z}_q[X]/(X^N+1)$ such that $\forall j \in [M],i \in [N], g_{\coeff}(h_{\varphi^{\coeff}_j(i)})= c_{j,i-1}$, the relation $\mathcal{R}_{\nttcoeffrelation}^{\batchgeneral}$ is defined as
\begin{align*}
\mathcal{R}_{\nttcoeffrelation}^{\batchgeneral} = \left\{ 
\begin{array}{c}
     \bigl( (M,N,H,\psi, \{\varphi_j^{\coeff}, \varphi_j^{NTT}\}_{j \in [M+1]}, \polyoracle{g_{\ntt}}, \polyoracle{g_{\coeff}}) ; (g_{\ntt}, g_{\coeff})\bigr): \\
    \forall i \in [N], j \in [M],  g_{\ntt}(h_{\varphi^{\ntt}_j(i)}) = \sum_{k \in [N]} g_{\coeff}(h_{\varphi^{\coeff}_j(k)})\cdot(\psi^{2(i-1)+1})^{k-1}; \\
    |H| \geq M\cdot N>1
\end{array}
\right\}
\end{align*}
where $\psi$ is the $2N$-th primitive root of unity for $\mathbb{Z}_q$, and M is the number of stacked polynomials.
\end{definition}

The relation above checks the generalized case where the encoding is captured by the maps $\varphi$.
We construct a PIOP protocol $\protocol_{\nttcoeffrelation}^{\batchgeneral}$ (described in \cref{proto:proto_ntt_general}) for $\mathcal{R}_{\nttcoeffrelation}^{\batchgeneral}$. The protocol is similar to \cref{proto:proto_ntt} except that we use the generalized challenge and interpolating polynomials $\sigma^{\batchgeneral}_{\alpha,\eta, \varphi}(X)$ and $\Lambda^{\batchgeneral}_{\alpha,\eta, \varphi}(X)$ defined in \cref{def:challenge_lagrange_poly_general}. 

\begin{definition}[Generalized Challenge Polynomial $\sigma^{\batchgeneral}_{\alpha, \eta, \varphi}(X)$ and Lambda Lagrange Basis Polynomial $\Lambda^{\batchgeneral}_{\alpha, \eta, \varphi}(X)$]
\label{def:challenge_lagrange_poly_general}
For a trace domain $H=\{h_1,\ldots,h_{n_H}\}$, challenges $(\alpha,\eta) \in \mathbb{F}\times \mathbb{F}$, a $2N$-th primitive root of unity $\psi$ and $N$-th primitive root of unity $\phi$, a disjoint and covering mapping set $\varphi=\{\varphi_1,\ldots,\varphi_{M+1}\}$, we denote generalized challenge polynomial as $\sigma^{\batchgeneral}_{\alpha, \eta, \varphi}(X)$ where $\forall i \in [N], \forall j \in [M], \sigma_{\alpha, \eta, \varphi}^{\batchgeneral}(h_{\varphi_j(i)})= \eta^{j-1} \cdot \alpha^{i-1}$ and for $j = M+1$, $\forall i \in [n_H-M\cdot N],\sigma_{\alpha,\eta,\varphi}^{\batchgeneral}(h_{\varphi_{M+1}(i)})= 0$.
In addition, we denote generalized lambda Lagrange basis polynomial as $\Lambda^{\batchgeneral}_{\alpha,\eta,\varphi}(X)$ where  $\forall i\in [N], \forall j \in [M],\Lambda_{\alpha,\eta,\varphi}^{\batchgeneral}(h_{\varphi_j(i)}) =\eta^{j-1} \cdot   \ell_{\psi \phi^{i-1}}(\alpha) = \eta^{j-1} \cdot \ell_{x_{i}}(\alpha)$ is the Lagrange basis for interpolation evaluated at $\alpha$. Similarly, for $j=M+1$, $\forall i\in [n_H-M\cdot N],\Lambda_{\alpha,\eta,\varphi}^{\batchgeneral}(h_{\varphi_{M+1}(i)}) = 0$. 
\end{definition}

{
\setlength{\intextsep}{0pt}
\setlength{\textfloatsep}{0pt}
\setlength{\floatsep}{0pt}
\begin{figure}
    \begin{adjustbox}{minipage=1.176\linewidth,scale=1, center}
    \begin{titledbox}{Protocol $\protocol_{\nttcoeffrelation}^{\batchgeneral}$ for $\mathcal{R}^{\batchgeneral}_{\nttcoeffrelation}$}
    \begin{pcvstack}
    \begin{enumerate}
        \item $\prover$ sends $\polyoracle{g_{\ntt}}, \polyoracle{g_{\coeff}}$
        \item $\verifier$ sends challenge $\alpha, \eta \sample (\mathbb{F}\times \mathbb{F})$
        \item $\prover$ sends $\polyoracle{R}, \polyoracle{Q}$
        \item $\verifier$ samples $z \sample \mathbb{F}$, queries $\polyoracle{g_{\ntt}}, \polyoracle{g_{\coeff}}, \polyoracle{Q}, \polyoracle{R}$ at point $z$;
        \\ \quad $\verifier$ computes $\sigma_{\alpha,\eta,\varphi^{\coeff}}^{\batchgeneral}(z)$, $\Lambda^{\batchgeneral}_{\alpha,\eta,\varphi^{\NTT}}(z), Z_H(z)$
    \end{enumerate}
    $\verifier$ accepts if $g_{\coeff}(z)\cdot \sigma^{\batchgeneral}_{\alpha,\eta,\varphi^{\coeff}}(z) - g_{\ntt}(z)\cdot \Lambda^{\batchgeneral}_{\alpha,\eta,\varphi^{\ntt}}(z)= R(z)z + Q(z)Z_H(z) \\
    \phantom{accepts if }\wedge \deg(R)\leq |H|-2 \wedge deg(Q) \leq |H|-2 \wedge |H|  \geq M\cdot N >1$
    \end{pcvstack}
    \end{titledbox}
    \end{adjustbox}
    \caption{
    PIOP Protocol $\protocol_{\nttcoeffrelation}^{\batchgeneral}$ for $\mathcal{R}_{\nttcoeffrelation}^{\batchgeneral}$
    }
    \label{proto:proto_ntt_general}
\end{figure}}

\begin{theorem}
\label{theo:main_general}
The protocol $\protocol^{\batchgeneral}_{\nttcoeffrelation}$ described in \cref{proto:proto_ntt_general} is complete, and knowledge-sound PIOP for the relation $\mathcal{R}_{\nttcoeffrelation}^{\batchgeneral}$ (\cref{def:ntt_relation_general}). 
\end{theorem}

\begin{proof}

\textit{Sketch.}  The full proof is provided in the Appendix \ref{app:proof}. It can be seen that the random factor $\eta$ is distributed to the correct polynomial instance among the encoded $[M]$ instances and that $\eta$ factor is factored out as a random linear combination for batched verification. For the case of $j=M+1$ mapping set encoding other information,
we zero out its contribution during verification. The previous proof arguments apply identically here with the changes.
\end{proof}

\subsubsection{Batching.}
\label{sec:poly_batch}

The generic batching strategy \cite{bulletproofs} applies here to multiple instances of $\mathcal{R}_{\nttcoeffrelation}^{\batchgeneral}$.
Given $T$ instances of non-uniform relations from before, batching is straightforward by separating each instance with a randomly sampled $\zeta \sample \mathbb{F}$,  $\zeta (g_{\coeff}(z)\cdot \sigma^{\batchgeneral}_{\alpha,\eta,\varphi^{\coeff}}(z) - g_{\ntt}(z)\cdot \Lambda^{\batchgeneral}_{\alpha,\eta,\varphi^{\ntt}}(z)) + \zeta^2 (\ldots)+\ldots$ + $\zeta^T(\ldots)$. 
The correctness trivially holds using similar arguments and soundness follows from the Schwartz-Zippel lemma \cite{Schwartz1980}. 

\noindent\textbf{Instantiation of PIOP Protocols.} 
We recall the instantiation
in Appendix \ref{sec:compilation}.
In particular for the use case at \cref{sec:application}, we use the FRI-based univariate PCS \cite{fri_pcs_2026} from STARK. 
Instead of sampling $z$ from the field, $z$ is sampled from an out-of-domain domain $\mathbb{F}\setminus (D \cup H)$ where $D$ is the FRI evaluation domain.

\section{Applications}
\label{sec:application}
\subsubsection{More Efficient Verifiable TFHE.}

We focus on the TFHE \cite{Chillotti2019} with programmable boostrapping (PBS). In particular, we compare our scheme instantiated with STARK for verifiable TFHE with the previous approaches from \cite{hasteboot,zama_TremblayThibault2025}. We recall TFHE scheme and provide the details of the verifiable FHE constraints in the \cref{sec:tfhe_details} and \cref{sec:tfhe_zkp_details}. 
We evaluate our technique on the same verifiable PBS operation task used as the benchmark in \cite{hasteboot,zama_TremblayThibault2025} to illustrate the efficiency of the proposed technique in practice. 
We extend the Plonky3\footnote{Note that the Plonky3 library implements a STARK-based univariate proving system with FRI PCS rather than PLONK-based Proof System with FRI PCS which the Plonky2 library implements.}
library with $\name$ and implement verifiable TFHE by generating proofs for TFHE PBS operation. 
The result is summarized in \cref{tab:compare_tfhe} and we only focus on post-quantum primitives and thus drop Dory-based PCS which relies on bilinear pairings. STARK with $\name$ achieves the best overall prover time, verifier time and communication size simultaneously while the SoTA Hasteboot \cite{hasteboot} suffers from inefficiency in at least one of the metrics depending on the exact PCS used. We focus only on one of the parameter sets that is the Goldilocks curve with $q=2^{64}-2^{32}+1,n=728,N=2^{10}, B=2^8, \ell=8$ whereby it takes $71.26ms$ to evaluate a single-FHE NAND operation on the CPU. Zama's number is as reported from \cite{zama_TremblayThibault2025} and Hasteboot's number is rerun on Macbook Pro M5 except BaseFold PCS as the implementation is not available as a script.

{
\setlength{\abovedisplayskip}{0pt}
\setlength{\belowdisplayskip}{0pt}
\setlength{\abovedisplayshortskip}{0pt}
\setlength{\belowdisplayshortskip}{0pt}
\begin{table}[ht!]
\begin{center}
\begin{adjustbox}{minipage=\linewidth,scale=0.85}
\begin{tabular}{r|l|r|r|r}
  \hline
  Work  & Proof System w/ PCS & Prover Time (s) & Verifier Time (ms) & Proof Size (MB) \\
  \hline
  Zama\cite{zama_TremblayThibault2025} & \cite{zama_TremblayThibault2025} & 25 \textbf{min}* & 4* & 0.2*\\
  \hline
  Hasteboot \cite{hasteboot} & Multilinear,Brakedown & 3.25 & 226.60 & 162.50 \\
    & Multilinear,BaseFold& 106.36* & 4.63* & 4.92* \\
  \hline
  $\name$ (This) & Univariate,FRI,STARK & 7.31 & 31.57 & 0.84\\
   \hline
\end{tabular}
\end{adjustbox}
\end{center}
\caption{ End-to-end performance comparions for verifiable TFHE PBS under different schemes and settings. * - As reported in \cite{hasteboot}. }
\label{tab:compare_tfhe}
\end{table}
}

\subsubsection{Lattice-based Encrypted Multi-Asset ERC-20 Token.}
{
\setlength{\intextsep}{0pt}
\setlength{\textfloatsep}{0pt}
\setlength{\floatsep}{0pt}
\begin{table}[ht!]
\begin{center}
% \rowcolors{1}{}{lightgray}
\begin{adjustbox}{minipage=\linewidth,scale=0.85}
\begin{tabular}{r|l|r|r|r|r}
  \hline
  Work  & System & Asset No. & Gas & Gas per Asset & Max Value
  \\
  \hline
  Zether\cite{bunz2020zether} & \makecell[l]{Account-based,DDH,\\Sigma Protocol}& 1 & 7.19M & 7.19M& $2^{32}$\\
  \hline
  $\tokenname$ (This) & \makecell[l]{UTXO-like, Lattice,\\$\name$, STARK} & $1022$ & 18.11M & 0.02M & $2^{64}$ \\
   \hline
\end{tabular}
\end{adjustbox}
\end{center}
\caption{Comparison of the $\mathsf{Transfer}$ operation with existing work. }
\label{tab:compare_tokens}
\end{table}
}

We defer the formalization of our construction to \cref{sec:erc20_complete}.
Our definition of encrypted multi-asset ERC-20 is based on the syntax of ERC-1155 which is a multi-asset token standard similar to ERC-20 where each token can encode values for multiple assets. We use a combination of hashed token and the BDLOP commitment scheme \cite{BDLOP16} with extractable message \cite{LNPS21} to form the coin as $\tokenname$:
% which has following forms:
\begin{align*}
    coin =
    \begin{bmatrix}
    \token\\
    \commitment \\
    \end{bmatrix}
    =
    \begin{bmatrix}
    \mathbf{0} \\
    \mathbf{A} \\
    \mathbf{B_1} \\
    \mathbf{B_2}
    \end{bmatrix}
    \mathbf{r}
    +
    \begin{bmatrix}
    \hash({m})\\
    \mathbf{0} \\
    m \\
    \sqrt{q} \cdot m \\
    \end{bmatrix}
\end{align*}
where $\hash(\cdot)$ is a hash function, $m := (v_1,\ldots,v_{\numasset},\vec r)$, $v$ is the coin value, $\vec r$ is the randomization factor, $\numasset$ is the number of assets.
We record the coin with the corresponding owner and will execute the transaction if the corresponding signature is produced and the well-formedness is proven through $\name$. 
We make use of $\name$ integrated with STARK to produce the proof that the coins are well-formed with respect to the above form. 
For the constraints, we use the standard base decomposition to enforce that the values are positive and summed to 0 for incoming and outgoing coin values.
For the arithmetic, we use $\name$ to obtain efficient polynomial multiplication in the NTT domain and norm check in the coefficient domain.
On the Ethereum testnet Hoodi (Fusaka), our proof-of-concept implementation
\footnote{PoC prototypes: \url{https://github.com/jpmorganchase/PADL}.}  
achieves $\approx 18$ million total gas used for on-chain verification for $1022$ number of asset transaction which amortizes to $\approx20$K gas units consumed per asset transferred (including asset-anonymity set). 
Thus, we are able to privately transact thousand(s) of assets while incurring a cost footprint similar to that of single-asset, discrete-log based scheme such as Zether \cite{bunz2020zether} and the comparison is summarized in \cref{tab:compare_tokens} while the details can be found in \cref{sec:compare_tokens}. 
The proof generation time is $\approx 500$ms. 
This enables practical confidential multi-asset transactions with asset anonymity that scale to thousand(s) type of assets in one token unit which facilitates institutional multi-asset use cases such as repo, DvP, and bond markets \cite{zhu2025towards,PADL}.

\section*{Disclaimer}
This paper was prepared for informational purposes by the Global Technology Applied Research center of JPMorganChase. This paper is not a product of the Research Department of JPMorganChase or its affiliates. Neither JPMorganChase nor any of its affiliates makes any explicit or implied representation or warranty and none of them accept any liability in connection with this position paper, including, without limitation, with respect to the completeness, accuracy, or reliability of the information contained herein and the potential legal, compliance, tax, or accounting effects thereof. This document is not intended as investment research or investment advice, or as a recommendation, offer, or solicitation for the purchase or sale of any security, financial instrument, financial product or service, or to be used in any way for evaluating the merits of participating in any transaction.

\bibliographystyle{splncs04}
\bibliography{ref.bib}

\appendix
\section{Appendix Organization}
We recall background on norms, polynomial IOP, NIZK, signature, module SIS/LWE, commitment scheme with extraction, and TFHE. 
Verifiable TFHE details is then given, and it is followed by the $\tokenname$ formalization and construction. Deferred Proofs is given in the last Appendix section.

\section{Additional Background}
We define various set norms as follows:
\begin{definition}[Norms \cite{LNPS21}]
\begin{itemize}
    \item For an element $w \in \mathbb{Z}_q$: $|w\|_\infty = |w \bmod^{\pm} q|$
    \item For a quotient ring polynomial $w = w_0 + w_{1} X + \cdots + w_{d-1} X^{d-1} \in {R}_q$:
    $$
    \|w\|_\infty = \max_j \|w_j\|_\infty, \quad \|w\|_p = \sqrt[p]{\|w_0\|_\infty^p + \cdots + \|w_{d-1}\|_\infty^p}.
    $$
    \item For $\mathbf{w} = (w_{1}, \ldots, w_k) \in {R}_q^k$:
    $$
    \|\mathbf{w}\|_\infty = \max_j \|w_j\|_\infty, \quad \|\mathbf{w}\|_p = \sqrt[p]{\|w_{1}\|^p + \cdots + \|w_k\|^p}.
    $$
\end{itemize}
By default, we denote $\|\mathbf{w}\| := \|\mathbf{w}\|_2$.
\end{definition}

\subsection{Polynomial IOP}
\label{app:piop_background}

\begin{definition}[Polynomial IOP]
A polynomial interactive oracle proof (PIOP) for an indexed relation $\mathcal{R}$ is an interactive protocol between a prover $\mathcal{P}$ and a verifier $\mathcal{V}$ denoted as $\langle \mathcal{P}(\mathbb{i}, \mathbb{x}, \mathbb{w}) ,\mathcal{V}(\mathbb{i,x})\rangle \rightarrow \{0,1\}$ where $\{0,1\}$ is the verifier output such that:
\begin{itemize}
    \item \textbf{Interaction.} In each round, the prover sends polynomial oracles and/or field elements, then the verifier sends challenges. Each oracle $\polyoracle{p}$ for polynomial $p(X)$ specifies the number of variable $\ell$ and the degree in each variable.
    The verifier can query all the oracles at arbitrary points $\mathbf{r} \in \mathbb{F}^{\ell}$, then the verifier $\mathcal{V}$ outputs accepts or rejects at the end of execution.
    \item \textbf{(Perfect) Completeness.} For any honest prover $\mathcal{P}$ and any $(\mathbb{i}, \mathbb{x}; \mathbb{w}) \in \mathcal{R}$, $$\Pr[\langle \mathcal{P}(\mathbb{i}, \mathbb{x}, \mathbb{w}) ,\mathcal{V}(\mathbb{i,x})\rangle = 1] = 1.$$
    \item \textbf{$\delta$-Soundness.} A PIOP is $\delta$-sound if for any unbounded adversary $\mathcal{A}$ and any $(\mathbb{i,x})\notin \mathcal{L}(\mathcal{R})$, $\Pr[\langle \mathcal{A}(\mathbb{i,x}), \mathcal{V}(\mathbb{i,x}) \rangle=1] \leq \delta$.
    \item \textbf{$\delta$-Knowledge Soundess.} A PIOP is $\delta$-knowledge sound if there exists a PPT extractor $\mathsf{Ext}$ with oracle access to all supplied polynomials, such that for any unbounded prover $\mathcal{A}$ and any $(\mathbb{i,x})$,
    $$
    \Pr[b=1 \wedge (\mathbb{i,x};\mathbb{w})\notin \mathcal{R}: b \gets \langle \mathcal{A}(\mathbb{i}, \mathbb{x}) ,\mathcal{V}(\mathbb{i,x})\rangle; \mathbb{w} \gets \mathsf{Ext}^{\mathcal{A}}(\mathbb{i,x}) ] \leq \delta
    $$
    The scheme is ``knowledge sound'' or is an ``argument of knowledge'', if the knowledge soundness error $\delta$ is negligible in the security parameter $\lambda$.
    \item \textbf{Honest-Verifier Zero-Knowledge.} 
    % This is from hyper plonk
    A PIOP has statistical zero-knowledge with query bound $t$ and query checker $C$ if there is a PPT simulator $\simulator$ such that for every field $\mathbb{F}$, index $\mathbb{i}$, instance $\mathbb{x}$, witness $\mathbb{w}$, and every $(t,C)$-admissible verifier $\verifier^*$, the following transcripts are within $\delta$ statistical distance:
    $$\mathsf{View}(\prover(\mathbb{F},\mathbb{i,x,w}), \verifier^*) \approx_{\delta} \simulator^{\verifier^*}(\mathbb{F},\mathbb{i},\mathbb{x}).$$
    The view consists of $\verifier^*$'s randomness, the non-oracle messages sent by $\prover$, and the list of answer to $\verifier^*$'s queries. A verifier is (t,C)-admissible if it makes no more than $t$ queries, and each query is accepted by the checker $C$. $\langle\prover, \verifier\rangle$ is honest-verifier zero-knowledge (HVZK) if there is a simulator for $\verifier$.
\end{itemize} 
\end{definition}

\begin{remark}
\label{remark:knowledge_soundness}
It is proven in Lemma 2.3 \cite{Chen2023} that any $\delta$-sound PIOP for indexed relation $\mathcal{R}$ is $\delta$-knowledge sound, if the witness $\mathbb{w}$ consist of only polynomials provided as oracle in the statement $\mathbb{x}$.
\end{remark}

\subsection{Instantiation of PIOP Protocols}
\label{sec:compilation}
In the standard compilation theorems \cite{Bnz2020,Chiesa2020}, polynomial oracle in the PIOP protocol can be replaced with polynomial commitment scheme (PCS) which enables the prover to commit to a polynomial and later open the committed polynomial to its evaluation points with proofs. Thus, a PIOP for an indexed oracle relation can be compiled into a public-coin interactive argument of knowledge 
using the standard compilation theorems \cite{Bnz2020,Chiesa2020}. 
A public-coin interactive argument of knowledge can be compiled into non-interactive argument using the Fiat-Shamir transform \cite{fiatshamir}.

\subsection{Non-interactive Zero-knowledge Argument of Knowledge (NIZK) Scheme}
\begin{definition}
Let $\mathcal{R}$ be an indexed NP relation. A non-interactive argument of knowledge for $\mathcal{R}$ is a tuple of PPT algorithms $\nizk := (\ppgen, \keygen, \allowbreak \prove, \verify)$ defined below:
\begin{itemize}
    \item $\publicparam \gets \ppgen(\secparam)$: On input a security parameter $\secparam$, outputs public parameter $\publicparam$.
    \item $(\provingkey, \verificationkey) \gets \keygen(\mathbb{i}, \publicparam)$: On input an index $\mathbb{i}$, outputs a proving key $\provingkey$ and a verification key $\verificationkey$.
    \item $\pi \gets \prove(\provingkey, \mathbb{x}, \mathbb{w}):$ On input a proving key $\provingkey$, a statement instance $\mathbb{x}$, a witness $\mathbb{w}$, outputs a proof $\pi$.
    \item $b\gets \verify(\verificationkey,\mathbb{x}, \pi)$: On input a verification key $\verificationkey$, a public statement instance $\mathbb{x}$, a proof $\pi$, outputs a decision bit $b$.
\end{itemize}

$\nizk$ satisfies the following properties:
\begin{itemize}
    \item Completeness. For all $(\mathbb{i,x};\mathbb{w}) \in \mathcal{R}$, 
    \begin{align*}
        \Pr \left[\verify(\verificationkey, \mathbb{x}, \pi)=1 : 
        \begin{array}{c}
             \publicparam \gets \ppgen(\secparam);  \\
             (\provingkey, \verificationkey) \gets \keygen( \mathbb{i}, \publicparam);\\
             \pi \gets \prove(\provingkey, \mathbb{x},\mathbb{w})
        \end{array} 
        \right] = 1.
    \end{align*}
    \item Knowledge Soundness. For all PPT adversary $\adversary$, there exists a PPT extractor $\extract$ such that 
    \begin{align*}
        \Pr \left[ 
       \begin{array}{c}
             \verify(\verificationkey, \mathbb{x}, \pi)=1 \wedge  \\
             (\mathbb{i,x};\mathbb{w}) \notin \mathcal{R}
        \end{array}
        :
        \begin{array}{c}
             \publicparam \gets \ppgen(\secparam);  \\
             (\provingkey, \verificationkey) \gets \keygen( \mathbb{i}, \publicparam);\\
             (\mathbb{x},\pi)\gets \adversary(\publicparam, \provingkey);\\
             \mathbb{w} \leftarrow \extract^{\adversary}(\publicparam, \provingkey, \mathbb{x},\pi) 
        \end{array}
        \right] \leq \negl.
    \end{align*}

    \item Zero-Knowledge. There exists PPT simulator algorithms $(\zksimulator_1, \zksimulator_2, \zksimulator_3)$ such that for all PPT adversaries $\mathcal{A}:=(\adversary_1, \adversary_2, \adversary_3)$, there exists a negligible function $\negl$ such that

    \begin{align*}
       \Big| 
       \Pr \left[ 
       \adversary_3(\mathsf{st}_{\adversary},\pi)=1
        :
        \begin{array}{c}
             \publicparam \gets \ppgen(\secparam);  \\
              (\mathsf{st}_{\adversary}, \mathbb{i})\gets \adversary_1(\publicparam); \\
              (\provingkey, \verificationkey) \gets \keygen( \mathbb{i}, \publicparam);\\
              (\mathsf{st}_{\adversary},\mathbb{x,w})\gets \adversary_2(\mathsf{st}_{\adversary}, \provingkey,\verificationkey); \\
              (\mathbb{i,x}; \mathbb{w})\in \mathcal{R};\\
                       \pi \gets \prove(\provingkey, \mathbb{x},\mathbb{w})
                       % \\ b \gets \adversary(\pi)
        \end{array}
        \right] 
        \\ - 
       \Pr \left[ 
       \adversary_3(\mathsf{st}_{\adversary}, \pi)=1
        :
        \begin{array}{c}
             (\publicparam, \mathsf{st}) \gets \zksimulator_1(\secparam);  \\
               (\mathsf{st}_{\adversary}, \mathbb{i})\gets \adversary_1(\publicparam); \\
                (\provingkey,\verificationkey, \mathsf{st}) \gets \zksimulator_2(\mathsf{st}, \mathbb{i}); \\
              (\mathsf{st}_{\adversary},\mathbb{x,w})\gets \adversary_2(\mathsf{st}_{\adversary}, \provingkey,\verificationkey); \\
              (\mathbb{i,x}; \mathbb{w})\in \mathcal{R};\\
                    \pi \gets \zksimulator_3(\mathsf{st},\mathbb{x});
        \end{array}
        \right] 
        \Big| \leq \negl.
    \end{align*}
\end{itemize}
\end{definition}

\subsection{Signature Scheme}
% Signature scheme is recalled here.

\begin{definition}[Signature Scheme]
 A signature scheme is a tuple of PPT algorithms $\DS:=(\keygen, \sign, \verify)$  defined as follows:
\begin{itemize}
    \item  $(\sk,\vk) \leftarrow\keygen(\secparam)$: On input a security parameter $\secparam$, this algorithm outputs a signing key $\sk$ and a verification key $\vk$. 
    \item  $(\sigma) \leftarrow \sign(\sk, m)$: On input a secret key $\sk$ and a message $m$, this algorithm outputs a signature $\sigma$. 
    \item  $b \leftarrow \verify(\vk, m, \sigma)$: On input a verification key $\vk$, a message $m$, and a signature $\sigma$, this algorithm outputs a bit $b$. 
\end{itemize}

A signature is correct, if for any $(\sk,\vk) \gets \keygen(\secparam)$ and any message $m$, it holds that $\Pr[\verify(\vk, m, \sign(\sk,m))]=1$.

A signature scheme satisfies EUF-CMA, if for any PPT adversary $\adversary$, it holds that 
$$
\Pr[\verify(\vk, m, \sigma)=1: (\sk,\vk) \gets \keygen(\secparam); (m,\sigma) \gets \adversary^{\sign(\sk, \cdot)}(\vk);m \notin Q_m] \leq \negl
$$
where $Q_m$ records all messages queried to the oracle $\sign(\sk, \cdot)$.
\end{definition}

\subsection{Module SIS/LWE}

\begin{definition}[$\msis_{\kappa,m,\beta}$ \cite{LNPS21}]
\label{def:msis} 
Given $\mathbf{A} \sample  R_q^{\kappa \times m}$, the homogeneous Module-SIS problem with parameters $\kappa, m > 0$ and $0 < \beta < q$ asks for $\vec z \in R_q^m$ such that $\mathbf{A} \vec z = 0$ and $0 < \|\vec z\| \leq \beta$. An algorithm $\adversary$ is said to have advantage $\epsilon$ in solving $\msis_{\kappa,m,\beta}$ if 
\begin{align*}
\Pr \left[ 0 < \|\vec z\| \leq \beta \land \mathbf{A} \vec z = 0 \mid \mathbf{A} \sample R_q^{\kappa \times m};\ \vec z \gets \adversary(\mathbf{A}) \right] \geq \epsilon.
\end{align*}
If the advantage $\epsilon$ is negligible, we say $\msis_{\kappa,m,\beta}$ is hard for $\adversary$.
\end{definition}

\begin{definition}[$\mlwe_{m,\lambda,\chi}$ \cite{LNPS21}]\label{def:mlwe} 
The Module-LWE problem with parameters $\lambda, m > 0$ and a distribution $\chi$ over $\{-1, 0, 1\}$ asks one to distinguish between the following two distributions:
\begin{enumerate}
    \item $(\mathbf{A}, \mathbf{A} \vec {s} + \vec {e})$, where $\mathbf{A} \sample  R_q^{m \times \lambda}$, $\vec {s} \sample  \chi^{\lambda d}$, and $\vec {e} \sample \chi^{m d}$;
    \item $(\mathbf{A}, \vec {u})$, where $\mathbf{A} \sample R_q^{m \times \lambda}$ and $\vec {u} \sample R_q^m$.
\end{enumerate}
An algorithm $\adversary$ is said to have advantage $\epsilon$ in solving $\mlwe_{m,\lambda,\chi}$ if
\begin{align*}
     &\left| \Pr \left[ b = 1 \mid \mathbf{A} \sample R_q^{m \times \lambda};\ \vec {s} \sample \chi^{\lambda d};\ \vec {e} \sample \chi^{m d};\ b \gets \adversary(\mathbf{A}, \mathbf{A} \vec {s} + \vec {e}) \right] \right. \\
    &\quad - \left. \Pr \left[ b = 1 \mid \mathbf{A} \sample R_q^{m \times \lambda};\ \vec {u} \sample R_q^m;\ b \gets \adversary(\mathbf{A}, \vec {u}) \right] \right| \geq \epsilon.
\end{align*}
If the advantage $\epsilon$ is negligible, we say $\mlwe_{m,\lambda,\chi}$ is hard for $\adversary$.
\end{definition}

\begin{definition}[$\mlwe'_{\lambda,\kappa,\chi}$ \cite{Nguyen22}]\label{def:mlwe1} 
The (knapsack) Module-LWE problem with parameters $\kappa, \lambda > 0$ and a distribution $\chi$ over $\{-1, 0, 1\}$ asks one to distinguish between the following two distributions:
\begin{enumerate}
    \item $(\mathbf{A}, \mathbf{A} \vec{s})$, where $\mathbf{A} \sample R_q^{\kappa \times (\kappa + \lambda)}$, $\vec {s} \sample \chi^{(\kappa + \lambda) d}$;
    \item $(\mathbf{A}, \vec {u})$, where $\mathbf{A} \sample R_q^{\kappa \times (\kappa + \lambda)}$ and $\vec {u} \sample R_q^\kappa$.
\end{enumerate}
An algorithm $\adversary$ is said to have advantage $\epsilon$ in solving the dual $\mlwe'_{\lambda,\kappa,\chi}$ problem if
\begin{align*}
     &\left| \Pr \left[ b = 1 \mid \mathbf{A} \sample R_q^{\kappa \times (\kappa + \lambda)};\ \vec{s} \sample \chi^{(\kappa + \lambda) d};\ b \gets \adversary(\mathbf{A}, \mathbf{A} \vec{s}) \right] \right. \\
    &\quad - \left. \Pr \left[ b = 1 \mid \mathbf{A} \sample R_q^{\kappa \times (\kappa + \lambda)};\ \vec{u} \sample R_q^\kappa;\ b \gets \adversary(\mathbf{A}, \vec {u}) \right] \right| \geq \epsilon.
\end{align*}
If the advantage $\epsilon$ is negligible, dual $\mlwe'_{\lambda,\kappa,\chi}$ is hard for $\adversary$.
\end{definition}

We refer to (Appendix C, \cite{Esgin2019}) that discusses the duality between the Knapsack (where we denote as $\mlwe'$) and MLWE (which we denote as $\mlwe$) in more details. 
% For convenient, we make use of both notations.

\subsection{Commitment Scheme with Extraction}
\label{sec:bdlop_background}

A commitment scheme with extraction\footnote{Commitment with verifiable extraction is different than trapdoor commitment\cite{trapdoorcommitment,cryptoeprint:2009/007} where the trapdoor commitment allows the secret key holder to open to any value while verifiable extraction allows secret key holder to recover the committed value.} is a commitment scheme that allows a secret key holder to extract the committed value committed under the commitment key assuming the commitment is well-formed.

\begin{definition}[Commitment Scheme]
A commitment scheme $(\commitscheme:=(\allowbreak \commitgen, \allowbreak \commit, \open))$ consists of the following set of algorithms:
	\begin{itemize}
	\item{ $\commitkey \gets \commitgen\textnormal{(}\secparam\textnormal{)}$:}  on input a security parameter $\secparam$, this algorithm outputs a commitment key $\commitkey$.
    \item{$\commitment \gets \commit\textnormal{(}\commitkey, m,r \textnormal{)}$:} on input a commitment key $\commitkey$, a message $m$ and a random value $r$, this algorithm outputs a commitment $\commitment$.
    \item{$b \gets \open\textnormal{(}\commitkey, \commitment, m,r \textnormal{)}$:} on input a commitment key $\commitkey$, a commitment $\commitment$, a message $m$ and a random value $r$, this algorithm outputs a bit $b\in \{0,1\}$.
	\end{itemize}
	\label{definition:com}

A $\mathsf{COM}$
% $(\commitgen, \commit, \open)$ i
is called a commitment scheme if it satisfies the hiding and binding properties defined below where $m$ is sampled from the message space $\mathcal{M}$ and $r$ is sampled from the masking space $\mathcal{\tilde M}$. 

\textbf{Hiding. } Let the advantage $\advantage_{\mathsf{Hiding}}(\adversary)$ of breaking the hiding property be the probability defined below. A scheme satisfies hiding if for all PPT adversary $\adversary$, the following probability is negligible:
\begin{align*}
\abs{ \Pr\left[
\begin{array}{cc}
    \multirow{3}{*}{$\adversary(\commitment)=b:$}  
    & \commitkey \gets \commitgen(\secparam); r \sample \tilde {\mathcal{M}} \\ 
    & (m_0,m_1)\gets \adversary(\commitkey);\ b\sample \{0,1\}; \\
    & \commitment \gets \commit(\commitkey, m_b, r)
\end{array}
\right] - \frac{1}{2}}
\end{align*}

\textbf{Binding. } Let the advantage $\advantage_{\mathsf{Binding}}^{}(\adversary)$ of breaking the binding property be the probability defined below. A scheme satisfies binding if for all PPT adversary $\adversary$, the following probability is negligible:

\[
\Pr\left[
\begin{gathered}
m_0 \neq m_1 \wedge \\
\open(\commitkey, \commitment,m_0,r_0)=\\
\open(\commitkey, \commitment,m_1,r_1)= 1
\end{gathered}
\ \middle| \ 
\begin{gathered}
\commitkey \gets (\commitgen(\secparam)); \\
(\commitment, m_0,r_0,m_1,r_1)\gets \adversary(\commitkey)
\end{gathered}
\right]
\]
\label{definition:com_properties}
\end{definition}

\begin{definition}[BDLOP Commitment Scheme \cite{BDLOP16,Nguyen22}] Let $R_q = \mathbb{Z}_q[X]/(X^N+1)$ and let $n$ be the maximum number of $R_q$ elements committed as message $\vec m$. BDLOP commitment scheme is defined by the following set of algorithms:
\begin{itemize}
	\item{ $\commitgen\textnormal{(}\secparam\textnormal{)}$:}  sample $\mathbf{A}\sample R_q^{\kappa_{\sis}\times (\kappa_{\lwe}+\kappa_{\sis}+n)}$, $\mathbf{B}\sample R_q^{n\times (\kappa_{\lwe}+\kappa_{\sis}+n)}$, then return $(\mathbf{A},\mathbf{B})$.
    \item{$\commit\textnormal{(}\commitkey, m,r \textnormal{)}$:} parse $(\mathbf{A},\mathbf{B}) \gets \commitkey$, parse $\vec r \gets r$ , compute $\commitment_0 =\mathbf{A}\vec r $, $\commitment_m = \mathbf{B}\vec r+m$, then return $\begin{bmatrix}
        \commitment_0\\ \commitment_m
    \end{bmatrix}$.
    \item{$\open\textnormal{(}\commitkey, \commitment, m,r \textnormal{)}$:} parse $(\mathbf{A},\mathbf{B}) \gets \commitkey$,  $\vec r \gets r, \vec m \gets m$,
    $\begin{bmatrix}
        \commitment_0\\ \commitment_m
    \end{bmatrix} \gets \commitment$, return $\begin{bmatrix}
        \commitment_0\\ \commitment_m
    \end{bmatrix}=^? 
    \begin{bmatrix}
        \mathbf{A}\vec r\\ \mathbf{B}\vec r+\vec m
    \end{bmatrix}$.
\end{itemize}

For $r\sample \chi^{(\kappa_{\sis}+\kappa_{\lwe}+n)N}$,    
the hiding property is established upon the hardness of $\mlwe'_{\kappa_{\lwe},\kappa_{\sis}+n,\chi}$ \cite{Nguyen22} and the BDLOP commitment scheme is binding if $\msis_{\kappa_{\sis},\kappa_{\sis}+\kappa_{\lwe}+n,\beta_{\sis}}$ is hard.
\end{definition}

\begin{remark}
    For the distinct $\vec r$ and $\vec r'$ together with $\vec m$ and $\vec m'$ that can be exactly extracted from the ZKPs for the same commitment $\commitment$ that proved the exact (non-relaxed) relation, $\beta_{\sis} = \norm{\vec r-\vec r'} \leq 2\beta $ where $\norm{\vec r} \leq \beta$. For $\beta_{\sis}$ obtained from relaxed opening, see \cite{ALS20,LNPS21}.
\end{remark}

\begin{definition}[Commitment Scheme with Extractability]

A commitment scheme $\mathsf{COM}_{\extract}:=(\commitgen,\commit,\open,\extract)$ has verifiable extractibility if it additionally supports the following sets of algorithms with modified $\commitgen$ and that it satisfies extractability defined below.
\begin{itemize}
    \item{ $(\commitkey,\sk) \gets \commitgen \textnormal{(}\secparam\textnormal{)}$:}  on input a security parameter $\secparam$, this algorithm outputs a commitment key $\commitkey$ and a secret extracting key $\sk$.
    \item{ $m \gets \extract\textnormal{(}\commitment,\sk\textnormal{)}$:}  on input a commitment $\commitment$, a secret key $\sk$, this algorithm outputs a message $m$.
\end{itemize}

\textbf{Extractable.} For all $(\commitkey,\sk)\gets \commitgen(\secparam)$, all message $m \in \mathcal{M}$, all random value $r \in \mathcal{\tilde M}$, $\extract(\commit(\commitkey,m,r),\sk)=m$.

The original commitment key generation is replaced with the modified version and the secret key $\sk$ is kept hidden.
    
\end{definition}

It is shown that in \cite{LNPS21}, BDLOP commitment scheme is extractable if $\mathbf{B}$ is a MLWE instance, that $m$ is committed alongside $\lfloor \sqrt q \rceil\cdot m$, and that the $q$ is large enough to tolerate errors. We use $\sqrt q$ to denote $\lfloor \sqrt q \rceil $ for brevity.

\begin{definition}[BDLOP Commitment Scheme with Extractability]
\label{def:bdlop_extractable}
BDLOP with verifiable extraction has the following set of algorithms:
    \begin{itemize}
    \item{ $(\commitkey,\sk) \gets \commitgen\textnormal{(}\secparam\textnormal{)}$:}  sample $\mathbf{A}\sample R_q^{\kappa_{\sis}\times (\kappa_{\lwe}+\kappa_{\sis}+2n)}$, $(\vec {s_1}, \vec {s}_1^*,\ldots)\sample (\chi^{\kappa_{\sis}\cdot N})^{2n}$, $(\vec e_1, \vec e_1^*, \ldots) \sample (\chi^{(\kappa_{\sis}+\kappa_{\lwe}+2n)N})^{2n}$, compute 
    $\mathbf{B} =\begin{bmatrix}
       \vec s_1^T\mathbf{A} + \vec e^T_1 \\
       \vec {(s^*_1)}^T\mathbf{A} + \vec {(e^*_1)}^T \\
       \vdots
    \end{bmatrix}$
    ,then returns $((\mathbf{A},\mathbf{B}), (\vec s_1,\vec {e}_1,\vec s_1^*,\vec {e}_1^*, \ldots))$.
    \item{$\commit\textnormal{(}\commitkey, m,r \textnormal{)}$:} parse $(\mathbf{A},\mathbf{B}) \gets \commitkey$, parse $\vec r \gets r$, parse $\vec m \gets (m_1,m_2,\ldots,m_n) := m$, compute $\vec m^* = (m_1,\sqrt qm_1,\ldots, m_n, \sqrt q m_n)$, compute $\commitment_0 =\mathbf{A}\vec r $, $\commitment_m = \mathbf{B}\vec r+\vec m^*$, then return $\begin{bmatrix}
        \commitment_0\\ \commitment_m
    \end{bmatrix}$.
    \item{$\open\textnormal{(}\commitkey, \commitment, m,r \textnormal{)}$:} parse $(\mathbf{A},\mathbf{B}) \gets \commitkey$,  $\vec r \gets r,$ 
    $\begin{bmatrix}
        \commitment_0\\ \commitment_m
    \end{bmatrix} \gets \commitment$, 
    parse $\vec m \gets (m_1,m_2,\ldots,m_n) := m$, compute $\vec m^* = (m_1,\sqrt qm_1,\ldots, m_n, \sqrt q m_n)$,
    return 
    $\begin{bmatrix}
        \commitment_0\\ \commitment_m
    \end{bmatrix}=^? 
    \begin{bmatrix}
        \mathbf{A}\vec r\\ \mathbf{B}\vec r+\vec m^*
    \end{bmatrix}$.    
    \item{ $m \gets \extract\textnormal{(}\commitment,\sk\textnormal{)}$:} parse 
    $\begin{bmatrix}
        \commitment_0\\
        \commitment_{1}\\ 
        \commitment_{1}^*\\
        \vdots
    \end{bmatrix} := \commitment$, parse $(\vec s_1,\vec e_1 ,\vec s_1^*,\vec e_1^*, \ldots) := \sk$, for $i\in [n]$, compute 
    \begin{align*}
    x_{i} = \commitment_{i} - \mathbf{s}_{i}^T\commitment_{0} = \mathbf{e}_{i}^T{\vec r}+m_i;\quad  x^*_i = \commitment_{i}^* - (\mathbf{s}_{i}^*)^T\commitment_{0} = (\mathbf{e}_{i}^*)^T{\vec r+\sqrt{q}m_i},
    \end{align*}
    compute $k = \sqrt q x_i - x_i^* \mod \sqrt {q}$, $m_i = \frac{x^*_i+k}{\sqrt q}$, finally return $\vec m = (m_1,\ldots,m_n)$.
\end{itemize}
\end{definition}

The (extractable) BDLOP commitment well-formness relation is defined as:
\begin{align*}
\mathcal{R}_{\mathsf{BDLOP}} = \left\{ 
\begin{array}{c}
     \bigl( (\mathbf{A},\mathbf{B}, \commitment) ; (\vec r, \vec m)\bigr): \\
     \vec m^* = (m_1,\sqrt qm_1, \ldots,m_n, \sqrt qm_n) \wedge \vec m = (m_1, \ldots,m_n)\\
    \wedge\ \commitment=\begin{bmatrix}
        \commitment_0\\ \commitment_m
    \end{bmatrix}= 
    \begin{bmatrix}
        \mathbf{A}\vec r\\ \mathbf{B}\vec r+\vec m^*
    \end{bmatrix} \wedge \norm{\vec r}_{\infty} \leq \eta
\end{array}
\right\}
\end{align*}
where $\eta$ is the bounded coefficient range for sample from $\chi$ distribution and $\eta$ is chosen such that $e_i^T\vec r$ is bounded within $\sqrt q/4$ as described below.

\textbf{Properties.} For all $((\mathbf{A}, \mathbf{B}),(\vec s_1,\vec e_1,\vec s_2,\vec e_2))\gets \commitgen(\secparam)$, if $(\mathbf{A},\mathbf{B}, \commitment) \in \mathcal{L}({\mathcal{R}_\mathsf{BDLOP}})$, the size of $e_i^T\vec r$ and $(e_i^*)^T\vec r$ are $< \sqrt q/4$ (no reduction modulo $q$ take place), then $\extract(\commit(\commitkey,\vec m,\vec r),\sk)=\vec m$ as shown in \cite{LNPS21}.
The binding property remains the same except that the random value needs to be sampled according to increased committed message size, and the hiding property remains the same with an additive factor which comes from replacing LWE instance with random instance during key generation. For $r\sample \chi^{(\kappa_{\sis}+\kappa_{\lwe}+2n)N}$,    
the hiding property is established upon the hardness of $\mlwe_{\kappa_{\sis}+\kappa_{\lwe}+2n,\kappa_{\sis},\chi} + \mlwe'_{\kappa_{\lwe},\kappa_{\sis}+2n,\chi}$, and the BDLOP commitment scheme is binding if $\msis_{\kappa_{\sis}, \allowbreak \kappa_{\sis}+\kappa_{\lwe}+2n,\beta_{\sis}}$ is hard where $\beta_{\sis}$ is the extracted norm of $\norm{r-r'}$.
% $\boxed{\eta_{\max}
% =\left\lceil\frac{q^{1/4}}{2\sqrt{\ell d}}\right\rceil-1}.$

\begin{remark}
\label{remark:commit_optimization}
In practice, $\mathbf{A}$ is shared by all participants in the system while $\mathbf{B}$ can be swapped in and out depending on the requirement on who can extract/decrypt the committed value. To ensure verifiable extractability for commitment key generated with secret key, commitment is proved to satisfy the relation $\mathcal{R}_{\mathsf{BDLOP}}$.
\end{remark}

\subsection{Torus Fully Homomorphic Encryption (TFHE)}
\label{sec:tfhe_details}

In this work, we focus on applying $\name$ to verifiable TFHE \cite{Chillotti2019} which is an FHE scheme with programmable bootstrapping (PBS). We follows the description from \cite{hasteboot} with minor differences.

\subsubsection{Ciphertext.} TFHE uses the learning with errors (LWE) \cite{regev_lattice} and the ring learning with errors (RLWE) \cite{ringlwe} ciphertexts. An \textbf{LWE ciphertext} encrypting a message $m \in \mathbb{Z}_t$ for some small positive integer $t$ has the following form:
$$
\mathsf{LWE}_{s}(m)=(\vec{a},b) = (\vec{a}, \langle{\vec a,\vec s}\rangle + \Delta \cdot m + e \mod q) \in \mathbb{Z}_q^{n+1}
$$
where $\Delta= \lfloor q /t\rceil$ is the scaling factor, $e$ is the noise term with short norm, $\vec{s}\in \{0,1\}^{n}$ is the secret key, and $\vec{a} \sample \mathbb{Z}_q^n$.
A \textbf{RLWE ciphertext} encrypting a message $m(X)$ has the form
$$\mathsf{RLWE}_z(m)=(a(X),b(X))= (a(X), a(X) \cdot z(X) + m(X) + e(X)) \in R_q^{2}$$ 
where $a(X) \sample R_q$ and $z(X)$ is a secret polynomial, and $e(X)$ is a error term polynomial. When the message needs a scaling factor $\Delta$, we use $\mathsf{RLWE}_z(\Delta m)$ to reflect it.
Similar to \cite{hasteboot}, we use $\rlwe$ to denote the space $R_q^2$.

TFHE further requires two more types of ciphertext definition. Given a fixed decomposition base $B$ and let $\ell = \lceil \log_{B}q\rceil$. An RLev ciphertext encrypting the message $m$ at each gadget level (scaled by successive powers of the decomposition base B) has the form
$$
\mathsf{RLev}_z(m)  =
\begin{bmatrix}
  k_0 \\ k_1 \\ \vdots \\ k_{\ell-1}.
\end{bmatrix} 
= \begin{bmatrix}
  \mathsf{RLWE}_z(m \cdot B^0)  \\ \mathsf{RLWE}_z(m \cdot B^1) \\ \vdots \\ \mathsf{RLWE}_z(m \cdot B^{\ell-1})
\end{bmatrix} \in \rlwe^{\ell}.
$$
An RGSW ciphertext encrypting message $m$ consists of two RLev ciphertext encrypting $m$, $-mz$ has the form
$$
\mathsf{RGSW}_z(m) =
\begin{bmatrix}
  k\\ k'
\end{bmatrix}
= \begin{bmatrix}
  \mathsf{RLev}_z(-mz)\\ \mathsf{RLev}_z(m)
\end{bmatrix} \in \rgsw
$$
where $\rgsw =\rlwe^{2\ell}$.

\subsubsection{Ciphertext Manipulations.} 
Let $t(X)$ be a $B$-bounded (plaintext) polynomial, let $\odot: R_q\times \rlwe \rightarrow \rlwe$ to be defined as $t(X) \odot c = (t(X)a(X), t(X)b(X))$ where $c = (a(X), b(X)) \in \rlwe$. The resulting ciphertext encrypts $t(X)m(X)$. Similarly, let $d(X) = \sum^{\ell-1}_{i=0} d_i(X) B^i$ where $d_i(X)$ is a $B$-bounded polynomial, let $\otimes: R_q\times \rlwe^{\ell} \rightarrow \rlwe$ to be defined as 
$d(X) \otimes k = \sum_{i=0}^{\ell-1}d_i(X) \odot k_i$ where $k \in \rlwe^\ell$. The resulting ciphertext is a RLWE ciphertext encrypting $d(X)m(X)$.

Finally, an external product $\diamond : \rlwe \times \rgsw \rightarrow \rlwe$ is defined as $c \diamond K =a(X) \otimes k + b(X) \otimes k'$ where $c = (a(X),b(X))$ encrypting $m(X)$ and $K = 
\begin{bmatrix}
    k \\ k'
\end{bmatrix}
$ encrypting $\mu$. The resulting ciphertext is a RLWE ciphertext encrypting $m(X) \cdot \mu$.

\SetKwComment{Comment}{/* }{ */}
\SetKwInOut{Input}{Input}
\SetKwInOut{Output}{Output}
\SetKwInOut{Params}{Params}
% \RestyleAlgo{ruled}
\begin{algorithm}[hbt!]
\caption{PBS Operation}\label{alg:pbs_algo}
\Params{$B \in \mathbb{Z}^+,$ $\ell = \lceil \log_B q \rceil $, $\bsk = \{K_i\}_{i=0}^{n-1} \in \rgsw^n$, $\ksk \in \rlwe^\ell$ }
\Params{$\lut \in R_q$, a lookup table with respect to a function $f$ }
\Input{$(\vec a, b) \in \mathbb{Z}_q^{n+1}$ encrypting message $m \in \mathbb{Z}_t$ }
\Output{$(\widehat {\vec a}, \widehat b) \in \mathbb{Z}_q^{n+1}$ encrypting message $f(m) \in \mathbb{Z}_t$ }
% \KwResult{$y = x^n$}
\inlinecomment{Modulus Switching}\;
$a_i' \gets \lfloor \frac{a_i \cdot 2N}{q} \rceil \mod 2N$ for $i \in [0,n)$\;
$b' \gets \lfloor \frac{b \cdot 2N}{q} \rceil \mod 2N$\;
\inlinecomment{Blind Rotation}\;
$acc \gets (0, \lut \cdot X^{-b'}) \in \rlwe$\;
\For{i = 0 to n-1}{
$h \gets X^{a'_i} \odot acc - acc$\;
$d \gets h \diamond K_i$\;
$acc \gets acc+ d $\;
}
\inlinecomment{Key Switching}\;
$(a'',b'') \gets acc$\;
$(a''',b''') \gets (0,b'') - (a'' \otimes \ksk)$\;
\inlinecomment{Sample Extraction}\;
$\alpha \gets (a'''_0, -a'''_{N-1}, \ldots, -a'''_{1})\in \mathbb{Z}_q^N$\;
\Return ($ (\alpha_0,\ldots, \alpha_{n-1}), b'''_0) \in \mathbb{Z}_q^{n+1}$  \;
\end{algorithm}

\subsubsection{Overview of TFHE PBS}
 
An algorithm overview is given in \cref{alg:pbs_algo}. We follows the description given by \cite{zama_TremblayThibault2025,hasteboot} where the key switching is done first before sample extraction. More details can be found in \cite{Chillotti2019}. 

TFHE programmable bootstrapping (PBS) shown in pseudo-code form at \cref{alg:pbs_algo} takes as input public parameters:
\begin{itemize}
    \item a decomposition base $B \in \mathbb{Z}^+$ and the resulting number of splits $\ell = \lceil \log_B q \rceil$,
    \item a bootstrapping key $\bsk = \{K_i\}_{i=0}^{n-1} = \{ \mathsf{RGSW}_z(s_i)\}_{i=0}^{n-1}$ where $z \in R_q$ is a bootstrapping secret key.
    \item a key switching key $\ksk = \mathsf{RLev}_{\widehat s}(z)$ where $\widehat s$ is the zero-padded LWE secret key $\vec s$ and $z$ is the bootstrapping secret key $z$. 
\end{itemize}
together with the input an LWE ciphertext $(\vec a,b) = \langle \vec a,\vec s\rangle+\Delta \cdot m + e$ under the secret key $\vec s \in \{0,1\}^n$, it outputs a LWE ciphertext encrypting $f(m)$ under the original LWE secret key $\vec s$ with a refreshed noise. The PBS operation consists of four stages.
\begin{description}
    \item[Modulus Switching:] This stage maps the input LWE ciphertext from $q$ to $2N$ to be compatible with the rotation operation during the blind rotation stage.
    \item[Blind Rotation:] This stage iteratively computes the CMUX gate $\mathsf{CMUX}(K_i, acc, X^{a_i'}\odot acc)$ which conditionally rotates the accumulator according to the encrypted secret-key bit $s_i$. At the end of the blind rotation stage, we have that the total rotation is captured by $\lut \cdot X^{-b'+\sum_i a_i' s_i}$ which rotates the $\lut$ by the modulus-switched encoding of $-(\Delta m)$ with accumulated error $\tilde e$. Assuming the lookup-table polynomial $\lut(X)$ is encoded with the negacyclic property and sufficient error-tolerant redundancy, the error $\tilde e$ will be absorbed and the constant coefficient of the resulting ciphertext encodes  $\Delta f(m)$.
    \item[Key Switching:] Given a ciphertext encrypted under the bootstrapping secret key $z$, key switching stage uses a key switching key $\ksk = \mathsf{RLev}_{\widehat s}(z) $ and the operation $\otimes$ to turn the given ciphertext into a new ciphertext encrypted under the new key $\widehat s(X)$ while retaining the same underlying encrypted message. For PBS, we set $\widehat s(X)$ to be a polynomial that encodes the original LWE secret key $\vec s$ with zero-padding.
    \item[Sample Extraction:] The key switching stage described above leaves the ciphertext as an RLWE ciphertext. To continue evaluation of an FHE circuit, one needs to turn the RLWE ciphertext back into an LWE ciphertext. Using the property of negacyclic polynomial multiplication where $[a'''\widehat s]_0 = a'''_0s_0 - \sum^{N-1}_{i=1} a'''_i s_{N-i}$ and the fact that the message is encoded in the constant coefficient $b_0'''=\langle \vec \alpha, \vec s\rangle + \mu_0+e_0$ where $\mu_0 = \Delta f(m)$, the sample extraction can then set $\alpha = (a'''_0, -a'''_{N-1},\ldots,-a'''_1)$ so that $\langle \alpha, \widehat s \rangle= a'''_0s_0 - a'''_{N-1}s_1 - \ldots-a'''_1 s_{N-1} = [a'''\widehat s]_0$. Since $s_i=0$ for $i \in [n, N)$, we can truncate $\alpha$ to $(\alpha_0, \ldots,\alpha_{n-1})$ without issue.
\end{description}

\section{Verifiable TFHE Details}
\label{sec:tfhe_zkp_details}
Since for the construction, we consider $q$ to be prime. Therefore, we use $\mathbb{F}_q$ which matches the proving field $\mathbb{F}$ of the underlying proving system instead to denote each element of $\mathbb{Z}_q$ used in the previous section.

\subsection{Non-Arithmetic Relations}
\subsubsection{Decomposition Proof}
\label{sec:decomposition}
In blind rotation and key switching, TFHE decomposes the polynomials into $\ell$-numbered B-bounded polynomials, that is given a coefficient vector $\vec a$, the decomposition splits $\vec a$ into $\vec a_0,\vec a_1,\ldots,\vec a_{\ell-1}$ such that I) each coefficient in $a_i=(a_{i,0},\ldots,a_{i,N-1})$ satisfies $a_{i,j} \in [0,B)$ and II) $\vec a = \sum_{i=0}^{\ell-1} \vec a_i \cdot B^i < q$.

LogUp \cite{habock2022multivariate} can be used to solve condition I) efficiently. For II) one can use the generic recursive lexicographic bit-comparison that compares bit one by one \cite{hasteboot} which has the following form for two bits:
$$\indicator{a<c}=\indicator{a_1<c_1} + \indicator{a_1=c_1}\cdot \indicator{a_0<c_0}$$
where $a=a_0 + a_1\cdot B$ and $c= c_0 +c_1\cdot B$.

Alternatively, for $q= 2^{64}-2^{32}+1=(d_0,d_1,d_2,d_3,d_4,d_5,d_6,d_7)=(1,0,0,0,255,\allowbreak 255,255,255)$ in byte representation and $B=2^8$ considered in this paper, the largest representation $q-1$ is $(0,0,0,0,255,\allowbreak 255,255,255)$. Therefore, it is sufficient to check that either III) the highest 4 digits are not all $255$ or IV) if the highest 4 digits are all $255$, then the bottom $4$ digits must be all zero. 

Let $\sigma_{top}= (255-d_4)+(255-d_5)+(255-d_6)+(255-d_7)$ and $\sigma_{bot}=d_0+d_1+d_2+d_3$. Note that condition I) already make sure that $\sigma_{top}$ and $\sigma_{bot}$ cannot overflow/underflow. We then verifies the following
\begin{itemize}
    \item $CP_{D1}= \sigma_{top} - \cdot w \cdot \sigma_{top} \cdot \sigma_{top}  = 0$,
    \item$ CP_{D2}=\sigma_{bot} - w\cdot\sigma_{top}\cdot \sigma_{bot}=0.$
\end{itemize}

For the case that III) holds ($\sigma_{top}\neq 0$), it immediately implies $w =  \sigma_{top}^{-1}$ for $CP_{DP1}=0$ to holds, then $CP_{D2}=0$ trivially holds. For the case that III) does not holds  ($\sigma_{top} =0$), we want IV) to holds. Note that for this case $CP_{D1}=0$ holds trivially and $CP_{D2}$ reduces to $\sigma_{bot} = 0$.
Since condition I) ensures each $d_i \in [0, 255]$, the integer sum $\sigma_{bot} \in [0, 1020] \ll q$, so $\sigma_{bot} = 0$ in $\mathbb{F}$ implies $\sigma_{bot} = 0$ as an integer, forcing all bottom digits to zero.

\subsection{TFHE Relations}
\subsubsection{Modulus Switching}
Modulus switching phase maps $a \in \mathbb{Z}_q$ to $b \in \mathbb{Z}_{2N}$ defined as follows,
$$b = \lfloor \frac{a\cdot 2N}{q}\rceil \mod 2N.$$
\cite{hasteboot} shows that, assuming $4N | q-1$, then let $k=\frac{q-1}{4N}$,  except the case that $a=k$, for each $b' \in [2N]$, the segment 
$$[(2b'-1)\cdot k +1, (2b'+1)\cdot k]$$  
is mapped to $b \in \mathbb{Z}_{2N}$ satisfying $b' \equiv b \mod 2N$. Therefore, it is sufficient to verify the following
\begin{itemize}
    \item $b \in \{0,\ldots,2N-1\}$, $b'\in [2N]$, $b\equiv b' \mod 2N$,
    \item $\indicator{a \neq k} = \indicator{a-(2b'-1)\cdot k - 1 \in [0,2k)}$,
    \item $\indicator{a = k} \rightarrow\indicator{b=0}$\footnote{The \cite{hasteboot} uses equality but the segment $[q-k,k-1] \mod q$ (obtained by setting $b'=2N$) maps to $b=0$}.
\end{itemize}

\subsubsection{Blind Rotation}
We use the fact that multiplication of polynomial ring $f(X)$ by the monomial $X^{a_i'}$ for some $a_i' \mod 2N$ is equivalent to multiplying to slot $j$ of NTT representation $\widehat f_j$ the term $\psi^{(2j+1)\cdot a_i'}$ which is the $j$-th NTT term for the monomial $X^{a_i'}$ where $\psi^{a_i'}$ is computed  by binary exponentiation chain $\psi^{a_{i}'}= \psi^{\sum^{\log N}_{k=0} b_k \cdot 2^k} = \prod^{\log N}_{k=0} b_k \cdot \psi^{2^k}$, $b_k$ is the bit decomposition of $a_i$ and $\psi^{2^k}$ is precomputed. 
Let $\rho = \psi^{2a_i'}$ and $t_0= \psi^{a_i'}$.
$\psi^{(2j+1)\cdot a_i'}$ can be computed by the simple relation $t_j=t_{j-1} \cdot \rho=\psi^{(2(j-1)+1)\cdot a_i'} \cdot \psi^{2a_i'} = \psi^{(2j+1)\cdot a_i'}$.

Using the linearity of NTT transform, we check the constraint that the decomposed NTT (decomposition correctness is checked by committing to another equivalent coefficient domain representation using $\name$ and then apply \cref{sec:decomposition}) can be re-composed back to $\widehat h \gets \NTT(X^{a'_i} \odot acc - acc)$. 
Then,  arithmetic operations such as $\diamond,+,-,\odot $ are straightforward since the NTT representations are encoded in the trace row.

\subsubsection{Key Switching.} Key switching requires proving the arithmetic relation of $(a''',b''')=(0,b'') - (a'' \otimes \ksk)$. Digits decomposition technique from \cref{sec:decomposition} is used here. Then, it is straightforward arithmetic constraint on ring operations using $\name$ to convert back and forth between the NTT representation and the coefficient representation whenever necessary. 

\subsubsection{Sample Extraction}
Given an RLWE ciphertext $(a(X),b(X))$, sample extraction is a simple permutation on the coefficients of $a(X)$ and masking on $b(X)$ so that only constant coefficient remains. Masking and permutation can be checked as part of public inputs to the next PBS invocation proof similar to \cite{zama_TremblayThibault2025} or be checked using a simple multiplication of masking polynomial inside AIR and (partial negated) equality check with the correct row indexing.

\section{Lattice-based Multi-Asset Encrypted ERC-20/ERC-1155 Tokens with Verifiability}
\label{sec:erc20_complete}

\newcommand{\createaccount}{\mathsf{CreateAddress}}
\newcommand{\ledger}{\mathsf{SC}}
\newcommand{\coinlist}{\mathsf{CoinList}}
\newcommand{\val}{\vec{\mathsf{v}}}
\newcommand{\balance}{\vec{\mathsf{bal}}}
\newcommand{\owner}{{\mathsf{owner}}}
\newcommand{\registeraccount}{\mathsf{Register}}
\newcommand{\setup}{\mathsf{Setup}}
\newcommand{\tx}{\mathsf{Tx}}
\newcommand{\supplytoken}{\mathsf{SupplyTk}}
\newcommand{\burn}{\mathsf{Burn}}
\newcommand{\burnval}{\mathsf{Burn}}
\newcommand{\spent}{\mathsf{Spent}}
\newcommand{\mint}{\mathsf{Mint}}
\newcommand{\createcoin}{\mathsf{CreateCoin}}
\newcommand{\local}{\mathsf{Local}}
\newcommand{\transfer}{\mathsf{Transfer}}
\newcommand{\balanceof}{\mathsf{BalanceOf}}
\newcommand{\preparetx}{\mathsf{PrepareTx}}
\newcommand{\decryptbalance}{\mathsf{DecryptBalance}}
\newcommand{\ethaddress}{\mathsf{Address}}
\newcommand{\sender}{\mathsf{Sen}}
\newcommand{\receiver}{\mathsf{Rec}}
\newcommand{\issuer}{\mathsf{Issuer}}
\newcommand{\original}{\mathsf{Ori}}
\newcommand{\inputlength}{\ell_{\mathsf{inp}}}
\newcommand{\findcoin}{{\mathsf{FindCoin}}}
\newcommand{\parseval}{{\mathsf{ParseVal}}}
\newcommand{\verifytx}{{\mathsf{VerifyAndExecuteTx}}}
\newcommand{\valueset}{{\mathcal{V^*}}}
\newcommand{\balancegame}{\mathsf{Balance}}
\newcommand{\privategame}{\mathsf{Privacy}}

\newcommand{\corruptlistoracle}{\mathsf{CorruptList}_{\mathcal{O}}}
\newcommand{\mintrequest}{\mathsf{MintReqList}}
\newcommand{\valuelistoracle}{\mathsf{ValList}_{\mathcal{O}}}
\newcommand{\txlistoracle}{\mathsf{TxList}_{\mathcal{O}}}
\newcommand{\txlistadversary}{\mathsf{TxList}_{\adversary}}
\newcommand{\txlistadversarysubmitted}{\mathsf{SubmittedTxList}_{\adversary}}
\newcommand{\accountlistoracle}{\mathsf{Account}_{\mathcal{O}}}
\newcommand{\totalburnoracle}{\mathsf{Total}_{\val_\burn}}
\newcommand{\reqvaloracle}{\mathsf{Req_{\val_\mint}}}
\newcommand{\totalvaloracle}{\mathsf{Total_{\val_\mint}}}
\newcommand{\corrupt}{\mathsf{Corrupt}}
\newcommand{\honest}{\mathsf{Honest}}
\newcommand{\instruction}{\mathsf{inst}}
\newcommand{\params}{\mathsf{params}^{\mathcal{O}}}
\newcommand{\fn}{\mathsf{function}}

ERC-20 \cite{vogelsteller2015erc20} is a widely adopted standard for fungible tokens on the Ethereum blockchain. It defines a common set of rules and functions that enable digital assets to interact seamlessly with wallets, exchanges, and smart contracts. However, each ERC-20 contract typically represents a single asset type, resulting in inefficiencies when hundreds or thousands of assets must be managed or traded.
ERC-1155 \cite{radomski2018erc1155} addresses this limitation by allowing a single smart contract to manage hundreds or thousands of distinct asset types. It also supports batch operations, enabling multiple asset types and quantities to be transferred within an atomic transaction. Note that ERC-1155 can also manage non-fungible tokens, such as by setting the total supply of a specific asset within an ERC-1155 to one unit. 
Multi-asset tokens that represent different asset classes are useful for financial institutions as they can be used for DvP, bond exchange\cite{zhu2025towards}, repo tokenization \cite{qpadl}, securities markets \cite{Avgouleas2019}, and others.

\textbf{Related Works.} In the classical setting, Zether \cite{bunz2020zether}, a single-asset confidential transaction scheme, provides a framework for achieving ERC-20-like transactions with confidentiality and demonstrates the practicality of the proposed system via concrete gas-cost measurements. In the post-quantum setting, most of the exiting works focus on lattice-based RingCT-like schmes such as MatRiCT \cite{matrict_2_2022, Esgin2019} for transacting with single-asset tokens. Recently, Lether \cite{lether} looks at account-based Zether-like \cite{bunz2020zether} lattice-based single-asset transaction scheme. However, it lacks any feasibility analysis for smart-contract based token such as ERC-20. Similarly, while PQ-TaDL \cite{qpadl} focuses on the multi-asset transaction scheme, it lacks gas cost analysis which is crucial for on-chain deployability.

\textbf{Challenges.}
Gas is a unit that measures the computational effort required to execute operations on the EVM.
Gas limits ensure blocks can be propagated and validated quickly across the network, preventing state bloat and keeping the blockchain feasible to run on commodity hardware which preserves decentralization.
In addition, gas limits prevent denial-of-service attacks by bounding the computation any single transaction or block can consume, ensuring the network remains usable for all participants.
As documented in \cite{ethereum_storage, evm_codes}, calldata costs about 16 gas per byte or 512 gas units per 32 bytes. A write to the persistent storage for Ethereum costs about 22,100 gas units per 32 bytes which is 43 times more expensive than calldata. Assuming the ideal scenario where the slot is already written to, the cost can drop to $2,900$ gas units per 32 bytes. Lether \cite{lether}, an account-based (single-asset) transaction scheme, homomorphically adds the ciphertext to update the account state. The ciphertext is around $17$KB and at minimum the number of account to update is $2$ (input and output) which costs around $3.155$ million gas units just to perform a state update for a single asset. Thus, the gas cost will exceed the single-transaction gas cost limit of around 16 million gas units ($2^{24}$) \cite{rebuffo2024eip7825} for asset size of around $6$ for state updates alone without accounting for any verification arithmetic. Therefore, to support up to thousand of assets using the approach from \cite{lether} would incur substantial gas cost.
Lether further make use of LNP22 \cite{LNP22} to prove various linear relations as well as performing the approximate range proof and the verifiable encryption proof. LNP22 proof requires both the coefficient form of the quotient ring polynomial for calculating the norm, as well as the NTT representation to perform polynomial multiplication efficiently. The cost of performing the zero-knowledge proof verification as well as the NTT transform quickly add up and make fitting the verification on-chain in single transaction to be not possible when combined with the account state updates.
PQ-TaDL \cite{qpadl}, while supporting multi-asset transactions in a more compact form rather than incurring a linear increase in cost, faces challenges similar to the above due to the use of LNP22\cite{LNP22} which is not gas-efficient because of the need to perform negacyclic polynomial arithmetic and have access to the coefficient form for short-norm checks.

\textbf{Our Work.}
We are interested in an encrypted version of ERC-1155 for a multi-asset setting in which transfer amounts are hidden (value confidentiality). Although our proposed construction does not provide account anonymity out of the box, output-account anonymity can be introduced by sending transactions to decoy output accounts. Input-account anonymity can be provided using a one-out-of-many proof \cite{Groth2015} or the encrypted table-based ledger approach presented in \cite{qpadl}. Both approaches require modifications to the underlying zero-knowledge proof (ZKP). Note that providing anonymity requires changes to the underlying transaction model as the ERC-20/ERC-1155 token is defined using a one-to-one transaction interface.
We focus on the core mechanisms required to support payments in an ERC-1155-like multi-asset setting. In particular, we omit delegated token spending, commonly implemented through approvals, as well as safe-transfer checks designed to prevent users from accidentally transferring tokens to (smart contract) addresses that do not support them. Both can be treated as extensions and both are not modeled by Zether \cite{bunz2020zether} as well. Our work resolves the gas limit issue by using $\name$ with STARK as well as using UTXO style token whose contract-callable state is succinctly representable by a single hash digest which only takes up 32 bytes of storage space. The verifiably extractable commitment material is communicated via calldata which is very cheap as it does not need to be callable within the smart contract and is only used during the initial verification function call. The wallet owner can always retrieve the calldata since it is stored permanently on-chain \cite{ethereum_storage}.

\textbf{Double-Spending, Replay Attack, and Front-running Protection. } 
Under the blockchain's consensus-safety assumptions (finality) \cite{buterin2019casperfriendlyfinalitygadget}, finalized blocks define a stable total order of transactions. The EVM applies these transactions sequentially as deterministic and atomic state transitions. Therefore, finalized blockchain state can be modeled as serial state-machine replication \cite{evm_state_machine}. Consequently, among competing transactions attempting to spend the same token, at most the first transaction in the finalized execution order can successfully remove it from $\ledger$. Every subsequent attempt observes that the token is no longer unspent and therefore reverts. Thus, no finalized ledger history can contain two successful spends of the same token. $\tokenname$ prevents double spending by combining this execution property with the invariant that every spending operation must verify that the token is unspent and then remove it from $\ledger$. We note that the token for $\tokenname$ is not totally unique (hash output collisions and input collisions) but the $\ledger$ does keep track of the tokens (even similar tokens) in an account-specific token list and spending one token will remove one such token from the corresponding list. The double-spending protection works over this non-unique structure due to the serial state-machine nature of $\ledger$.  

Similarly, $\tokenname$ is not susceptible to \textit{replay} attacks where the transaction is replayed similar to the underlying blockchain transaction as the forger needs to essentially forge a signature on a new message due to incrementing account nonce. In Zether \cite{bunz2020zether}, the ZKP is generated with respect to the account state which makes denial-of-service front-running possible by continuously triggering a state update to the user's account before the transaction from the user is processed. However, a similar denial-of-service front-running is not possible on $\tokenname$ as the account state is a collection of currently possessed unspent tokens.

\subsection{$\tokenname$ Algorithms}
\label{sec:northstaralgo}
At the EVM protocol level, there are two types of account namely the externally owned account (EOA) and the contract account. An EOA is controlled by the possession of the corresponding private key and can initiate
transactions by producing a valid signature, while a contract account (smart contract) is controlled by code, has no private key, and only acts when triggered by an EOA or by other smart contract. When an EOA invokes  a smart contract functionality that alters the blockchain state\footnote{It is different from a local call that only executes the smart contract functionality locally such as reading the state.}, the EOA needs to use its corresponding private key to sign the transaction invoking the smart contract functionality. If the smart contract function is called by an EOA, the EOA address invoking the smart contract functionality can be retrieved by the smart contract function using the variable $msg.sender$ in Solidity or $caller$ is other languages. For the remainder of the paper, we will use Solidity's syntax. If the contract is called by another contract, then \textit{msg.sender} is the calling contract's address. 
For contract-based callers, such as smart contract wallets \cite{multisign_smartcontract}, authorization is determined
by the contract's own code rather than by a protocol-level EOA signature. For example, a multi-signature wallet must implement the required signature-checking and threshold logic internally. We do not formally analyze such contract-based authorization mechanisms in this work. 
Note that the function selector and function parameters are encoded in the transaction's calldata, which is part of the signed transaction message.
The property presented above applies to general EVM-based blockchains such as Ethereum \cite{wood2014ethereum}, Avalanche \cite{avalanche}, Polygon \cite{polygon}, and others.

Moreover, we use the issuer's ``$\mint$'' instead of user's ``$\mathsf{Fund}$'' from Zether as our scheme is meant to be useful for multi-asset tokenization schemes while $\mathsf{Fund}$ is more useful to ties the intrinsic value of the underlying blockchain token to the smart contract token. In this particular use case of multi-asset tokenization, $\mint$ will be triggered by the smart contract issuer who has its own business logic for issuing multi-asset token while $\burn$ will also trigger custom business logic. We do not model any business logic here but we still model the case that no extra token value is created outside the $\mint$ process and token value consumed by $\burn$ is equal to the claimed value.
First, we define the $\tx_\eoa$ transaction object formally below that captures the properties described above. Note that for Zether \cite{bunz2020zether}, the property for $\tx_{\eoa}$ described below is implicitly assumed and embedded into the Zether description.
\begin{definition}[Smart Contract Call using Blockchain Transaction $\tx_\eoa$]
\label{def:tx_eoa}
$\tx_{\eoa}$ is defined to be $\tx_{\eoa} := (\vk_\sender,\ethaddress_{\sender}\footnote{For certain signature schemes, the sender public key and its derived address can be recovered from a valid signature. For compatibility with general signature scheme, we include the sender public key and the derived address directly. }, \ethaddress_{\ledger},\mathsf{function},\allowbreak \mathsf{params}, \aux, \sigma)$ with respect to some digital signature scheme $\DS$. $ret \gets \verifytx(\tx_{\eoa})$ is an algorithm that performs the following:
\begin{itemize}
    \item $\deriveaddress(\vk_\sender) =\ethaddress_{\sender}$,
    \item $\DS.\verify(\vk_\sender, (\ethaddress_{\ledger},\mathsf{function},\mathsf{params}, \aux), \sigma)=1$.
    \item If the verification above passes, the smart contract function is invoked $ret \gets \mathsf{function}(\mathsf{params})$ and is implicitly passed in additional inputs $msg.sender$ and $\ledger$ where $msg.sender = \ethaddress_{\sender}$ and $\ledger$ is the smart contract ledger, then the return value $ret$ is returned,
    \item otherwise return $\bot$, 
\end{itemize}
where $\deriveaddress(\vk_\sender) := \text{0x}||\mathsf{ToHexChar}(\mathsf{Keccak_{256}}(\vk))[24:]$ is a deterministic algorithm that takes in a verification key $\vk$ and outputs an address $\ethaddress$ by computing Keccak hash function \cite{keccak_sha3} and takes the rightmost 20 bytes (40 hex characters). 
$\aux$ contains all other values contained in the signed messages such as $nonce$, $chainID$, amount and other auxiliary information. $\ethaddress_\ledger$ is the address identifier for the smart contract ledger $\ledger$.

\end{definition}

Let $\hash: \mathbb{F}_q^* \rightarrow \mathbb{F}_q^2$ be a random oracle. Let $\numasset$ be the number of assets such that $\numasset < N-\lceil \frac{\secpar}{\log_2(q)}\rceil$. 
We define $< $ on two vectors $\vec a = (a_1,\ldots, a_n)$ and $\vec b = (b_1, \ldots, b_n)$, $\vec{a} < \vec{b} \;\iff\; \forall\, i \in \{1,\dots,n\},\; a_i < b_i$. 
The algorithms to interact with $\tokenname$ including $\setup,\transfer,\burn$ are described below splitted into four categories, namely user (off-chain), honest issuer (off-chain), honest smart contract (on-chain), and helper subroutines. 

We define two relation $\mathcal{R}_{\transfer}$ and $\mathcal{R}_{\burn}$ as follows:
\begin{align*}
\mathcal{R}_{\burn} = \left\{ 
\begin{array}{c}
     \bigl( (\numasset,\commitkey, \token, \commitment') ; (\vec r, \vec m)\bigr): \\
    \vec m = (m_1, \ldots,m_N) \wedge \vec m^* = (m_1, \ldots,m_{\numasset}, 0_{1},\ldots, 0_{N-\numasset})  \\
    \wedge\ \commitment' = \commitscheme_{\extract}.\commit(\commitkey, \vec m^*, \vec r)
    \wedge \norm{\vec r}_{\infty} \leq  \eta \\
    \wedge \ \token = \hash(\vec m) 
\end{array}
\right\},
\end{align*}

\begin{align*}
\mathcal{R}_{\transfer} = \left\{ 
\begin{array}{c}
     \bigl( (\numasset, \commitkey_{\receiver}, \vec \token_{\original},  \token_{\sender}, \token_{\receiver}, \commitment_{\receiver}); \\
      (\vec M_{\original}, \vec m_{\sender}, \vec r_{\receiver}, \vec m_{\receiver})\bigr): \\
    \vec \token_{\original} = (\token_1,\ldots,\token_{\inputlength}) \wedge \vec M_{\original} = (\vec m_{1},\ldots, \vec m_{\inputlength}) \\
    \wedge \ \forall i \in \{\sender,\receiver\}\cup [\inputlength], \vec m_{i} = (m_{i,1}, \ldots,m_{i,N})\\
    \wedge \ \forall j \in [\numasset], m_{i,j} \in [0,2^{64}) 
    \wedge \ \forall j \in [\numasset],  \sum_{i=1}^{\inputlength} m_{i, j} = m_{\sender,j}+m_{\receiver,j}\\
    \wedge\ \forall i \in \{\sender,\receiver\}\cup [\inputlength], \token_i= \hash(\vec m_{i}) \\
    \wedge\ \commitment_{\receiver} = \commitscheme_{\extract}.\commit(\commitkey_{\receiver}, \vec m_{\receiver}, \vec r_{\receiver})
    \wedge\ \norm{\vec r_{\receiver}}_{\infty} \leq \eta
\end{array}
\right\}.
\end{align*}

As an optimization, note that $\mathcal{R}_{\transfer}$ does not prove the correctness of secondary commitment $\commitment_{\sender}$ for efficiency purpose, it does not affect the security of the scheme as the sender's  $\commitment_{\sender}$ is here to recover the token message/value and the sender created the commitment for its own account.
If the sender's $\commitment_{\sender}$ is malformed, the only downside is that the sender (who submitted the transaction) cannot decrypt its own token balance if it is stateless. 
The object stored inside $\ledger.\coinlist[\ethaddress]$ indexed by $\ethaddress$ is a multiset. 
% \yeoh{formalize this?}

\medskip\noindent\textbf{Setup Algorithm:}
\begin{itemize}
\item  $\publicparam \gets \ppgen(\secparam):$ $\ppgen$ algorithm computes  
        \begin{itemize}
            \item $\publicparam_\nizk \gets \nizk.\ppgen(\secparam)$
            \item $(\provingkey_{\mathcal{R}_{\transfer}}, \verificationkey_{\mathcal{R}_{\transfer}}) \gets \nizk.\keygen_{\mathcal{R}_{\transfer}}(\mathbb{i}_{\mathcal{R}_{\transfer}},\publicparam_\nizk) $
            \item $(\provingkey_{\mathcal{R}_{\burn}}, \verificationkey_{\mathcal{R}_{\burn}}) \gets \nizk.\keygen_{\mathcal{R}_{\burn}}(\mathbb{i}_{\mathcal{R}_{\burn}},\publicparam_\nizk)$ 
            \item returns $\publicparam := (\publicparam_{\nizk}, \provingkey_{\mathcal{R}_{\transfer}}, \verificationkey_{\mathcal{R}_{\transfer}},\provingkey_{\mathcal{R}_{\burn}}, \verificationkey_{\mathcal{R}_{\burn}})$
        \end{itemize}
\end{itemize}

Public parameter $\publicparam$ is passed in implicitly to algorithms below.

\medskip\noindent\textbf{User Off-Chain Algorithms:}
\begin{itemize}
    \item $(\sk, \vk, \ethaddress, \extractorkey, \commitkey_{\ethaddress})
          \gets \createaccount(\secparam)$:  $\createaccount$ algorithm generates a key pair $(\sk,\vk) \gets \DS.\keygen(\secparam)$, parses to address $\ethaddress \gets \deriveaddress(\vk)$, generates $(\commitkey_{\ethaddress}, \extractorkey) \gets \commitscheme_{\extract}.\commitgen(\secparam)$, returns $(\sk, \vk, \allowbreak \ethaddress, \extractorkey, \commitkey_{\ethaddress})$.
    \item $\tx_{\registeraccount}
          \gets \registeraccount.\local(\sk, \vk, \ethaddress, \commitkey_{\ethaddress})$:
          $\registeraccount.\local$ algorithm, constructs the register transaction $\tx_{\registeraccount}:= (msg, \DS.\sign(\sk, msg))$ where $msg := (\vk, \ethaddress, \ethaddress_{\ledger}, \registeraccount.\verify, (\commitkey_{\ethaddress}), \aux)$  with $\aux$ from \cref{def:tx_eoa}, outputs a register transaction $\tx_{\registeraccount}$.
    \item $\tx_{\transfer}
    % \vec \token_{\sender}
          \gets \transfer.\local(\sk, \vk, \extractorkey, \ethaddress_{\sender}, \ethaddress_{\receiver}, \val_{\receiver})$:
          $\transfer.\local$ algorithm computes
          \begin{itemize}
          \item $(\vec \token_{\original}, \vec M_{\original}) \gets 
          \findcoin(\ethaddress_{\sender}, \val_{\receiver}, \extractorkey)$,
          \item $\val_{\sender} = \sum_{\vec m_i \in \vec M_{\original}}\parseval(\vec m_i)$ - $\val_{\receiver}$,
          \item  $(\token_{\sender}, \commitment_{\sender}, (\vec r_{\sender},\vec m_{\sender})) \gets \createcoin(\ethaddress_{\sender}, \val_{\sender}) $,
          \item  $(\token_{\receiver}, \commitment_{\receiver}, (\vec r_{\receiver},\vec m_{\receiver})) \gets \createcoin(\ethaddress_{\receiver}, \val_{\receiver}) $,
            \item $\pi \gets \nizk.\prove_{\mathcal{R}_{\transfer}}(\provingkey_{\mathcal{R}_{\transfer}},(\numasset, \ledger.\commitkey[\ethaddress_{\receiver}], \vec \token_{\original}, \allowbreak \token_{\sender}, \allowbreak \token_{\receiver},\commitment_{\receiver}),(\vec M_{\original}, \vec m_{\sender},\vec r_{\receiver}, \vec m_{\receiver}))$,
          \end{itemize}
            constructs a transfer transaction $\tx_{\transfer}:= (msg, \DS.\sign(\sk, msg))$ where $msg := (\vk, \ethaddress, \ethaddress_{\ledger}, \transfer.\verify, (\ethaddress_\receiver, \vec \token_{\original}, \token_{\sender}, \token_{\receiver}, \allowbreak \commitment_{\sender},\allowbreak \commitment_{\receiver},\pi), \aux)$  with $\aux$ from \cref{def:tx_eoa}, outputs the transfer transaction $\tx_{\transfer}$.
    \item $\tx_{\burn}
          \gets \burn.\local(\sk, \vk, \extractorkey, \ethaddress, \token, \commitment)$:
           $\burn.\local$ algorithm computes
           \begin{itemize}
              \item $\vec m \gets \commitscheme_{\extract}.\extract(\commitment,\extractorkey)$,
              \item $\vec m^* := (m_1,\ldots m_{\numasset}, 0_1,\ldots, 0_{N-\numasset})$ where $\vec m :=(m_1,\ldots,m_N)$,
               \item $\commitment' \gets \commitscheme_{\extract}.\commit(\ledger.\commitkey[\ethaddress_{\issuer}], \vec m^*, \vec r )$ where $\vec{r} \sample  \chi^{(\kappa_{\sis}+\kappa_{\lwe}+2n)N}$, 
               \item $\pi \gets \nizk.\prove_{\mathcal{R}_{\burn}}(\provingkey_{\mathcal{R}_{\burn}},(\numasset, \ledger.\commitkey[\ledger.\issuer], \token, \commitment'),(\vec r, \vec m))$,
           \end{itemize}
           constructs a burn transaction $\tx_{\burn}:= (msg, \DS.\sign(\sk, msg))$ where $msg := (\vk, \ethaddress, \ethaddress_{\ledger}, \burn.\verify, (\token,\commitment',\pi), \aux)$  with $\aux$ from \cref{def:tx_eoa}, outputs the burn transaction $\tx_{\burn}$.
    \item $\val \gets \balanceof(\ethaddress, \extractorkey)$: $\balanceof$ algorithm
    \begin{itemize}
        \item     reads $(\vec \token,  \vec  \commitment):= \ledger.\coinlist[\ethaddress]$, 
        \item computes $\val:= \parseval(\sum_{i=1}^{|\ledger.\coinlist|} \commitscheme_{\extract}.\extract(\commitment_i,\extractorkey))$ where  $(\token_i, \commitment_i) \in (\vec{\token}, \vec \commitment)$,
        \item  returns $\val$.
    \end{itemize}
\end{itemize}

\medskip\noindent\textbf{Issuer Off-Chain Algorithms:}
\begin{itemize}
    \item $(\ledger, \ethaddress_{\ledger})
          \gets \setup(\ethaddress, \sk, \commitkey)$:
          $\setup$ algorithm constructs a ledger with an empty state $\ledger$ except with the issuer address set as $\ledger.\issuer := \ethaddress$ and the issuer commit key set as $\ledger.\commitkey[\ethaddress]:= \commitkey$, deploys the contract $\ledger$ onto the chain by first constructing a deployment transaction $\tx_{\mathsf{deploy}}$ signed with $\sk$, sends the $\tx_{\mathsf{deploy}}$ to the chain, and returns the deployed ledger ($\ledger,\ethaddress_{\ledger}$).
     \item $(\token, \commitment, \tx_{\mint}) \gets \mint.\local(\sk, \vk, \ethaddress_{\issuer},\ethaddress_{\receiver}, \val)$:
        $\mint.\local$ algorithm 
        \begin{itemize}
            \item returns ($\emptyset$, $\emptyset,\emptyset$) if $\val$ contains a non-positive value,
            \item otherwise mints a multi-asset token $(\token, \commitment, \cdot) \gets \createcoin(\ethaddress_{\receiver}, \val)$,
        \end{itemize}
         constructs a minting transaction $\tx_{\mint} := (msg,\DS.\sign(\sk,msg))$ where $msg := (\vk, \ethaddress_{\issuer}, \allowbreak \ethaddress_{\ledger}, \allowbreak \mint.\verify, (\ethaddress_{\receiver},\token, \commitment),\aux)$ with $\aux$ from \cref{def:tx_eoa}, then returns $(\token, \commitment, \tx_{\mint})$.
\end{itemize}

\medskip\noindent\textbf{Smart Contract On-Chain  Algorithms:}

\noindent The following algorithms (except $\registeraccount.\verify$) automatically return $b=0$ when any address $\ethaddress$ passed in as a parameter are not registered $\ledger.\commitkey[\ethaddress] = \emptyset$. For brevity, this common preliminary condition check is not written in the algorithms.
Each on-chain algorithm implicitly has access to $msg.sender$ as defined in \cref{def:tx_eoa}.
For any indexable/mappable state variable $state$, it can be read via $\ledger.\mathsf{state}[\cdot]$.

\begin{itemize}
    \item $b \gets \registeraccount.\verify(\commitkey_{\ethaddress})$:
          $\registeraccount.\verify$ algorithm parses $\ethaddress := msg.sender$, checks $\ledger.\commitkey[\allowbreak \ethaddress]$ is
          not already registered, stores $\commitkey_{\ethaddress}$ as $\ledger.\commitkey[\ethaddress]:= \commitkey_{\ethaddress}$, outputs a bit $b:=1$ if there it is a new registration and $b:=0$ otherwise.
    \item $b \gets \mint.\verify(\ethaddress, \token, \commitment)$:
          $\mint.\verify$ algorithm verifies the caller is the issuer $\ledger.\issuer =^? msg.sender$, returns $0$ if it is not the issuer, otherwise appends to the list
          $\ledger.\coinlist[\ethaddress] = \ledger.\coinlist[\ethaddress]+ (\token, \commitment)$, and then returns $1$. 
    \item $b \gets \transfer.\verify(\ethaddress_{\receiver},  \vec {\token}_{\original}, \token_{\sender}, \token_{\receiver}, \commitment_{\sender}, \commitment_{\receiver},\pi )$:
        $\transfer.\verify$ algorithm parses the sender $\ethaddress_{\sender}:= msg.sender$, $\commitkey_{\receiver}:= \ledger.\commitkey[\ethaddress_{\receiver}]$, % =^? \ethaddress_{\sender}
        \begin{itemize}
            % \item 
            \item verifies the proof $1 =^? \nizk.\verify_{\mathcal{R}_{\transfer}}(\verificationkey_{\mathcal{R}_{\transfer}},(\numasset, \commitkey_{\receiver}, \vec \token_{\original}, \token_{\sender}, \allowbreak \token_{\receiver}, \commitment_{\receiver}),\pi)$, 
            \item verifies that each token is available to be spent $(\vec{ \token}_{\original},\cdot) \subseteq \ledger.\coinlist[\ethaddress_{\sender}]$, 
            \item returns 0 if any of the verification fail, 
        \end{itemize}
        otherwise 
        \begin{itemize}
            \item for each $\token \in \vec {\token}_{\original}$, remove from the list $\ledger.\coinlist[\allowbreak \ethaddress_{\sender}]:= \ledger.\coinlist[\allowbreak \ethaddress_\sender] \setminus (\token, \cdot)$ one instance of $(\token,\commitment)$ tuple,
            \item for $i \in \{\sender,\receiver\}$, appends to $\ledger.\coinlist[\ethaddress_{i}] =\ledger.\coinlist[\ethaddress_{i}]+ (\token_{i}, \commitment_{i})$,
            % \item for $i \in \{\sender,\receiver\}$, records the commitment $\ledger.\cmlist[\ethaddress_i] = \ledger.\cmlist[\ethaddress_i] + \commitment'_{i}$
            \item            and outputs $1$.
        \end{itemize}
        % (\numasset,\commitkey, \token, \commitment') 
    \item $b \gets \burn.\verify(\token, \commitment', \pi)$: Algorithm $\burn.\verify$ parses the owner address $\ethaddress := msg.sender$, verifies proof $1 =^?\nizk.\verify_{\mathcal{R}_{\burn}}(\verificationkey_{\mathcal{R}_\burn}, (\numasset, \commitkey_{\issuer} , \token, \commitment'), \pi)$ where $\commitkey_{\issuer} := \ledger.\commitkey[\ledger.\issuer ]$, checks $(\token, \cdot)\in \ledger.\coinlist[\ethaddress] $, returns $0$ if any of the checks fail, otherwise 
    \begin{itemize}
        \item removes one instance of $\token$ from the list $\ledger.\coinlist[\ethaddress]:= \ledger.\coinlist[\ethaddress] \setminus (\token, \cdot)$,
        \item emits an event $(\burn, \ethaddress, \token, \commitment')$, 
        \item returns $1$.
    \end{itemize}
\end{itemize}

\medskip\noindent\textbf{Helper Algorithms:}
\begin{itemize}
    \item $(\token, \commitment, (\vec r, \vec m))\gets \createcoin(\ethaddress_{\receiver}, \val_{\receiver}):$ $\createcoin$ algorithm 
    \begin{itemize}
        \item parses $(v_1,\ldots, v_{\numasset}) := \val_{\receiver}$,       
        \item computes $\vec m := (v_1, \ldots,  v_{\numasset}, \mu_1, \ldots, \mu_{N-\numasset})$ where $(\mu_1, \ldots, \mu_{N-\numasset})\sample  \mathbb{Z}_q^{N-\numasset}$,
        \item $\token \gets \hash(\vec m)$,
        \item $\commitment \gets \commitscheme_{\extract}.\commit(\ledger.\commitkey[\ethaddress_{\receiver}], \vec m, \vec r)$ where $\vec r \sample \chi^{(\kappa_{\sis}+\kappa_{\lwe}+2n)N}$,
        \item returns $(\token, \commitment, (\vec r, \vec m))$.
    \end{itemize}
    \item $(\vec \token, \vec M)\gets \findcoin(\ethaddress, \val, \extractorkey):$ $\findcoin$ algorithm 
    \begin{itemize}
        \item returns $\bot$ if $\balanceof(\ethaddress, \extractorkey) < \val$,
        \item otherwise retrieves $(\vec \token,  \vec  \commitment):= \ledger.\coinlist[\ethaddress]$, 
        finds $\ell$
        where  $(\token_i, \commitment_i) \in (\vec{\token}, \vec \commitment)$ and for any $\ell' < \ell, \parseval(\sum_{i=1}^{\ell'} \commitscheme_{\extract}.\extract(\commitment_i,\extractorkey)) < \val$,
        \item $\vec \token := (\token_1, \ldots, \token_{\ell})$,
        \item $\vec M := (\vec m_{1}, \ldots, \vec m_{\ell})$ where $\vec m_i \gets \commitscheme_{\extract}.\extract(\commitment_i, \extractorkey)$, 
        \item returns $(\vec \token, \vec M)$.
    \end{itemize}
    \item $\val \gets \parseval(\vec m):$ $\parseval$ algorithm parses $(v_1,\ldots,v_{\numasset},\allowbreak \mu_1,\ldots,\mu_{N-\numasset}):= \vec m$, $\val := (v_1,\ldots,v_{\numasset})$, returns $\val$.
\end{itemize}

\begin{remark}
In practice, the components of $(\token,\commitment)\in\ledger.\coinlist$ are stored separately. Since $\token$ must be readable by the smart contract, it is maintained in persistent EVM storage. In contrast, $\commitment$ does not need to be accessed by the smart contract and is recorded in transaction history through calldata or an event log. Any participant may retrieve it from blockchain history, subscribe to the corresponding event, and maintain an indexed off-chain copy for efficient access. This separation saves on gas consumption since $\token$ is only $32$ bytes and $\commitment$ is around KB range. As noted in \cref{remark:commit_optimization}, $A$ from the commitment key $(A,B_{\ethaddress}) := \commitkey_{\ethaddress}$ can be shared across all participants in the same contract to further reduce storage space and $A$ can be expanded from a common seed. Given $\ethaddress_\ledger$, anyone can read the state of $\ledger$ and see all transactions $\tx_{\eoa}$ posted to the blockchain.
\end{remark}

\subsubsection{$\tokenname$ Summary.}

With the algorithms defined, we are now ready to describe the $\tokenname$ use cases. We describe a generic multi-asset tokenization use case while omitting most inputs/outputs of algorithms for conciseness since the inputs/outputs are clear from context. We start with an issuer with address $\ethaddress_{\issuer}$ who sets up a $\tokenname$ smart contract using the algorithm $\setup$. The participant registers its commitment key to the contract with $\registeraccount.\local$ or the issuer could also register the corresponding key for the participant. On deposit from a participant to the issuer's system, the issuer mints using $\mint$ a coin of the following form:

\begin{align}
    coin =
    \begin{bmatrix}
    \token\\
    \commitment \\
    \end{bmatrix}
    =
    \begin{bmatrix}
    \mathbf{0} \\
    \mathbf{A} \\
    \mathbf{pk}_{1} \\
    \mathbf{pk}_{2}
    \end{bmatrix}
    \mathbf{r}
    +
    \begin{bmatrix}
    \hash({m})\\
    \mathbf{0} \\
    m \\
    \sqrt{q} \cdot m \\
    \end{bmatrix}
\end{align}
where $m := (v_1,\ldots,v_{\numasset},r_1,\ldots,r_{N-\numasset})$, $v_i$ is the asset value deposited by the participant for an asset $i$, $r_i$ is a masking factor, $(\mathbf{A},\pk_1,\pk_2)$ is the commitment-key tuple $\commitkey$ associated with the coin's owner.
The participants can now trade the $coin(s)$ using the $\transfer$ operation or redeem the asset from the $coin(s)$ using the $\burn$ operation. $\transfer$ operation will take sender's coin(s) as input and then outputs one sender balance coin and one new coin for the receiver. Note that the spent input coin(s) are removed and the newly minted output coin(s) are recorded in an atomic manner. Since the system operates on top of a replicated state machine (EVM) with strong consistency\footnote{Distributed nodes process the operations in the exact same sequential order.}, the atomic transaction where coin(s) are removed and new coins are admitted prevents any double-spending from happening. On the other hand, $\burn$ takes the specified $coin(s)$ and notifies the issuer the withdrawal token values. Throughout the lifecycle of the system, the asset values remain hidden from observers.

\subsection{Correctness and Security Requirements}

We derive our correctness and security requirements from \cite{bunz2020zether} but adapted to the use case of multi-asset tokenization. 

\textbf{Oracle.}
The adversary can interact with the smart contract using the user algorithms defined previously.
For convenience, we define $\valueset$ to be a set of value such that a value list $\val \in \valueset$ if and only if $\forall v_i \in \val, v_i \in [0,2^{64})$.

We define the following bookkeeping states for the experiment:
\begin{itemize}
    \item $\txlistadversarysubmitted$: It records transactions $\tx$ sent by the adversary to the blockchain that has been verified in a list regardless of whether the corresponding smart contract function is executed successfully.
    \item $\txlistadversary$: It records transactions $\tx$ sent by the adversary into a list if the corresponding smart contract function has been executed successfully. Note that $\txlistadversarysubmitted \subseteq \txlistadversary$.
    \item $\txlistoracle$: It records transactions $\tx$ generated by the oracle.
    \item $\accountlistoracle$: It records the account information generated in the system.
    \item $\mintrequest$: It records the mint request in the system.
    \item $\reqvaloracle$: It records the total amount of requested mint token value.
    \item $\totalvaloracle$: It records the total amount of minted token value issued by the issuer.
    \item $\burn_{\adversary/\honest}$: It records the total amount of token redeemed out of the system using the burn operation by an adversary $\adversary$ and the honest parties respectively.
\end{itemize}

\noindent In addition, we define the oracles that the adversary may have access to as follows:
\begin{itemize}
     \item $(\vk, \ethaddress, \commitkey_{\ethaddress}) \gets \mathcal{O}_{\createaccount}()$: The oracle computes $\mathsf{account} := (\sk, \vk, \ethaddress, \extractorkey, \commitkey_{\ethaddress}) \gets \createaccount(\secparam)$, records $\mathsf{account}$ to $\accountlistoracle$, then returns $(\vk, \ethaddress, \commitkey_{\ethaddress})$.
    \item $\tx_{\transfer} \gets \mathcal{O}_{\transfer}(\ethaddress_{\sender}, \ethaddress_{\receiver}, \val)$: The oracle checks that if the value list is valid $\val \in \valueset$,  it has access to the account $\ethaddress_\sender \in \accountlistoracle$, 
    then it calculates the current balance $\balance := \parseval(\balanceof(\allowbreak \ethaddress_{\sender}))$, it checks there is enough balance $\val \leq \balance$, it merges all owned tokens $\verifytx( \transfer.\local(\sk_{\sender}, \vk_{\sender},\extractorkey_{\sender}, \allowbreak \ethaddress_{\sender}, \ethaddress_{\sender}, \allowbreak  \balance))$, it transacts using $ \tx_{\transfer} \gets \transfer.\local(\sk_{\sender}, \vk_{\sender},\extractorkey_{\sender}, \allowbreak \ethaddress_{\sender}, \ethaddress_{\receiver}, \allowbreak  \val)$ from the merged balance, records $\tx_{\transfer}$ to $\txlistoracle$ and outputs $\tx_{\transfer}$, otherwise it outputs $\bot$ if any of the checks fail.

    % Value-based burn
    \item $\tx_{\burn}\gets \mathcal{O}_{\burnval}(\ethaddress_\sender, \val)$: 
    The oracle checks that if the value list is valid $\val \in \valueset$,  it has access to the account $\ethaddress_\sender \in \accountlistoracle$, 
    then it calculates the current balance $\balance := \parseval(\balanceof(\allowbreak \ethaddress_{\sender}))$, it checks there is enough balance $\val \leq \balance$,
    it merges all owned tokens $(\cdot,\token_\receiver,\cdot, \commitment_\receiver, \cdot):= \tx_{\transfer}$ where $\tx_{\transfer} \gets \transfer.\local(\sk_{\sender},  \vk_{\sender},\extractorkey_{\sender}, \allowbreak \ethaddress_{\sender}, \ethaddress_{\sender}, \allowbreak  \balance)$, 
    it executes the merge transfer transaction $\verifytx(\tx_{\transfer})$,
    it then proceed to burn $ \tx_{\burn} \gets \burn.\local(\sk_{\sender}, \vk_{\sender}, \allowbreak \extractorkey_{\sender}, \allowbreak  \ethaddress_{\sender}, \token_\receiver, \commitment_\receiver)$, then outputs $\tx_\burn$, otherwise it outputs $\bot$.
    
    \item $\tx_\mint \gets \mathcal{O}_{\mint}(\ethaddress, \val)$: The oracle 
    checks that if the value list is valid $\val \in \valueset$ and the total potential token values will be safe $\reqvaloracle$ + $\val < 2^{64}$,
    \begin{itemize}
        \item it executes $(\token,\commitment, \tx_{\mint}) \gets \mint.\local(\sk_{\issuer}, \vk_\issuer,\ethaddress_{\issuer}, \allowbreak  \ethaddress, \val)$, 
        \item records $\tx_{\mint}$ to $\txlistoracle$, 
        \item records $(\tx_\mint,\val)$ to $\mintrequest$,
        \item records $\reqvaloracle$ = $\reqvaloracle$ + $\val$, 
        \item and outputs $\tx_\mint$,
    \end{itemize}
    otherwise it outputs $\bot$.

\end{itemize}
The oracle implicitly use account information from $\accountlistoracle$ created by oracle query $\mathcal{O}_{\createaccount}$ when answering the query and has access to issuer account information.

The experiment will additionally keep track of the ledger status over adversary-submitted transactions as follows when they are initially submitted to the smart contract $\ledger$. Given a submitted transaction $\tx_\adversary$,
\begin{itemize}
    \item computes $b_{\verifytx} := \verifytx(\tx_{\adversary})$,
    \item if $\tx_{\adversary}$ submitted by $\adversary$,
    appends $\tx_{\adversary}$ to $\txlistadversarysubmitted$,
    \item if the submitted transaction has been successfully executed $b_{\verifytx} =1$, appends $\tx_\adversary$ to $\txlistadversary$, otherwise aborts the remaining operations.
    \item For a transaction that is a burn operation, computes $\val_\burn := \parseval(\commitscheme_\extract\allowbreak.\extract(\commitment, \extractorkey_\issuer))$, if $\ethaddress$ for the burned token $\token$ is an honest account $(\cdot, \ethaddress,\cdot) \in \accountlistoracle$, updates $\burn_{\honest} = \burn_{\honest}  + \val_\burn$. Otherwise, the $\ethaddress$ for the burned token $\token$ is an adversarial account, updates $\burn_\adversary = \burn_\adversary + \val_\burn$. 
    \item For a transaction that is a mint operation, looks up $(\tx_{\adversary}, \val) \in \mintrequest$, records minted value total $\totalvaloracle := \totalvaloracle + \val$.
\end{itemize}

\textbf{Balance.} Similar to \cite{bunz2020zether, lether}, we define a \textit{balance} security requirement that preserves the integrity of the ledger. 
Double-spending protection and transfer/burn balance is implicitly captured by overdraft-safety requirement.
Intuitively, for a correctly set up ledger by an honest issuer, 
the adversary cannot redeem more token values than have been minted to the adversary and the token values it received. In other words, the total token values available to be redeemed cannot exceed the total minted value. 
% the token value minted by the issuer. 
Next, for any transaction submitted to the chain, it is authorized by the owner with a corresponding signature. Finally, for every transfer transaction, the receiver can always recover enough information about the encrypted token to be able to spend the received token and the balance of the token is positive with respect to some bound. 
Note that due to the UTXO-like nature of the construction, there is no epoch-locking requirement present in \cite{bunz2020zether,lether}.

\begin{definition}[Balance]
\label{def:balance}
The $\tokenname$ is balanced if it holds that every PPT adversary $\adversary$ has at most negligible advantage in the following experiment, where we define the advantage as $\advantage_{\balancegame}({\adversary}) := \prob{\balancegame_{\adversary}(\secpar) = 1}$.

\begin{adjustbox}{minipage=\linewidth,scale=0.85, center}
% \begin{minipage}{\linewidth}
    \procb{$\balancegame_{\adversary}(\secpar)$}{
       \publicparam \gets \ppgen(\secparam)\\
        \mathsf{account}_\issuer:= (\sk_{\issuer}, \vk_{\issuer}, \ethaddress_{\issuer}, \extractorkey_{\issuer}, \commitkey_{\issuer})\gets \createaccount(\secparam) \\
        (\ledger,\ethaddress_{\ledger}) \gets \setup(\ethaddress_\issuer, \sk_\issuer, \commitkey_{\issuer}) \\
        \accountlistoracle := \accountlistoracle \cup \mathsf{account}_\issuer\\
         (\cdot) \gets \adversary^{\mathcal{O}_{\burnval}, \mathcal{O}_{\transfer}, \mathcal{O}_{\createaccount}, \mathcal{O}_{\mint}}(\ethaddress_{\ledger}, \publicparam)\\
        \text{The experiment returns $1$ if any of the following is true:}\\
        \text{I) (Overdraft-Safety)}  \\
         \circ \totalvaloracle -  \burn_\honest - \mathsf{HonestBal} < \burn_\adversary, \text{where}\\
         \circ \mathsf{HonestBal} := \sum_{(\cdot,\ethaddress_i,\extractorkey_i,\cdot) \in \accountlistoracle} \sum_{(\token_j, \commitment_j) \in \ledger.\coinlist[\ethaddress_i]} \parseval(\commitscheme_\extract(\commitment_\receiver,\extractorkey_\receiver)) \\
        \text{II) (Unforgeability)}\\
        \circ \exists \tx := (\vk_\sender, \ethaddress_{\sender}, \ethaddress_{\ledger},\mathsf{function},\allowbreak \mathsf{params}, \aux, \sigma) \in \txlistadversarysubmitted, \text{such that}\\
        \circ (\vk_\sender,\ethaddress_{\sender},\cdot) \in \accountlistoracle \ \wedge \ \tx \notin \txlistoracle\\
        \circ \wedge\ \DS.\verify(\vk_\sender, (\ethaddress_\ledger, \mathsf{function}, \mathsf{params},\aux)= 1) 
        \\
        \text{III) (Spendability) } 
        \\
        \circ \exists (\cdot, \ethaddress_\receiver, \extractorkey_\receiver, \cdot) \in \accountlistoracle, \\
        \circ \exists \tx := (\cdot,\transfer.\verify,(\ethaddress_\receiver,\cdot,\token_{\receiver}, \cdot, \commitment_\receiver,\cdot),\cdot) \in \txlistadversary, \text{such that} \\
        \circ \vec m := \hash(\commitscheme_\extract(\commitment_\receiver,\extractorkey_\receiver)),\\
        \circ \hash(\vec m) \neq \token_\receiver \vee \parseval(\vec m) \notin \valueset
    }

% \end{minipage}
\end{adjustbox}

\label{def:balance_game}
\end{definition}

\textbf{Privacy (Transaction Confidentiality). }
The privacy notion is defined similarly to \cite{bunz2020zether} which is based on the left-or-right  indistinguishability-based definition but without sender or receiver anonymity. The adversary can send two \textit{consistent} instructions ($\instruction_{t,0},\instruction_{t,1}$) at some point $t$ and send one instruction for the rest of the time where \textit{consistency} is defined below to rule out trivial cases. Instruction here is defined to be a request to oracles $\mathcal{O}_\burnval,\mathcal{O}_{\mint},\mathcal{O}_{\transfer}$. 
The account owner actively maintains a central coin balance where the received coins are merged into one and the burning or transfer operation occurs from the merged coin. 
Intuitively, transaction values are hidden in mint, transfer, and burn operations and the observing adversary obtain no information about the hidden values. The experiment will execute $b$-th instruction at point $t$ based on the hidden bit $b$ which the adversary attempts to guess. We now define consistency below.

The instruction is a tuple of $\instruction:=(\mathsf{function}, \params)$.
Two instructions $\instruction_{t,0}:=(\mathsf{function}_{t,0},\params_{t,0}),  \instruction_{t,1} :=(\mathsf{function}_{t,1}, \params_{t,1})$ are \textit{consistent} with respect to a vector of instruction $\vec \instruction$, a point $t$ and initial ledger state $\ledger$ denoted as $\mathsf{IsConsistent}(\ledger, t, \allowbreak \vec \instruction, \instruction_{t,0},\instruction_{t,1})= 1$ if they satisfy the following:
\begin{itemize}
        \item The functions specified by the instructions at point $t$ are the same $\mathsf{function}_{t,0} = \mathsf{function}_{t,1}$. For $t^* \neq t$, we have that $\instruction_{t^*,0}= \instruction_{t^*,1}$, i.e. except for point $t$, all other points $t^*\neq t$ execute the same instruction regardless of the hidden bit $b$.
    % for point $t$.
    \item The instructions produce the same behavior regardless of the hidden bit $b$, $\verifytx(\mathcal{O}_{\fn_{t^*,0}}(\allowbreak \params_{t^*,0})) = \verifytx(\mathcal{O}_{\fn_{t^*,1}}(\allowbreak \params_{t^*,1}))$.
    \item  Instructions at point $t$ are contained in the instruction list given, $(\instruction'_{0,0},\instruction'_{0,1}, \instruction'_{1,0}, \allowbreak \instruction'_{1,1},\instruction'_{2,0}, \instruction'_{2,1} \ldots):= \vec \instruction$, $\instruction_{t,0}= \instruction'_{t,0} \wedge \instruction_{t,1} =\instruction'_{t,1}$.
    \item $\fn_{t,b} = \mint$:  Let $(\ethaddress_{t,b},\val_{t,b}):= \params_{t,b}$ for $b \in \{0,1\}$. 
    The receiver account must be the same $\ethaddress_{t,0} = \ethaddress_{t,1}$ but the amount can be different. If the receiver account is an adversarial controlled account ($\ethaddress \notin \accountlistoracle$), then the mint amount must be the same.
    \item $\fn_{t,b} = \transfer$: Let $(\ethaddress_{\sender,t,b},\ethaddress_{\receiver,t,b},\val_{t,b}):= \params_{t,b}$ for $b \in \{0,1\}$. The sender and receiver account are the same across the transfer instructions ($\ethaddress_{\sender,t,0}=\ethaddress_{\sender,t,1}$ and $\ethaddress_{\receiver,t,0}=\ethaddress_{\receiver,t,1}$), the sender/receiver account within the instruction can be different. If the receiver account is an adversarial controlled account ($\ethaddress_{\receiver,t,b} \notin \accountlistoracle$), then the receiver amount must be the same across instructions ($\val_{t,0}=\val_{t,1}$).
    \item $\fn_{t,b} =\burnval:$ Let $(\ethaddress_{t,b},\val_{t,b}):= \params_{t,b}$ for $b \in \{0,1\}$. The account address $\ethaddress_{t,b}$ are the same $\ethaddress_{t,0}= \ethaddress_{t,1}$. 
\end{itemize}
$\verifytx$ is computed with respect to the sequential state update to $\ledger$ caused by the execution of the instruction $\vec \instruction$ given a chosen bit $b$.

\begin{definition}[Privacy]
\label{def:privacy}
The $\tokenname$ is private if it holds that every PPT adversary $\adversary$ has at most negligible advantage in the following experiment, where we define the advantage as $\advantage_{\privategame}({\adversary}) := \prob{\privategame_{\adversary}(\secpar) = 1}$ conditioned on non-aborting $\privategame_{\adversary}(\lambda)$.

\begin{adjustbox}{minipage=\linewidth,scale=0.85, center}
    \procb{$\privategame_{\adversary}(\secpar)$}{
       \publicparam \gets \ppgen(\secparam)\\
        \mathsf{account}_\issuer:= (\sk_{\issuer}, \vk_{\issuer}, \ethaddress_{\issuer}, \extractorkey_{\issuer}, \commitkey_{\issuer})\gets \createaccount(\secparam) \\
        \text{Copy}\ \ledger\ \text{into}\ \ledger_{\mathsf{init}} \\
        (\ledger,\ethaddress_{\ledger}) \gets \setup(\ethaddress_\issuer, \sk_\issuer, \commitkey_{\issuer}) \\
        \accountlistoracle := \accountlistoracle \cup \mathsf{account}_\issuer\\
         b \sample \{0,1\}, \vec \instruction := ()\\
         (b',t,\state) \gets \adversary^{\mathcal{O}_{\createaccount}}(\ethaddress_{\ledger}, \publicparam)\\
         i := 0 \\
        \text{While } b'  = \emptyset \vee t = \emptyset: \\
        \circ (b',t,\instruction_{i,0}, \instruction_{i,1},\state) \gets \adversary^{\mathcal{O}_{\createaccount}}(\state)\\
        \circ \text{Appends }\instruction_{i,0}, \instruction_{i,1} \text{ to } \vec \instruction  \\
        \circ (\fn_{}, \params_{}) := \instruction_{i,b} \\
        \circ  \verifytx(\mathcal{O}_{\fn}(\params)) \\
        \circ i = i + 1\\
        if\ \mathsf{IsConsistent}(\ledger_{\mathsf{init}},t,\vec \instruction, \instruction_{t,0},  \instruction_{t,1}) \neq 1, \text{abort with } \bot \\
        \text{else return } b=b'
    }
    
\end{adjustbox}

\end{definition}

\begin{definition}[Correctness]
\label{def:correctness}
For all public parameters $\publicparam \gets \ppgen(\secparam)$, 
% all NIZK keys $(\provingkey, \verificationkey) \gets \nizk.\keygen(\mathbb{i},\publicparam)$, 
all account $(\sk, \vk, \ethaddress, \extractorkey, \commitkey_{\ethaddress})
          \gets \createaccount(\secparam)$, all ledger $(\ledger, \allowbreak \ethaddress_{\ledger}) \gets \setup(\ethaddress_{\issuer}, \allowbreak \sk_\issuer, \commitkey_{\issuer})$, all registered account $b_{\registeraccount} \gets  \verifytx(\registeraccount.\local(\sk, \vk, \ethaddress, \commitkey_{\ethaddress}))$ where $b_{\registeraccount}=1$, all minted token $b_{\mint} \gets \verifytx(\tx_\mint)$ where $(\token_\mint, \commitment_\mint, \tx_\mint)\gets \mint.\local(\allowbreak \sk_\issuer, \vk_\issuer,\ethaddress_{\issuer}, \allowbreak \ethaddress_{\receiver}, \val)$ and $b_{\mint}=1$, total minted token values is $<2^{64}$, a token $\token$ is valid if it fulfills one of the following,
\begin{itemize}
    \item the token is the issuer minted token $\token = \token_\mint$,
    \item or the token is accepted by transfer such that $1 \gets \transfer.\verify(\ethaddress_{\receiver},  \allowbreak \vec {\token}_{\original}, \token_{\sender}, \token_{\receiver}, \commitment_{\sender}, \commitment_{\receiver},\pi )$, and $\token = \token_{\sender}\vee \token=\token_{\receiver}$.
\end{itemize}
The valid token $ \token$ belonging to an address $\ethaddress$ that is unspent $(\token, \commitment) \in \ledger.\coinlist[\ethaddress]$\footnote{All token $\token \in \ledger.\coinlist[\ethaddress]$ is a valid token.} when consumed with either of the following, 
it holds that $b_{\burn}=1$ and $b_{\transfer}=1$ with probability $1$.

\begin{itemize}
    \item Transfer: $b_{\transfer} \gets \verifytx(\transfer.\local(\sk, \vk, \extractorkey, \allowbreak \ethaddress, \allowbreak \ethaddress_{\receiver}, \allowbreak \val_{\receiver}))$ where all token used in $\transfer.\local$ is a valid token $\token$.
    \item Burn: $b_{\burn}\gets \verifytx(\burn.\local(\sk, \vk, \extractorkey, \ethaddress, \token, \commitment))$, $\token = \hash(m)$ for $m= \commitscheme_{\extract}.\extract(\extractorkey_{\ethaddress})$, and the token values match with the issuer received event $(\burn, \ethaddress, \allowbreak \token', \commitment')$, $\parseval(m) = \parseval(\commitscheme_{\extract}.\extract(\commitment', \allowbreak \extractorkey_\issuer))$.
\end{itemize}

\end{definition}

\subsection{Security and Privacy Proofs}
\begin{theorem}[Correctness]
$\tokenname$ scheme with algorithms defined in \cref{sec:northstaralgo} satisfies the correctness property from \cref{def:correctness}.
    
\end{theorem}

\begin{proof}
We show the correctness by showing that a valid token $\token$ can be spent via $\transfer$ or $\burn$, and that when it is burned, the value redeemed is consistent with the user's view.
For either case of a valid $\token$ that is minted $\token=\token_{\mint}$ from $\mint$ operation or that it is transferred $\token=\token_{\sender} \vee \token=\token_{\receiver}$ using the $\transfer$ operation, the token $\token$ has a correctly generated accompanying $\commitment$. By the extractable property of the extractable commitment scheme, the token value extracted by the owner is $\val := \parseval(\vec m) = ( \commitscheme_{\extract}.\extract(\commitment,\extractorkey))$, and that $\val \in \valueset$ and $\token = \hash(\vec m)$.

By the correctness of $\DS$, the underlying blockchain signature verification will go through. Since $\transfer.\local$ uses $\vec \token_{\original}\in \ledger.\coinlist[\ethaddress]$, it will pass the verification that checks for token availability. 
We now focus on showing the last condition of verification where the $\transfer.\local$ algorithm can generate a transaction that fulfills the relation $\mathcal{R}_{\transfer}$. 
The algorithm correctly generates $\commitment_{\receiver}$ and thus fulfills the commitment arithmetic relation and the norm bound for $\vec  r_{\receiver}$.
The extracted message $\vec m_i$ fulfills $\token_i = \hash(\vec m_i)$ by the valid token property stated above. 
If the token $\token' \in \vec \token_{\original}$ is minted, then it fulfills $\val'_{i} < 2^{64}$ for $i \in [\numasset]$ since the total minted token values must be $<2^{64}$. Otherwise the token $\token' \in \vec \token_{\original}$ is from the transfer operation. Since the transfer enforces $\mathcal{R}_{\transfer}$ relation, the token $\token'$ values must be $<2^{64}$. 
In the $\transfer.\local$, the sender token's value is computed as  $\val_{\sender} = \sum_{\vec m_i \in \vec M_{\original}}\parseval(\vec m_i)$ - $\val_{\receiver}$, it is equivalent to $\val_{\sender} + \val_{\receiver} = \sum_{\vec m_i \in \vec M_{\original}}\parseval(\vec m_i)$ as needed by the relation as the relation only enforces the first $\numasset$-numbered slots for the quotient polynomial ring similar to $\parseval$.
By the correctness of $\nizk$ proof system, $\nizk.\verify_{\mathcal{R}_{\transfer}}$ must return 1.
Since all the conditions are fulfilled, it must be the case that $b_{\transfer}=1$.

Similar to above, we focus on showing that the $\burn.\local$ algorithm can generate a burn transaction that fulfills the relation $\mathcal{R}_{\burn}$.
Given a valid token $\token \in \ledger.\coinlist[\ethaddress]$, we have that $\token = \hash(\vec m)$ as stated in the valid token property above. Since the algorithm correctly generates the commitment, we can fulfill both the arithmetic relation and the norm bound. Since the committed message for issuer is correctly parsed where the last $N-\numasset$ non-value slots are zeroed out, it fulfills the last remaining conditions for $\mathcal{R}_{\burn}$.
Since all the conditions are satisfied, it must be the case that $b_{\burn}=1$ from the correctness of the $\nizk$ proof system.

Lastly, it can be seen that if the $\mathcal{R}_{\burn}$ relation proof has been verified, then by the relation and the extractable property of the extraction scheme, it must be the case that $\token = \hash(\vec m)$ and that $\val = \parseval(\vec m)$,
while for the issuer
$\val' := \parseval(\vec m') = ( \commitscheme_{\extract}.\extract(\commitment',\extractorkey_{\issuer}))$. 
Since the $\parseval$ algorithm extracts the first $\numasset$-numbered values, it follows that $\val = \val'$ as required. 

\end{proof}

\begin{theorem}[Balance]
$\tokenname$ scheme with algorithms defined in \cref{sec:northstaralgo} satisfies the balance property from \cref{def:balance} in the random oracle model.
\end{theorem}

\begin{proof}
We will show this theorem via a series of small changes made to the experiment.

\gamedes{0} The original balance experiment in \cref{def:balance}.
\\

\gamedes{1} The experiment aborts when there is a collision in $\hash$, the probability of collision is negligible.
\\

\gamedes{2} The experiment aborts when condition II) (Unforgeability) occurs. It can be seen that when the event occurs, it is a signature forgery against the underlying signature sheme $\DS$ and the message has not be queried to the signing oracle before. Therefore, assuming $\DS$ satisfies EUF-CMA, the probability of the abort event happening is negligible. 
\\

\gamedes{3} The experiment aborts when condition III) (Spendability) occurs. 
The extractability of the commitment scheme \cref{def:bdlop_extractable} states that if $\mathcal{R}_{\mathsf{BDLOP}}$ is satisfied (which is also checked by the relation $\mathcal{R}_{\transfer}$), then the extracted message $\vec m_\receiver$ is the committed value. Then, if $\mathcal{R}_{\transfer}$ is satisfied, we have that $\token_\receiver= \hash(\vec m_\receiver)$ and that $\parseval(\vec m_\receiver) \in \valueset$. 
Since it is either that $\token_\receiver \neq \hash(\vec m_\receiver)$ or $\parseval(\vec m_\receiver) \notin \valueset$, running the witness extractor of NIZK returns witness for the proof statement that does not satisfies the relation $\notin \mathcal{R}_{\transfer}$, thus breaking the knowledge-soundness of the NIZK scheme. Therefore, assuming $\nizk$ is knowledge-sound, the probability of the abort event happening is negligible. 
\\

\gamedes{4} The only condition left for the adversary to win the experiment is to redeem more coin values than the total values available to the adversary (Condition I). Since there is no collision (aborted earlier), the adversary must have generated a proof for a statement that does not satisfy $\mathcal{R}_{\burn}$ (which causes over-redeemed burn amount), or for statement that does not satisfy the $\mathcal{R}_{\transfer}$ where the balance $\parseval(\vec m_{input-sum}) \neq  \parseval(\vec m_\sender) + \parseval(\vec m_\receiver)$ does not equal, thus creating more value to be redeemed. 
Therefore, assuming that the $\nizk$ is knowledge-sound, the probability that the adversary wins $\GAME_4$ by passing condition I) (Overdraft-safety) is negligible. 
\end{proof}

\begin{theorem}[Privacy]
$\tokenname$ scheme with algorithms defined in \cref{sec:northstaralgo} satisfies the privacy property from \cref{def:privacy}.
\end{theorem}

\begin{proof}
    
We will show this theorem via a series of small changes made to the experiment.

\gamedes{0} The original privacy experiment in \cref{def:privacy}.
\\

\gamedes{1} The experiment aborts when there is a collision in $\hash$ output, the probability of collision is negligible. 
\\ 

\gamedes{2} 
If the parameters $\params$ for the submitted instruction $\instruction$ contain subset of $(\ethaddress, \ethaddress_\sender, \allowbreak \ethaddress_\receiver,\val)$ that can satisfy the relation $\mathcal{R}_\fn$ for $\fn = \burn/\transfer$, the experiment will now use a simulator to generate the proof without using any witness instead of generating proof with the witnesses. Otherwise for $\params$ that cannot satisfy the relation $\mathcal{R}_{\fn}$, the experiment returns an invalid proof.
Let ($\zksimulator_1, \zksimulator_2, \zksimulator_3$) be the simulator for the zero-knowledge experiment of the underlying proof system.
The experiment when creating transaction $\tx$ for operation $\burn,\transfer$, instead of computing the proof normally using $\nizk.\prove_{\mathcal{R}}(\provingkey,\mathbb{x},\mathbb{w})$, we use the simulator $(\pi,\cdot)\gets \zksimulator_3({\sf st},\mathbb{x})$ to generate the proof where $(\publicparam, {\sf st}) \gets \zksimulator_1(\secpar)$ and $\zksimulator_2(\mathsf{st}, \mathbb{i})$. Since the simulator is only used to generate the proof when the relations hold, this game is indistinguishable from the previous game by the NIZK zero-knowledge property. 
\\

\gamedes{3}
When the commitment $\commitment$ accompanying the token $\token$ is generated by any of the oracles $\mathcal{O}$ for the honest address $\ethaddress \in \accountlistoracle$, it replaces the original message $m$ with a randomly sampled message $m' \sample \mathcal{M}$ and records the function-token-commitment-message tuple ($\burn/\transfer,\token,\commitment,m$) to a list $\mathsf{CommitMessageList}$.
When the experiment needs to extract the value, rather than running $\commitscheme_{\extract}.\extract(\commitment)$ corresponding to a token $\token$, it looks up the commitment value pair from $\mathsf{CommitMessageList}$ and uses $m$ from it instead, the lookup is unique because of the token uniqueness and we aborted upon any random oracle collision earlier. 
This game is indistinguishable from the previous game by the hiding property of the commitment scheme. 

At this point, if the instructions given by the adversary are \textit{consistent}, the adversary gains no information about the bit $b$ before or after the chosen point $t$ since the same instruction is applied regardless of bit $b$ and the observable effect remains the same regardless of bit $b$. The adversary can only distinguish bit $b$ based on $\params_{t,0}$,$\params_{t,1}$ where the difference allowed is the values minted, transferred or burned. Since the probability of random value collision for $\vec r $ used in creating a new token $\token = \hash(\vec v||\vec r)$ by the experiment is negligible and the output from the random oracle $\hash$ is uniformly random conditioned on new input, $\token$ offered no information on the bit $b$ based on the output. Similarly, $\commitment$ created by the experiment for an honest account address is for a random message $m'$, it leaks no information about the bit $b$. Finally, the proof is generated without using the witness bit $b$ at all. Regardless of bit $b$, the same token set is being transferred or burned after being merged. Therefore, it follows that adversary cannot observe any information about bit $b$ and thus the probability of $b'=b$ is $1/2 + \negl$.

\end{proof}

\subsection{Instantiation}
\label{sec:compare_tokens}

We instantiate $\commitscheme_\extract$ with the extractable BDLOP from \cite{LNPS21} as recalled in \cref{def:bdlop_extractable} and $\hash$ with Rescue-Prime from \cite{rescue_prime}. We instantiate $\nizk$ with a STARK-based solution from \cite{ben2018scalable, deepfri, zkstark_habock2024note} with the $\name$ extension described by the \cref{proto:proto_ntt}. 
We first set the common parameters of the system to reach target security levels of $\approx 100$ bits for the proof which is checked online and $\approx 128$ bits of security for offline security where the data remains on-chain for a long period of time. We choose $q=2^{128} - 2^{35} -2^{32} + 1$ to be around $128$ bits. Then, the output of the hash function is then $\mathbb{F}_{128}^2$. We set the number of assets to be $\numasset = 1024-2 = 1022 \approx1000$. 
Consequently, we work over the ring of $R_q=\mathbb{Z}_q/(X^{N}+1)$ where $N=1024$ for the lattice-based commitment scheme and $\mathbb{F}_q$ for the STARK field.

\subsubsection{BDLOP Commitment.}
We need to determine $q,\kappa_\sis,\kappa_\lwe,\beta_{\sis},\chi, n_{message}, N$. $q$ is as determined above which is larger than the usual smaller choices of $64$ bits or $32$ bits for lattice-based schemes. The larger choice of $q$ is chosen to balance the amount of computation on-chain which is a 256-bit EVM machine. 
$n_{message}$ is set to be $1$ since we only need to commit to a singular message ring element $\vec m \in R_q$ per token.
$N=1024$ is as defined above to store up to $1022$ number of assets.
Following the work of \cite{crystal_dilithium}, we set $\chi$ to be a bounded uniform distribution with the range $[-\eta,\eta]$ for some positive $\eta$. Then, we have that $\beta_{\msis} = 2\eta\sqrt{(\kappa_{\sis}+\kappa_{\mlwe}+2n)N}$ as $\norm{\vec r - \vec r'} \leq 2\eta\sqrt{(\kappa_{\sis}+\kappa_{\mlwe}+2n)N}$ and $\vec r \sample \chi^{(\kappa_{\sis}+\kappa_{\mlwe}+2n) N}$.
With these parameters, we run the lattice estimator \cite{lattice_estimator} similar to prior works \cite{matrict_2_2022,lether, qpadl} to reach approximately $128$ bits of security with a target Hermite factor of Hermite factor of $\approx 1.0043$ based on the hardness required for the extractable BDLOP scheme from \cref{def:bdlop_extractable}, and determined $\kappa_\sis = 4$, $\kappa_\lwe=4$, and $\eta = 2^{15}$. With the parameter set, we note  that $ \norm{\vec e^T\vec r}_{\infty}  \leq 2^{44} \ll \sqrt q/4 \approx 2^{62}$ satisfying the requirement of \cref{sec:bdlop_background} for extractability.

% Prime 
\newcommand{\mds}{\mathsf{MDS}}
\newcommand{\roundconstant}{\mathsf{const_{round}}}

\subsubsection{Rescue-Prime Hash Function.} 
% \yeoh{todo}
Let $\mathbb{F}_q$ be the target prime field for which rescue-prime is operating over. We set the security level to be $128$. 
Using algorithms 3-7 from the rescue-prime specification \cite{rescue_prime}, we computed the public parameters MDS-matrix $\mds$,  and round constants $\roundconstant$ as specified in \cite{rescue_prime}. The chosen prime field $\mathbb{F}_q$ satisfies $\mathsf{gcd}(q-1,3)=1$ and thus we can set the power of s-box $\alpha$ to be $3$ and $\alpha^{-1}=\frac{2q-1}{3}$. We set the state width to be $w_\hash$ to be $12$ and the hash rate $r_\hash$ to be $8$. The capacity of the hash function $c_\hash$ is determined as $c_\hash = w_\hash - r_\hash = 4$. The number of rounds $round_\hash = 8$ is chosen based on the previously selected parameters such that the Gröbner basis attack reaches the desired complexity and the hash function $\hash:\mathbb{F}_q^* \rightarrow \mathbb{F}_q^{\mathsf{output}_\hash}$ achieves generically have at least $\log_2(\sqrt q)\cdot \min(\mathsf{output}_\hash,r_p) = 128$ bit of security against finding collisions, preimages and second-preimages as specified in \cite{rescue_prime}. We refer to \cite{rescue_prime} for details on the exact hash algorithm.

\textit{Rescue-Prime constraint.} 
First, note that we encode the $i$-th coefficient of the polynomial ring into the trace polynomial $v(X)$ as $c_i= v(\omega^i)$ where all coefficients $c_i$ are the target hash messages. We further define a selector polynomial 
$f_{abs}(X)$ that evaluates to $1$ when  $X \in \{\omega^{round_\hash \cdot (r- 1)}\}$ for $r \in [N/round_\hash]$ and $0$ otherwise on the remaining set $H$.
Let $\gamma^{\{1,2\}}_{j}(X)$ be a $round_\hash$-th periodic polynomial where $j$ is the $j$-th lane, $(1)$ represents the first half of the round constant and $(2)$ representing the second half. Let $\mathsf{MDS}_{i,j}$ be the MDS constant where $i$ is the $i$-th row and $j$ is the $j$-th column of the original MDS matrix.
Let $s_i(X)$ be the state trace polynomial and $\sigma_i(X)$ be the post-injection state trace polynomial. The set of state trace polynomials is $s=(s_1,\ldots,s_\ell)$ and the set of post-injection state traces $\hat \sigma=(\sigma_1,\ldots,\sigma_{r_p},s_{r_p+1},\allowbreak \ldots,s_{\ell})$. We enforce the following constraint that is designed to minimize the constraint degree,
$$\sigma_i(X)-s_i(X)-f_{abs}(X)\cdot v(\omega^{i-1}X)=0, \forall i \in [r_p];$$
$$ (\sum_{j}^{\ell} \mathsf{MDS}_{i,j}^{-1} \cdot (s_j(\omega X) - \gamma^{(2)}_{j}(X)))^{\alpha}= \sum_{j}^{\ell} (\mathsf{MDS}_{i,j} \cdot (\hat \sigma_j(X)^{\alpha})) + \gamma^{(1)}_{i}(X), \forall i \in [\ell].$$
where we take the next round output from $s_j(\omega X)$ and perform the backward half-round permutation and check the equality against the forward half-round permutation on the current round input from $\widehat \sigma_j(X)$ that has been conditionally mixed with the hash input.
Now, we only need to constrain the hash digest to match, state is initialized correctly using the boundary condition constraint, and the round permutation constraint to exclude the last row.

\subsubsection{$\name$ with ZK-STARK.}
% \yeoh{todo}
We prove the relation of $\mathcal{R}_{\transfer}$ and $\mathcal{R}_{\burn}$ by encoding the relation checks as AIR traces in a straightforward manner. The interesting part is that we commit to both the NTT representation and the coefficient representation simultaneously so that polynomial multiplication can be enforced using element-wise multiplication with $O(N)$ operations rather than performing the NTT computation in-circuit. The coefficient representation is here to assist in checking the relation $\norm{r}_{\infty} \leq \eta$. 
We then use $\name$ protocol to enforce the relation between coefficient representation and NTT representation.
The polynomial ring arithmetic specified in relation of $\mathcal{R}_\transfer$ and  $\mathcal{R}_\burn$  is checked using simple element-wise multiplication and addition since they are available as NTT representation. 
We use a standard base decomposition technique with table-lookup via the Logup \cite{habock2022multivariate} technique to enforce the 64-bit range for the value. In particular, we use a $2^{8}$-sized table and decompose the 64-bit into $8$ smaller representations where it is enforced that they recomposed back to the original value and is strictly smaller than $2^{8}$ (by table lookup). Similarly for the bounded $\norm{r}$, we first add $\eta$ to the $r$ coefficients so that centered representation is shifted to a bounded positive check only. Since $\eta=2^{15}$ and after the shift the coefficients must be $[0,2^{16})$ bounded, we can now perform the same byte decomposition and perform lookups on the $2^{8}$ table using LogUp. Rescue-Prime constraint design is as specified in the above. We set the parameters for the STARK scheme using conjecture following the line of \cite{cryptoeprint:2026/532}.
Therefore, we set the blow-up factor $2^\beta=16$ with $\beta$ to 4, the query count to $23$ and grinding to $2^{23}$ for a targeted security level of  $\approx100$ which is a common target for proof systems \cite{ethstark, stwo}.

\subsubsection{Encoding of Value in the NTT form.} 
For convenience and clarify of the construction, we describe our scheme by encoding the $\val$ in the standard coefficient form. However, it is slightly more efficient for the value $\val$ to be in the NTT form since we do not need to separately commit to $\vec m$ in both coefficient and NTT forms which save another invocation of $\name$ as well. That is instead of performing  as $\commitment \gets \commitscheme_{\extract}.\commit(\vec m,\vec r)$ where $\vec m := (\vec v || \vec \mu)$ and $\mu \sample \mathbb{Z}_q^{N-\numasset}$, we replace $\vec m$ with $\ntt^{-1}(\vec v||\vec \mu)$ and consequently the range check happens over the NTT domain-encoded values. Similarly for other interpretations of the value, it must first be converted back into the NTT domain. Note that this differs from the random value used in the commitment scheme, where the coefficient form is needed to check the norm and the NTT form is needed for efficient arithmetic. 

\subsubsection{Smart Contract.} 
Due to contract-size limitations, we split the smart contract such that a wrapper  smart contract manages the state and delegate the verification of proofs against the state to secondary smart contract(s) which contain the main verification logic for the proof.
For optimizing gas usage, the smart contract stores both the LDE merkle hash tree digest and the per-party commit key $\commitkey_{\ethaddress}$ as $\ledger.\commitkey\mathsf{digest}[\ethaddress] := \hash_{LDE}(\commitkey_\ethaddress)$ and $\ledger.\commitkey[\ethaddress] := \commitkey_{\ethaddress}$ where $\hash_{LDE}$ is defined to be the Merkle tree hash of the LDE evaluation of the interpolated polynomial of the input. The verification checks the equality of the proof's commit key hash digest used against the registered digest. The implementation does not check that the commit key is well-formed as it does not affect the security property (balance) and only affects the privacy of the registering party's received token value while the other honest party's privacy is preserved regardless of it. Note that any recipient can always simply publish its token balance to unmask itself even if the commit key is correctly formed.
Since the $\commitment_\sender$ is not verified directly by the verification function, we drop it as part of the main transaction verification to save on gas consumption. For retrieval convenience, the sender can separately send this as another transaction so that it can retrieve the data from the blockchain data if needed from the historical block data.  

\subsubsection{Evaluation.} 
We evaluate our token proposal using a proof-of-concept implementation on an EVM-based testnet for the $\transfer$ operation since the burn operation is cheaper than $\transfer$. Since the EVM internally uses 256-bit words, we choose our prime field mod $q$ to be closer to $128$ bits rather than $64$ bits or smaller such that the challenges are sampled from $128$-bit field rather than $64$-bit quadratic extension field.
For geth-based \cite{geth} node operators which has a size limit of 128 KB for the transaction mempool, we can simply split the verification call to be able to be processed by geth-based node operators. We evaluate $\tokenname$ on the Ethereum Hoodi testnet (Fusaka) and report the total gas consumed to be about $18.11$ million gas units for the operation $\transfer$ with one input coin and two output coins. The sender $\commitment$ storage is omitted in the $\transfer$ and it is assumed that the sender keeps track of its sender $\commitment$ locally or using any other off-chain storage. At the time of writing, the gas price is $0.132$ Gwei per gas unit, and for the  which is about $0.0024$ ETH and costs about $6$\footnote{Using the price of 2500 USD per ETH pricing at the time of writing.} USD in total for $\approx1000$ asset confidential transaction. We benchmark the most computationally intensive operation for the prover that is the time to generate the transaction's proof and it takes about $491$ms to generate the proof on a M5 Macbook Pro machine.

\begin{table}[ht!]
\begin{center}
% \rowcolors{1}{}{lightgray}
% \begin{adjustbox}{minipage=\linewidth,scale=1}
\begin{tabular}{r|l|r|r|r|r}
  \hline
  Work  & System & Asset No. & Gas & Gas per Asset & Max Value
  \\
  \hline
  Zether\cite{bunz2020zether} & \makecell[l]{Account-based,DDH,\\Sigma Protocol}& 1 & 7.19M & 7.19M& $2^{32}$\\
  \hline
  $\tokenname$ (This) & \makecell[l]{UTXO-like, Lattice,\\$\name$, STARK} & $1022$ & 18.11M & 0.02M & $2^{64}$ \\
   \hline
\end{tabular}
% \end{adjustbox}
\end{center}
\caption{Comparison of the Most Computational Intensive Operation $\transfer$ with existing work. }
\label{tab:compare_tokens_main}
\end{table}

As a point of reference, we now compare $\tokenname$ against Zether \cite{bunz2020zether}, a single-asset account-based confidential transfer with anonymity extension. The results are compared against the basic confidential Zether over the precompiled supported $\mathbb{G}_1$ group from the BN-128 curve without the anonymity extension as reported in \cite{bunz2020zether}. The gas comparison is for rough reference only as the gas cost might have changed between the upgrade fork in \cite{bunz2020zether} when compared to used testnet with Fusaka upgrade. The comparisons results are summarized at \cref{tab:compare_tokens_main}. Zether is able to achieve $7.19$M gas used per asset in a transfer transaction while supporting up to maximum of $2^{32}$ value. Zether's construction relies on the DDH assumption which is vulnerable to quantum computer attacks. In contrast, the $\tokenname$ relies on lattice-based and hash-based assumptions which are both assumed to be post-quantum secure. In addition, $\tokenname$ uses less gas per asset when $\numasset \geq 3$ and supports a larger range of $[0,2^{64})$ rather than only 32-bits in asset value.  

\textbf{Summary.} We show that the $\tokenname$ is able to support multi-asset transaction with confidentiality preserved for thousand(s) of assets while costing similar to that of single asset DDH-based scheme Zether \cite{bunz2020zether}. The practical confidential multi-asset transactions facilitated by $\tokenname$ enables institutional multi-asset use cases such as repo, DvP, and bond markets as discussed in \cite{qpadl,zhu2025towards}. 

\subsection{$\tokenname$ Discussion}
The main objective of $\tokenname$ is to provide a confidential mutli-asset ERC-20 token that is based on post-quantum primitives and demonstrate that it is feasible to be deployed with the current blockchain infrastructure. However, the current blockchain infrastructure lacks support for post-quantum signature, the current construction lacks input/output anonymity, and there is an opportunity for further verifier optimization.

\subsubsection{Advantage of $\tokenname$ Deployment with Current Blockchain Infrastructure.}
It can be seen that the spending authorization is tied with the $\DS$ scheme of the underlying blockchain while all other property (including balance) is based on post-quantum primitives. While it is possible to use knowledge of a preimage for a hashed-based public key to remove the signature's reliance on the underlying blockchain infrastructure, the underlying consensus protocol also needs to be post-quantum secure so that an adversary cannot maliciously modify the $\ledger$ state. Despite this, the $\tokenname$ still offers protection against harvest-now-decrypt-later \cite{newhouse2023pqc} style of attacks while simultaneously offering better performance when the number of assets in the transaction is large enough when compared to a classical scheme like Zether.

\subsubsection{Extension to Input/Output Anonymity.} 
Output-account anonymity can be introduced by sending transactions to decoy output accounts (sending a value of zero). Input-account anonymity can be provided using a one-out-of-many proof \cite{Groth2015} or the encrypted table-based ledger approach presented in \cite{qpadl}. Asset anonymity is provided out-of-the-box by $\tokenname$ and can be scaled up to thousand(s) of asset types while achieving a performance profile comparable to that of the current confidential single-asset transaction schemes.

\subsubsection{Replacing Extractable Commitment with Verifiable Encryption.}
We note that it is sufficient but not necessary to use an extractable commitment to produce $\commitment$ as we do not make use of the commitment binding property and we rely instead on $\hash$ to bind the value with $\token$. Therefore, it should be permissible to replace an extractable commitment scheme with an IND-CPA-secure verifiable encryption scheme. 

\subsubsection{Recursive SNARK Approach.} The verification can be made more efficient by using the recursive SNARK approach \cite{recursive_snark}. However, it incurs additional prover time when compared to a non-recursive approach. $\tokenname$ is efficient enough to be run without any recursion and can be made more verifier efficient when recursion is used at the expense of prover's computational time.

\section{Deferred Proofs}
\label{app:proof}

\subsection{Proofs for Univariate Sumcheck}
\begin{proof}[\cref{lemma:sum_over_h_constant}]
We have that $f \in \mathbb{F}[X]$ be a degree $\leq |H|-1$ polynomial, and we are going to show that $\sum_{h_i \in H} f(h_i)=|H| \cdot f(0)$. Let $f(X) = \sum_{j=0}^{|H|-1}a_jX^j$ and by definition $\{h_i\}_{i\in [ \ |H| \ ]}  = \{\omega^i\}_{i\in [ \ |H|\ ]}$ and $h_i=\omega^i$. It can be seen that

$$\sum_{h_i \in H} f(h_i) = \sum_{h_i \in H} \sum_{j=0}^{|H|-1}a_j h_i^j = \sum_{j=0}^{|H|-1}a_j \sum_{\omega^i \in H} (\omega^i)^j.$$

For the case $j=0$, we have
$$a_0 \sum_{\omega^i \in H} (\omega^i)^0= a_0 \cdot |H|=f(0) \cdot |H|.$$

For the case $1\leq j \leq |H|-1$, we have the following via geometric sum
%  \sum_{i \in [\ |H|\ ]} (\omega^j)^i = 
$$\sum_{\omega^i \in H} (\omega^i)^j = \sum_{i \in [\ |H|-1 \ ]\cup \{0\}} (\omega^j)^i = \frac{1-(\omega^j)^{|H|}}{1-\omega^j} = \frac{1-(\omega^{|H|})^{j}}{1-\omega^j}=0$$
where $\omega^{|H|}= \omega^0=1$ and $\omega^j\neq 1$ since $1\leq j \leq |H|-1$. Applying the above, we have $\forall j \in [1,|H|-1] ,a_j \sum_{\omega^i \in H} (\omega^i)^j= 0$. Thus, summing using the result from both cases, we have $\sum_{h_i \in H} f(h_i)=|H| \cdot f(0)$.
% Using geometric sequence
\end{proof}

\begin{proof}[\cref{lemma:uni_sumcheck}]
Let the polynomial divisor be $Z_H(X)$ with $\deg(Z_H) = |H|$.
By the polynomial division algorithm, there exists a unique decomposition of the form $f(X)= Q(X)Z_H(X) + r(X)$ where $\deg(Q) \leq d-|H|$ and $\deg(r) \leq |H|-1$. Write $r(X)$ as $r(X)=R(X)X+c$ to obtain the required form $Q(X)Z_H(X) + R(X)X + c$ where $\deg(R)\leq |H|-2$. 
Note that we have  $$\beta = \sum_{h_i \in H} f(h_i)= \sum_{h_i \in H} (Q(h_i)Z_H(h_i) + r(h_i)) = \sum_{h_i \in H} r(h_i)$$ since $Q(X)Z_H(X)$ vanishes over $H$.  Since $\deg(r)\leq |H|-1$, we apply \cref{lemma:sum_over_h_constant} to yield $\sum_{h_i \in H} f(h_i) =  \sum_{h_i \in H} r(h_i) =|H| \cdot r(0)$ and consequently $\frac{\beta}{|H|}=r(0)=c$.
Thus, we have $c= \frac{\beta}{|H|}$ as required.

If $R(X) \in \mathbb{F}[X]$ and $Q(X) \in \mathbb{F}[X]$, such that $f(X)= \frac{\beta}{|H|} + R(X)X + Q(X)Z_H(X)$ where $\deg(R) \leq |H|-2$ and $\deg(Q) \leq d - |H|$, then $$\sum_{h_i \in H} f(h_i)= \sum_{h_i \in H}\frac{\beta}{|H|} + R(X)X + Q(X)Z_H(X) = \sum_{h_i \in H}r(X)+ 0 = |H|\cdot \frac{\beta}{|H|}=\beta$$ where $r(X)= \frac{\beta}{|H|} + R(X)X$.

\end{proof}

\subsection{Proof for \cref{theo:main_general}}

\begin{proof}
\textbf{Completeness.} 
The completeness proof proceeds similarly to that of the \cref{theo:main_nongeneral}. Let
$\bigl( (M,N,H,\psi, \{\varphi_j^{\coeff}, \varphi_j^{NTT}\}_{j \in [M+1]}, \polyoracle{g_{\ntt}}, \polyoracle{g_{\coeff}}) \allowbreak ; (g_{\ntt}, g_{\coeff})\bigr) \in \mathcal{R}^{\batchgeneral}_{\nttcoeffrelation}$, and thus  $\forall i \in [N], j \in [M],  g_{\ntt}(h_{\varphi^{\ntt}_j(i)}) = \sum_{k \in [N]} g_{\coeff}(h_{\varphi^{\coeff}_j(k)})\cdot(\psi^{2(i-1)+1})^{k-1};|H| \geq M\cdot N$. Recall that for $\forall j\in [M],g_j(X) = \sum_{i=0}^{N-1} c_{j,i}X^{i} \in \mathbb{Z}_q[X]/(X^N+1)$, 
we have the encoding $\forall j \in [M],i \in [N], g_{\coeff}(h_{\varphi^{\coeff}_j(i)})= c_{j,i-1}$.

By definition, NTT representation for the $j$-th polynomial $g_j(X)$ can be written as $\vec g_{j,\ntt}=(\widehat g_{j,1},\cdots,\widehat g_{j,N})=(g_{\ntt}(h_{\varphi_j^{\ntt}(1)}),\ldots, g_{\ntt}(h_{\varphi_j^{\ntt}(N)}))$ where $\widehat g_{j,i} = g_{j}(\psi \phi^{i-1})= g_j(\psi^{2(i-1)+1})= \sum_{k=0}^{N-1} c_{j,k}(\psi^{2(i-1)+1})^{k} = \sum_{k \in [N]} g_{\coeff}(h_{\varphi_j^{\coeff}(k)})\cdot(\psi^{2(i-1)+1})^{k-1}$. Now, we show that $\sum_{h_i \in H} g_{\coeff}(h_i)\cdot \sigma^{\batchgeneral}_{\alpha,\eta,\varphi^{\coeff}}(h_i) - g_{\ntt}(h_i)\cdot \Lambda^{\batchgeneral}_{\alpha,\eta,\varphi^{\NTT}}(h_i) =0$ below.

\begin{align*}
& \sum_{h_i \in H} [ g_{\coeff}(h_i)\cdot \sigma^{\batchgeneral}_{\alpha,\eta,\varphi^{\coeff}}(h_i) - g_{\ntt}(h_i)\cdot \Lambda^{\batchgeneral}_{\alpha,\eta,\varphi^{\NTT}}(h_i) ]   \\
&= \sum_{i \in [n_H]}  g_{\coeff}(h_i)\cdot \sigma^{\batchgeneral}_{\alpha,\eta,\varphi^{\coeff}}(h_i) - \sum_{i \in [n_H]} g_{\ntt}(h_i)\cdot \Lambda^{\batchgeneral}_{\alpha,\eta,\varphi^{\NTT}}(h_i) \\
&= \sum_{j \in [M+1]} \Bigg[\sum_{i \in I_{j}^{\coeff}} g_{\coeff}(h_i)\cdot \sigma_{\alpha,\eta,\varphi^{\coeff}}(h_i) - \sum_{i \in I_{j}^{\NTT}} g_{\ntt}(h_i)\cdot \Lambda_{\alpha,\eta,\varphi^{\NTT}}(h_i) \Bigg]\\
&=  \sum_{j \in [M]} \Bigg[ \sum_{i \in [N]} c_{j,i-1} \cdot \eta^{j-1} \cdot  \alpha^{i-1} - \sum_{i \in [N]}  [\sum_{k \in [N]} g_{\coeff}(h_{\varphi^{\coeff}_{j}(k)})\cdot(\psi^{2(i-1)+1})^{k-1}] \cdot \eta^{j-1} \cdot \ell_{\psi \phi^{i-1}}(\alpha) \Bigg]\\
&\quad + \bigg[\sum_{i \in I_{M+1}^{\coeff}} g_{\coeff}(h_i)\cdot 0 - \sum_{i \in I_{M+1}^{\NTT}} g_{\ntt}(h_i)\cdot 0\bigg] \\
&=  \sum_{j \in [M]} \eta^{j-1} \Bigg[ \sum_{i \in [N]} c_{j,i-1}  \cdot  \alpha^{i-1} - \sum_{i \in [N]}  g_j(\psi^{2(i-1)+1}) \cdot \ell_{\psi^{2(i-1)+1}}(\alpha) \Bigg]\\
&= \sum_{j\in [M]} \eta^{j-1} (g_j(\alpha) - g_j(\alpha))  = 0.
\end{align*}

We now apply \cref{lemma:uni_sumcheck} and the above result, and since the polynomials on both side are equal, the verifier will always accept at the end of execution when $|H|\geq M\cdot N>1$. 

\textbf{Knowledge Soundness.}  It is sufficient to show $\delta$-soundness since we can apply \cref{remark:knowledge_soundness}, and the protocol will be $\delta$-knowledge sound as required. 

When $\bigl( (M,N,H,\psi, \{\varphi_j^{\coeff}, \varphi_j^{NTT}\}_{j \in [M+1]}, \polyoracle{g_{\ntt}}, \polyoracle{g_{\coeff}}) \allowbreak ; (g_{\ntt}, g_{\coeff})\bigr) \notin \mathcal{R}^{\batchgeneral}_{\nttcoeffrelation}$, we have that $\exists j\in [M],\exists i \in [N],\widehat g_{j,i} \neq g_j(\psi^{2(i-1)+1})$ where $(\widehat g_{j,1},\cdots,\widehat g_{j,N})=(g_{\ntt}(h_{\varphi^{\ntt}_j(1)}),\ldots, g_{\ntt}(h_{\varphi^{\ntt}_j(N)}))$ since $|H|\geq M\cdot N >1$ is checked directly. 

Note that the derivation from \textit{Completeness} proof is now $\sum_{j\in[M]} \eta_j (g_j(\alpha)-g'_j(\alpha)) = 0$ where $g'(X)$ is the polynomial interpolated in the general case. 
Since there exists at least a point $X=\psi^{2(i-1)+1}$ and a $j$-th for some $i \in [N], j\in[M]$ where $\widehat g_{j,i} =g'_j(\psi^{2(i-1)+1}) \neq g_j(\psi^{2(i-1)+1})$ 
% and $f'(\psi^{2(i-1)+1})$
, therefore $g_j(X) \neq g'_j(X)$. 
We now show that $g_j(\alpha)-g'_j(\alpha)$ is zero with negligible probability. Let $\bar g_j(X)=g_j(X)-g_j'(X)$ be a non-zero polynomial, we now apply Schwartz-Zippel (\cref{lemma:schwartz-zippel}) to obtain that $\Pr[\bar g_j(\alpha)=0] \leq \frac{N-1}{|\mathbb{F}|}$ where $\alpha \sample \mathbb{F}$. Taking union bound, we have that $\Pr[\exists j, \bar g_j(\alpha)=0] \leq \frac{M(N-1)}{|\mathbb{F}|}$.

We now show the probability of canceling out with $\eta$ factor. Let $P_{\alpha}(X)$ be the non-zero polynomial defined by the $\sum_{j\in [M]}  (g_j(\alpha) - g'_j(\alpha))X^{j-1}$. Then on random $\eta\sample \mathbb{F}$ point evaluation, $\Pr[P_{\alpha}(\eta)=0] \leq \frac{M-1}{|\mathbb{F}|}$ by \cref{lemma:schwartz-zippel}.

From \cref{lemma:uni_sumcheck}, we can rewrite $g_{\coeff}(h_i)\cdot \sigma^{\batchgeneral}_{\alpha,\eta,\varphi^{\coeff}}(h_i) - g_{\ntt}(h_i)\cdot \Lambda^{\batchgeneral}_{\alpha,\eta,\varphi^{\NTT}}(h_i) = R(X)X + Q(X)Z_H(X)$ where $\deg(R)\leq |H|-2$ and $\deg(Q) \leq |H|-2$ iff $\sum_{h_i \in H} [g_{\coeff}(h_i)\cdot \sigma^{\batchgeneral}_{\alpha,\eta,\varphi^{\coeff}}(h_i) - g_{\ntt}(h_i)\cdot \Lambda^{\batchgeneral}_{\alpha,\eta,\varphi^{\NTT}}(h_i)]=0$. 
Let $P_{LHS}(X)$ be the polynomial at the left hand side and $P_{RHS}(X)$ be the polynomial at the right hand side. We use Schwartz-Zippel lemma to conclude that $\Pr[P_{LHS}(z) - P_{RHS}(z)=0: P_{LHS}(X) \neq P_{RHS}(X)  \wedge z \sample \mathbb{F}] \leq \frac{ 2|H|-2}{|\mathbb{F}|}$.
% $\delta = \frac{N-1}{|\mathbb{F}|}$ 
Thus, it follows that $\protocol_{\nttcoeffrelation}$ is knowledge sound for $\delta = \frac{(2|H|-2) + (M-1) + (M(N-1))}{|\mathbb{F}|}=\frac{2|H| + MN -3}{\mathbb{F}}$ using union bound as required.

\end{proof}

\end{document}